\pdfoutput=1

\documentclass{raa}
\usepackage[pass,paperwidth=8.27in,paperheight=11.69in]{geometry}
\usepackage{graphicx,times}
\usepackage{amssymb,amsmath}
\usepackage{natbib}
\usepackage{a4pdf}

\bibpunct{(}{)}{;}{a}{}{,}

\begin{document}

\title{Comparing the behavior of orbits in different 3D
dynamical models for elliptical galaxies}

 \volnopage{ {\bf 2012} Vol.\ {\bf 12} No. {\bf 4}, 383--399}
   \setcounter{page}{383}

\author{Euaggelos E. Zotos}

\institute{Department of Physics, Section of Astrophysics, Astronomy and Mechanics, Aristotle University of Thessaloniki 541 24, Thessaloniki, Greece; {\it evzotos@astro.auth.gr}\\
\vs\no
{\small Received~~2011 June 13; accepted~~2012~~January 9}}

\abstract{ We study the behavior of orbits in two different
galactic dynamical models, describing the motion in the central
parts of a triaxial elliptical galaxy with a dense nucleus.
Numerical experiments show that both models display regular motion
together with extended chaotic regions. A detailed investigation
of the properties of motion is made for the 2D and 3D Hamiltonian
systems, using a number of different dynamical parameters, such as
the Poincar\'{e} surface of {a }section, the maximal Lyapunov
Characteristic Exponent, the $S(c)$ spectrum, the $S(w)$ spectrum
and the $P(f)$ indicator. The numerical calculations suggest that
the properties of motion in both potential{s} are very similar.
Our results show that one may use different kind{s} of
gravitational potentials in order to describe the motion in
triaxial galaxies {while }obtaining quantitatively similar
results.  }

\authorrunning{Euaggelos E. Zotos}
\titlerunning{Comparing the Behavior of Motion in 3D Dynamical Models}

\maketitle

\section{Introduction}
\label{sect:intro}

About forty years ago, the prevailing view was that
elliptical galaxies were oblate spheroids flattened by rotation (see
\citealt{Sandage+etal+1970}).
It was the pioneer{ing} work of
\cite{Bertola+Capaccioli+1975} and
\cite{Illingworth+1977} that led astronomers to abandon the
assumption that elliptical galaxies are necessarily oblate.

It is well known that observations of elliptical galaxies
yield only the projected isophotes and, thus, determination of their
intrinsic shapes requires statistical analysis, based on large
samples (see \citealt{Ryden+1996}; \citealt{Alam+Ryden+2002};
\citealt{Vincent+Ryden+2005}) or mapping of potentials via detailed
kinematical data for individual galaxies (see
\citealt{Davies+etal+2001}; \citealt{Rest+etal+2001};
\citealt{Statler+etal+2004}).

Today it is believed that the shapes of elliptical galaxies are
prolate or triaxial rather than oblate (see
\citealt{Benacchio+Galletta+1980}; \citealt{Binggeli+1980};
\citealt{Alam+Ryden+2002}). On the other
hand, kinematical studies of elliptical galaxies show
evidence that triaxial galaxies do exist. Moreover{,} observational
data indicate that most of the triaxial elliptical galaxies
host a black hole or a dense nucleus  in their centers (see
\citealt{Bak+Statler+2000};
\citealt{Statler+etal+2004}). On this basis, we believe that it
would be of interest to investigate the dynamical properties of a
triaxial elliptical galaxy{,} particularly in its central region.

In order to describe the motion in the triaxial elliptical galaxy
we use the well known logarithmic potential
\begin{equation}
V_{\rm g}=\frac{\upsilon _0^2}{2} \ln\left[x^2+ay^2+bz^2+c_{\rm
b}^2\right] \, ,
\end{equation}
where $\upsilon _0$ is used for the consistency of the galactic
units, $a$ and $b$ are flattening parameters, {and} $c_{\rm b}$ is
the scale length of the bulge component (see
\citealt{Binney+Tremaine+2008}). Expanding potential (1) in{ a}
Taylor series about the origin and keeping terms up to the fourth
degree in the variables{,} we find
\begin{equation}
V_{\rm g} = \frac{\upsilon _0^2}{2}\ln c_{\rm b}^2 +
\frac{\upsilon _0^2}{2c_{\rm b}^2}\left(x^2+ay^2+bz^2\right)
-\frac{\upsilon _0^2}{4c_{\rm
b}^4}\left(x^2+ay^2+bz^2\right)^2=\upsilon _0^2 \ln c_{\rm b} +
V_{\rm l}\,.
\end{equation}
Thus the polynomial potential $V_{\rm l}$ is
\begin{equation}
V_{\rm l} = \frac{\upsilon _0^2}{2c_{\rm
b}^2}\left(x^2+ay^2+bz^2\right) - \frac{\upsilon _0^2}{4c_{\rm
b}^4}\left(x^2+ay^2+bz^2\right)^2  \, ,
\end{equation}
where it was assumed that
\begin{equation}
\frac{x^2+ay^2+bz^2}{c_{\rm b}^2} \ll 1 \, .
\end{equation}
The reader can find more details about the Taylor expansion of the
logarithmic potential in \cite{Zotos+2011b}. To potentials (1) and
(3) we add the potential of a spherically symmetric nucleus
\begin{equation}
V_{\rm n}=\frac{-M_{\rm n}}{\sqrt{x^2+y^2+z^2+c_{\rm n}^2}} \, ,
\end{equation}
where $M_{\rm n}$ is the {nuclear }mass,
while $c_{\rm n}$ is the scale length of the nucleus. We apply a
Plummer sphere, in order to increase the central mass of the galaxy.
This method has been applied several times in previous works, having
as an objective to study the effects of the introduction of a
central mass component in a galaxy (see \citealt{Hasan+Norman+1990};
\citealt{Hasan+etal+1993}).

The aim of the present article is to investigate the
properties of motion near the center of a triaxial elliptical galaxy
described by potentials (1) and (3) with an additional dense nucleus
described by potential (5). In particular, we are interested to
study the regular or chaotic character of motion in both{ of} the above
described potentials, in order to be able to compare the
corresponding results. Furthermore, we shall compare the density in
the central parts of the triaxial galaxy derived using the two
potentials{ described above}. In order to
achieve a better picture for the properties of motion, we
first investigate the 2D system, that is when $z=0$, and
then we will use the corresponding results to study the dynamical
system of three (3D) degrees of freedom.

From the pioneer{ing} work of \cite{Henon+Heiles+1964} there {has
been an ongoing} interest in finding new methods, in order to
distinguish between ordered and chaotic motion in dynamical systems.
The Poincar\'{e} surface of section (PSS) for the two dimensional
(2D) systems and the Lyapunov Characteristic Exponent (LCE)
(\citealt{Benettin+etal+1976}; \citealt{Froeschle+1984};
\citealt{Lichtenberg+Lieberman+1992}) for dynamical systems with any
degree of freedom are two well known methods to characterize an
orbit as regular or chaotic. Over the last thirty years, an effort
has been made in order to find new, modern and also reliable and
fast ways to detect the chaotic behavior in galactic systems. One
could mention the frequency map analysis developed by Laskar
(\citealt{Laskar+etal+1992}; \citealt{Laskar+1993}), the dynamical
spectra of stretching numbers (the distribution of values of a given
parameter along the obit), introduced by \cite{Froeschle+1984} (see
also \citealt{Froeschle+etal+1993};
\citealt{Voglis+Contopoulos+1994}; \citealt{Contopoulos+etal+1995};
\citealt{Contopoulos+etal+1997}) and the $P(f)$ spectral method
applied and used by \cite{Karanis+Vozikis+2008}. In the present
research, we use{,} apart from the classical {PSS} technique and the
LCE, some modern methods such as the $S(c)$ and $S(w)$ dynamical
spectra and the $P(f)$ indicator.

Here, we must provide some additional theoretical information
regarding these new dynamical methods. We use the $S(c)$ spectrum in
order to characterize the nature of an orbit in a 2D dynamical
system. This spectrum has been proved{ to be} a very reliable tool
in several cases (see \citealt{Caranicolas+Papadopoulos+2007};
\citealt{Zotos+2011a}). The nature of a 2D orbit can be revealed by
looking at the shape of the $S(c)$ spectrum. If the shape of the
spectrum is a well defined $U$-type structure, then the
corresponding orbit is regular. On the other hand, if the shape is
complicated and highly asymmetric, with a lot of large and small
abrupt peaks, then the orbit is chaotic. Moreover, the $S(c)$
spectrum can help us identify resonant orbits of higher
multiplicity, as it produces {as many} $U$-type structures as the
total number of islands of the invariant curves on the $x-p_x$ phase
plane. One more advantage of this spectrum is that it can be
deployed in order to calculate the sticky period of a 2D orbit and
also to follow its time evolution towards the chaotic sea (see
fig.~5 in \citealt{Zotos+2011a}).

For the study of 3D orbits, we use the $S(w)$ spectrum. By
definition this spectrum is based on a complicate{d} combination of
the coordinates and the momenta of the 3D orbit. In particular, this
spectrum is an advanced form of the $S(c)$ spectrum and therefore
carries all the characteristics mentioned in the previous paragraph
regarding the pattern of the spectrum for regular and chaotic
orbits. The only difference is that in this case the $S(w)$ spectrum
produces as {many} $U$-type structures as the total number of
invariant 3D tori in the $\left(x, p_x, z\right)$ phase space. We
introduced this new spectrum definition in \cite{Zotos+2011a}, in
order to construct a new spectral definition appropriate for the
study of 3D orbits.

The Fourier Transform is usually defined as a transformation of a
quantity $q$ which is a function of time, $q(t)${, with regard} to
its respective function of amplitude $p$, which is a function of
frequency $P(f)$. We can define a series of time intervals between
successive crossing{s} over a section. Then we calculate the Power
Spectrum of these time intervals, using a Discrete Fast Fourier
Transformation (FFT) algorithm. Looking at the $P(f)$ spectrum of a
regular 2D or 3D orbit, we expect to observe a smooth curve, with
some additional peaks corresponding to the ``periodicities" of the
time series. On the contrary, in a 2D or 3D chaotic orbit no such
``periodicities" exist and therefore its $P(f)$ spectrum will
produce a very ``noisy" pattern with a large number of peaks which
would be very densely distributed. One of the main advantages of
this spectral method is that is uses only one orbit and we do not
need to trace the behavior of any nearby orbit. Furthermore, the
detection can be made quite early using less iterations, compared to
the iterations needed to reach a conclusive result using the LCE.
More detailed information regarding this method and its applications
can be found in \cite{Karanis+Vozikis+2008}.

Here we must remind the reader that the $S(c)$ spectrum is the
distribution function of the parameter $c$
\begin{equation}
S(c) = \frac{\Delta N(c)}{N \Delta c} \, ,
\end{equation}
where $\Delta N(c)$ are the numbers of the parameters $c$ in the
interval $\left(c, c + \Delta c \right)$ after $N$ iterations. The
parameter $c$ is defined~as
\begin{equation}
c_i = \frac{x_i - p_{xi}}{p_{yi}} \, ,
\end{equation}
where $\left(x_i, p_{xi}, p_{yi} \right)$ are the successive values
of the $\left(x, p_x, p_y \right)$ elements of the 2D orbits, on the
Poincar\'{e} $x-p_x, y=0, p_y>0$ phase plane. More details regarding
the $S(c)$ spectrum and its applications can be found in
\cite{Caranicolas+Papadopoulos+2007}{ and}
\cite{Caranicolas+Zotos+2010}.

The Hamiltonian corresponding to the potential (1) or (3) is
written{ as}
\begin{equation}
H=\frac{1}{2}\left(p_x^2+p_y^2+p_z^2\right) + V_{\rm
t}\left(x,y,z\right)=E \, ,
\end{equation}
where $V_{\rm t}$ represents $V_{\rm tg}=V_{\rm g}+V_{\rm n}$ or
$V_{\rm tl}=V_{\rm l}+V_{\rm n}$. Here $p_x,p_y${ and }$p_z$ are the
momenta per unit mass conjugate to $x,y$ and $z${ respectively},
while $E$ is the numerical value of the Hamiltonian (8), which is
conserved.

In this article, we use a system of galactic units, where the unit of length is 1kpc, the unit of mass is $2.325 \times 10^7 M_{\odot}$ and the unit of time is $0.97748 \times 10^8$ yr. The velocity unit is 10~km~s$^{-1}$, while $G$ is equal to unity. In the above units we use the values: $\upsilon _0 = 10, c_{\rm b} = 3, M_{\rm n} = 10, c_{\rm n} = 0.1, a = 1.5$ and $b = 1.7$.

The results of the present research are based on the numerical integration of the equations of motion
\begin{eqnarray}
\ddot{x} &=& - \frac{\partial V_{\rm t}(x,y,z)}{\partial x} \, , \nonumber \\
\ddot{y} &=& - \frac{\partial V_{\rm t}(x,y,z)}{\partial y} \, , \nonumber \\
\ddot{z} &=& - \frac{\partial V_{\rm t}(x,y,z)}{\partial z} \, ,
\end{eqnarray}
which was made using a Bulirsh-St\"{o}er routine in Fortran 95, with
double precision in all subroutines. The accuracy of the
calculations was checked by the constancy of the energy integral
(8), which was conserved up to the eighteenth significant figure.

This paper is organized as follows. In Section~2 we present and compare the results for the 2D systems. In Section~3 we compare the mass density near the center derived using the two 3D potentials. Moreover, we compare the properties of the 3D orbits in the two dynamical models. In Section~4, the conclusions and the discussion of our results are presented.

\section{Results for the 2D dynamical systems}
\label{sect:res}

\begin{figure}[h!!!]

 \centering

\includegraphics[width=80mm,angle=-90]{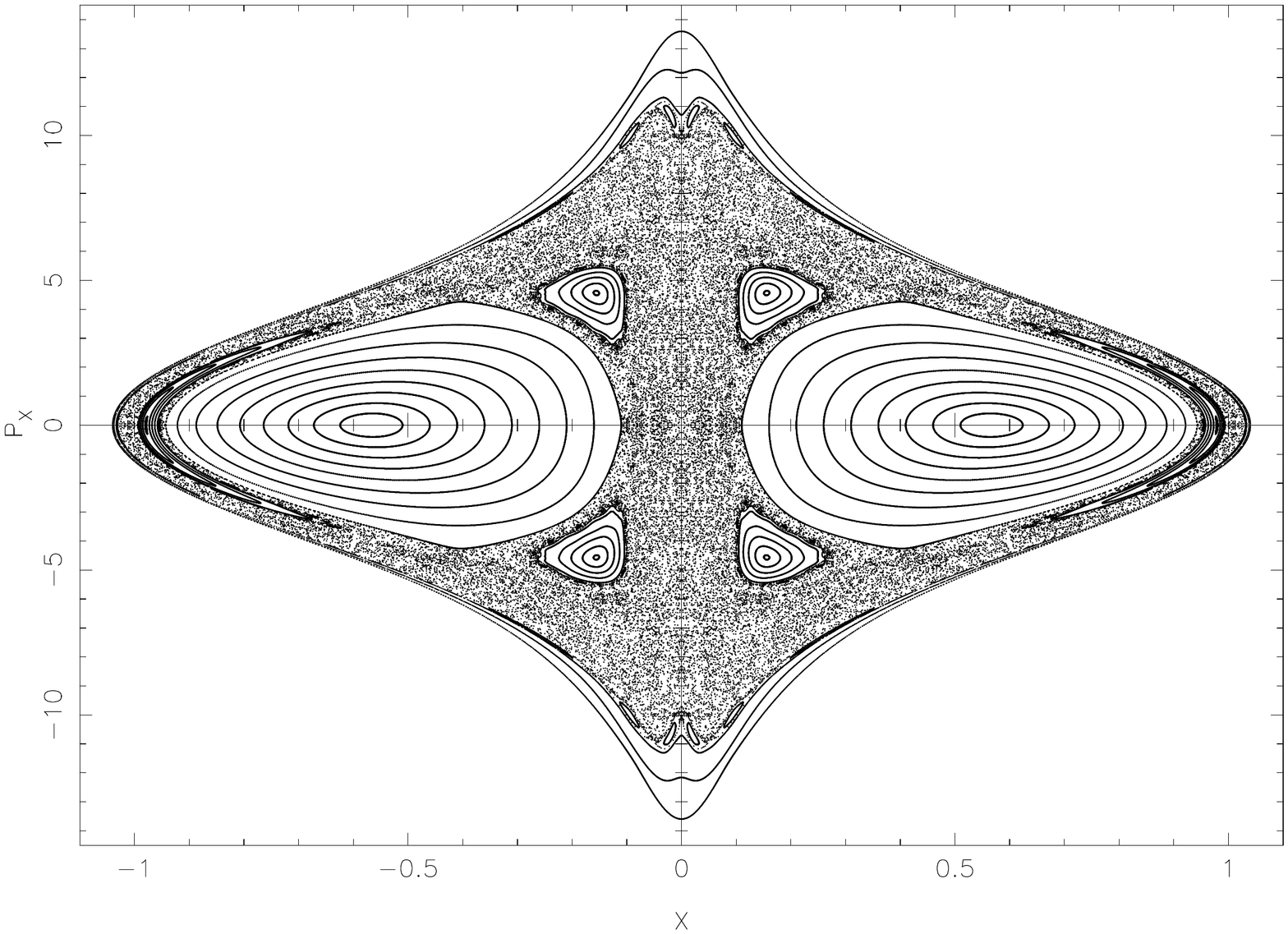}

\vspace{-3mm}

\caption{\baselineskip 3.6mm
The
 $x-p_x$ phase plane for the potential $V_{\rm tg}$,
when $\upsilon _0=10, c_{\rm b}=3, M_{\rm n}=10, c_{\rm n}=0.1,
a=1.5, b=1.7$ and $E_{\rm 2tg}=105.16$. \label{fig1}}
\end{figure}

\begin{figure}

 \centering

\includegraphics[width=80mm,angle=-90]{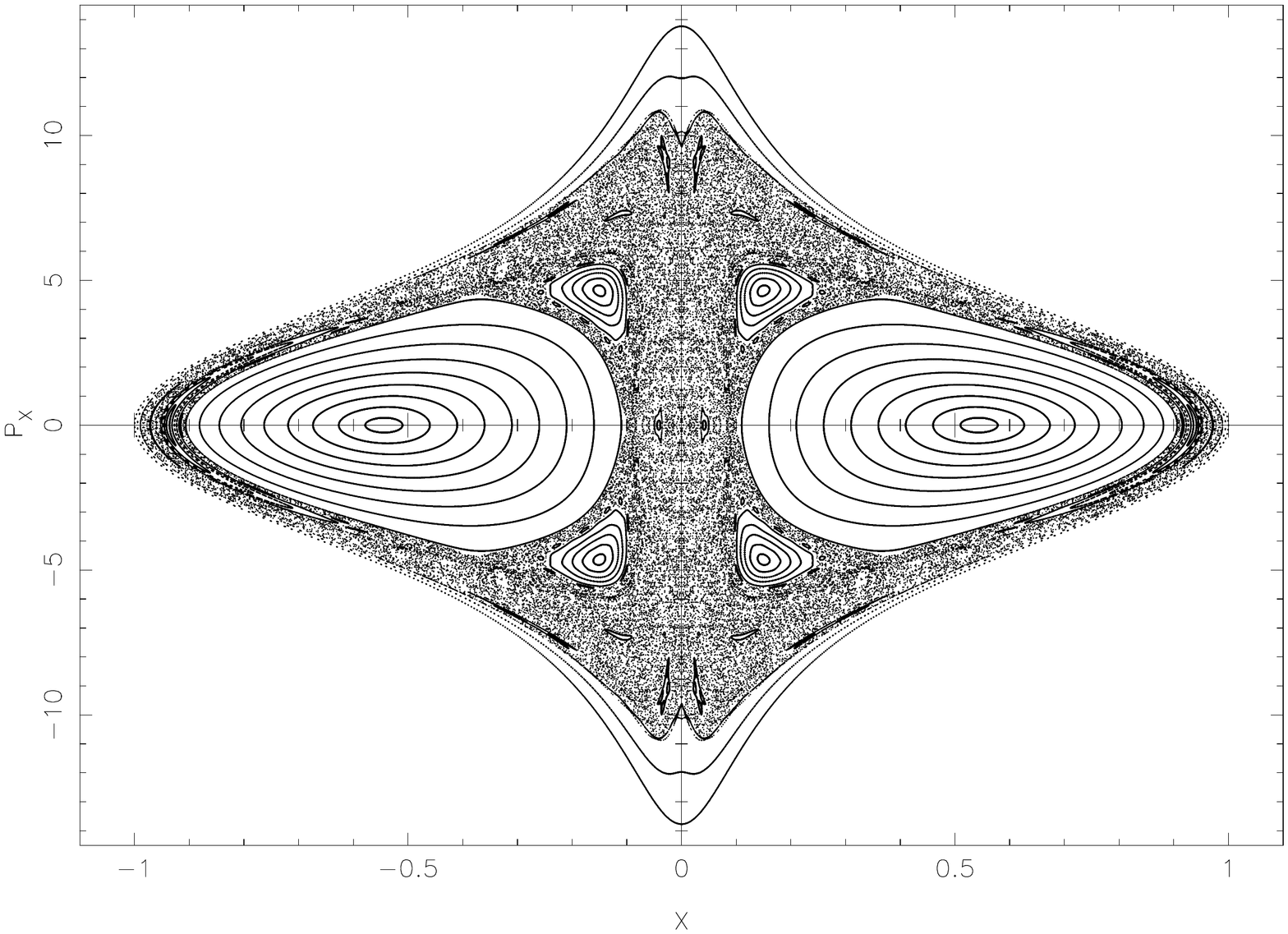}

\vspace{-3mm}
\begin{minipage}{120mm}
\caption{Similar {to} Fig.~\ref{fig1},  but for the potential
$V_{\rm tl}$. The value of energy is $E_{\rm
2tl}=-4.70$.\label{fig2}}\end{minipage}
\end{figure}

In this section, we study the character of orbits in the 2D
dynamical systems. In this case, we set $z=p_z=0$ in (8) and the
corresponding 2D Hamiltonian is
\begin{equation}
H_2=\frac{1}{2}\left(p_x^2+p_y^2\right) + V_{\rm
t}\left(x,y\right)=E_2 \, ,
\end{equation}
where $E_2$ is the numerical value of the Hamiltonian. As the
phase space of the system is four dimensional we use the $x-p_x,
y=0, p_y>0$ Poincar\'{e} phase plane. The results are presented in
Figures~\ref{fig1} and \ref{fig2}. 

Figure~\ref{fig1} shows the phase plane for potential $V_{\rm
tg}$, while Figure~\ref{fig2} shows the phase plane for potential
$V_{\rm tl}$. Here, we must emphasize that the two phase planes were
constructed for values of energies connected by the relation
\begin{equation}
E_{\rm 2tg}=E_{\rm 2tl}+\upsilon _0^2 \ln c_{\rm b}
\end{equation}
and the same initial conditions, in order to be able to make the
comparison. Here we {took} $E_{\rm 2tl}=-4.70, \upsilon _0^2 \ln
c_{\rm b}=109.86$, which gives the value 105.16 for $E_{\rm 2tg}$.
As one can see, the two phase planes are almost identical. In both
{f}igures, we see areas of regular motion and extended chaotic
regions. There are three main families of regular orbits. (i) Orbits
producing  invariant curves around each of the two stable 1:1
resonant periodic points. (ii) Orbits producing a set of three
islands - one of them is on the $x$-axis, while the other two are
symmetric with respect to the $p_x$ axis. These orbits are
characteristic of the 3:3 resonance. (iii) Box orbits producing
invariant curves surrounding the{ whole} chaotic sea. Note that the
area on the $x-p_x$ phase planes occupied by each of the above
families of regular orbits is quantitatively the same in both
Figures~\ref{fig1} and \ref{fig2}. In addition to the regular orbits
there are also a large number of irregular orbits producing a large,
unified chaotic sea. Note that the extent of the chaotic sea is
almost the same in both phase planes. The differences between the
two phase planes produced by potentials $V_{\rm tg}$ and $V_{\rm
tl}$ are  negligible and they are confined to some tiny islands,
embedded inside the chaotic sea. These tiny islands are produced by
secondary resonances.


In order to investigate and compare in detail the properties of
motion in both potentials, we present and compare in the
following a number of orbits belonging {to} different families of
orbits.

\begin{figure*}[h!!]

\vs \centering

\resizebox{0.8\hsize}{!}{\rotatebox{0}{\includegraphics*{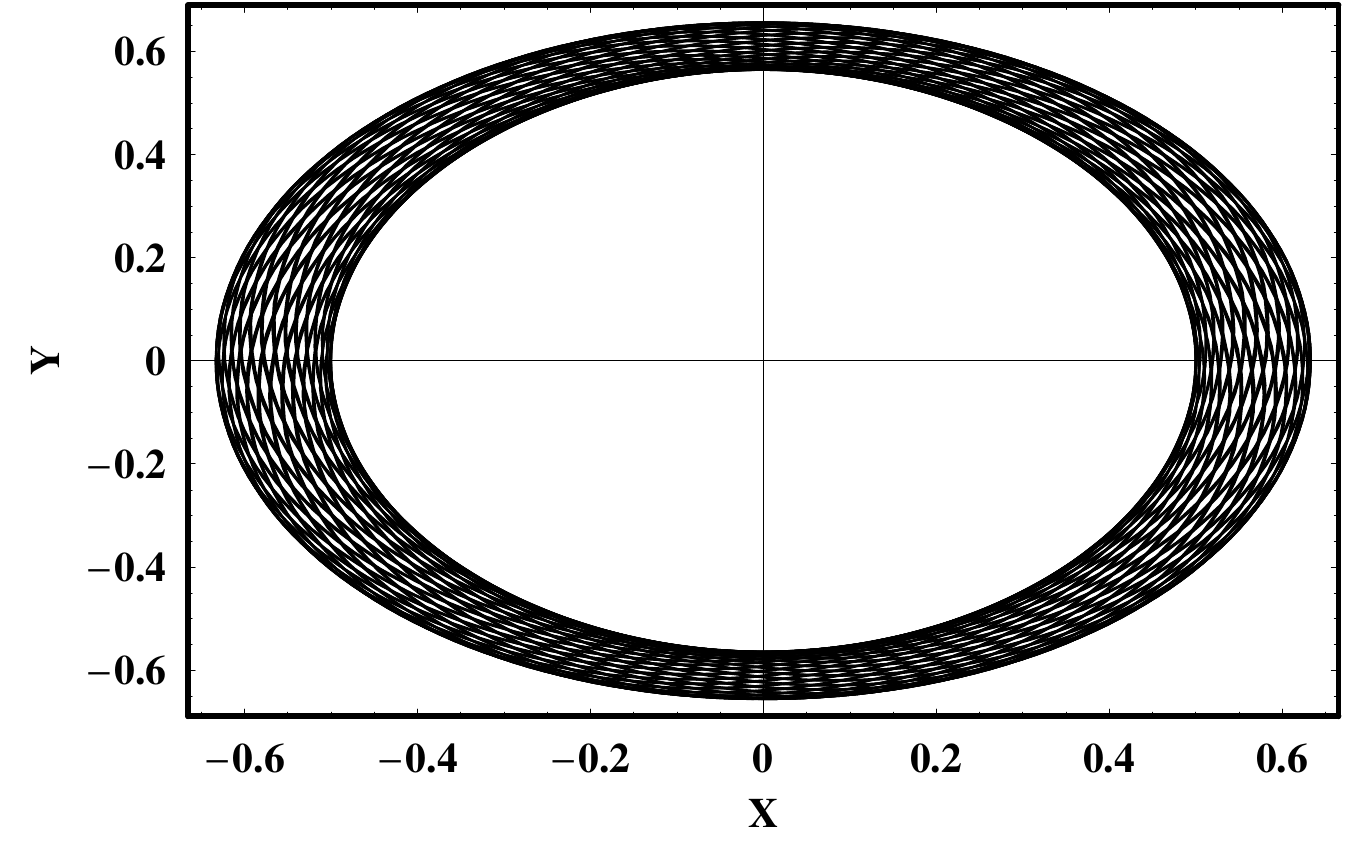}}
                         \rotatebox{0}{\includegraphics*{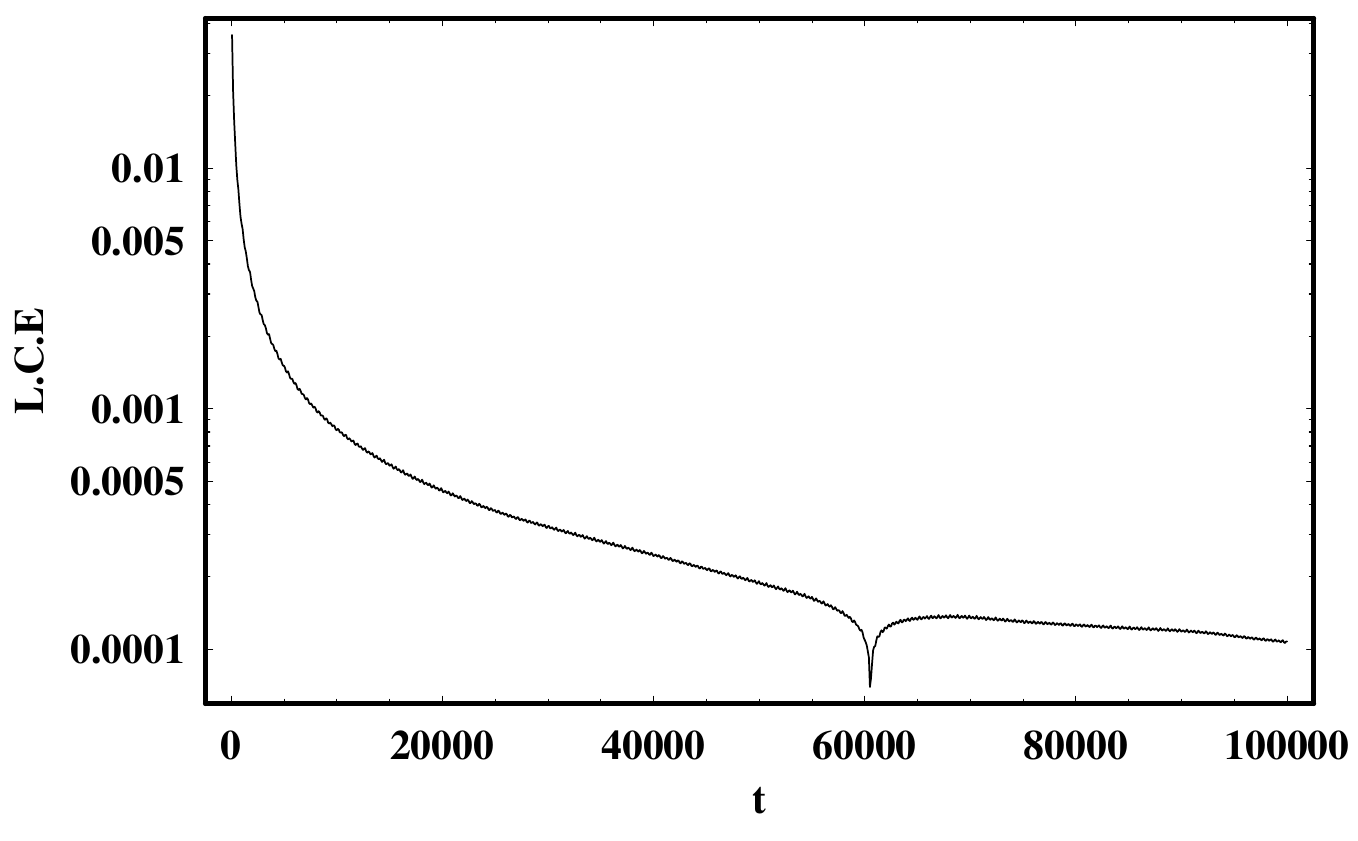}}}

\begin{minipage}{30mm}

\hspace{15mm}
 {\fns(a)}
\end{minipage}\hspace{25mm}
\begin{minipage}{30mm}
\hspace{16mm}{\fns(b)}\hs\hs\end{minipage}

\vs

\resizebox{0.8\hsize}{!}{\rotatebox{0}{\includegraphics*{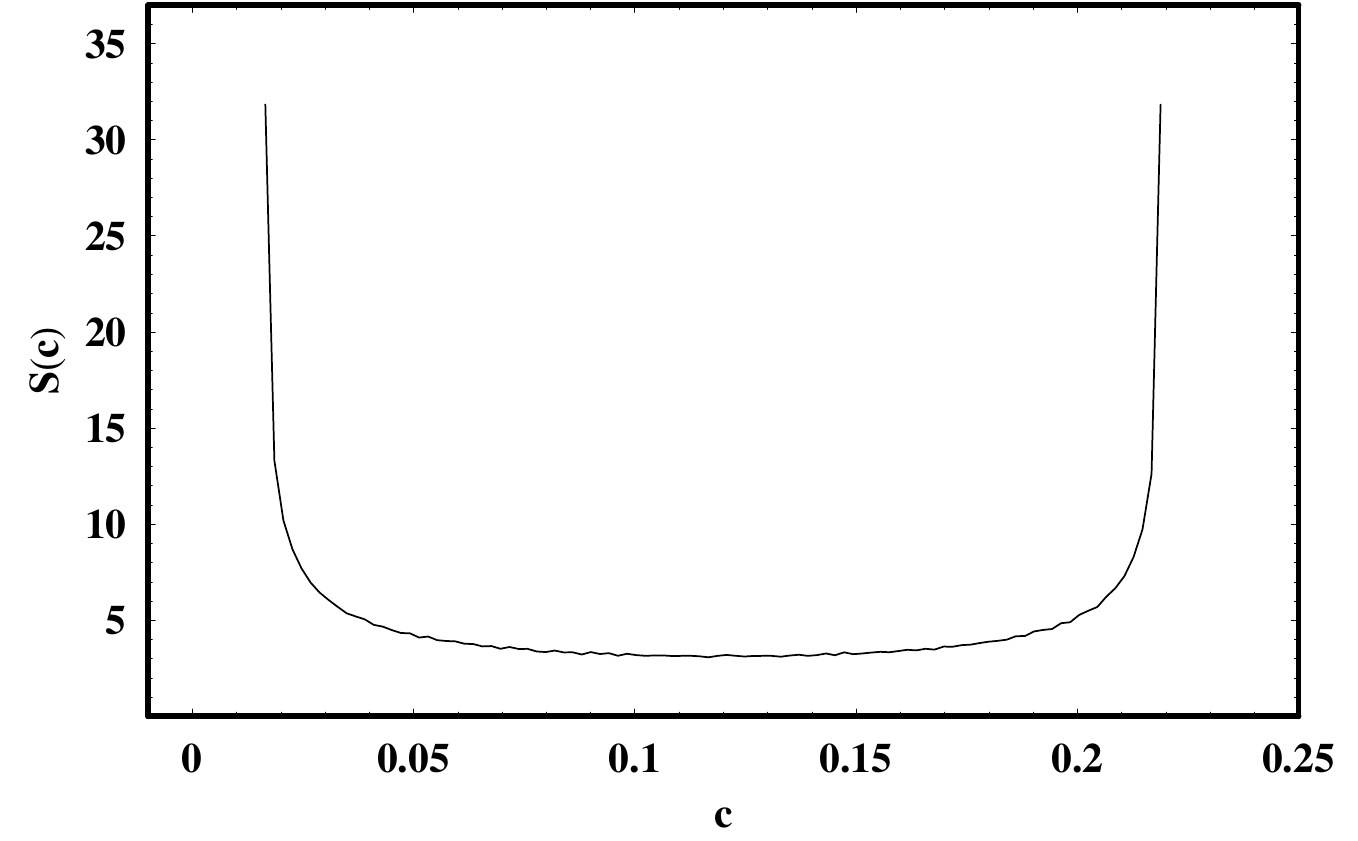}}
                         \rotatebox{0}{\includegraphics*{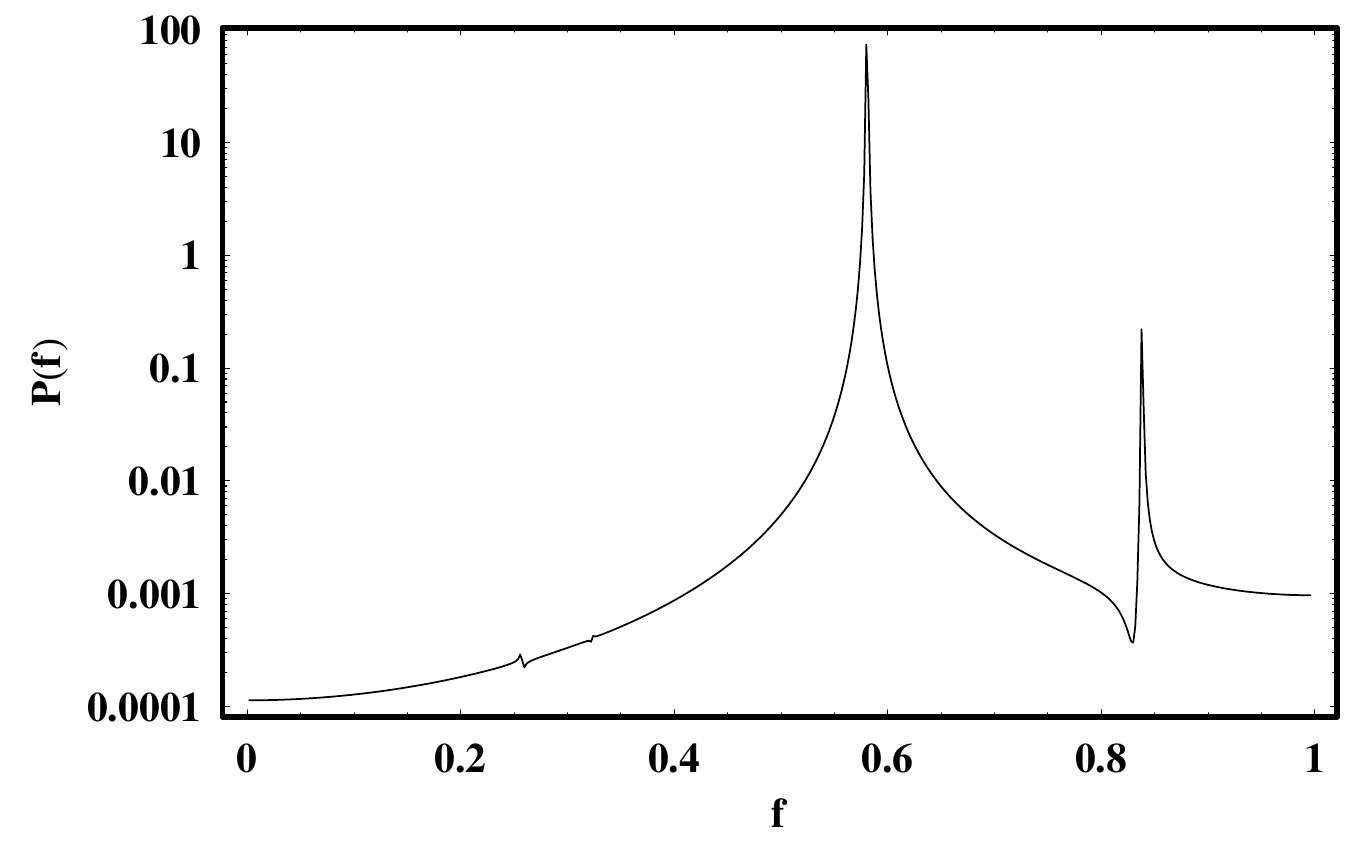}}}

\begin{minipage}{30mm}

\hspace{14mm}
 {\fns(c)}
\end{minipage}\hspace{25mm}
\begin{minipage}{30mm}
\hspace{18mm}{\fns(d)}\hs\hs\end{minipage}

\vspace{-4mm}
 \caption{\baselineskip
3.6mm 
(a) A regular orbit in the 2D potential $V_{\rm tg}$. Initial
conditions are: $x_0=0.5, y_0=0,p_{x0}=0$, while $p_{y0}$ is found
from the energy integral. The values of all other parameters and
energy are as in Fig.~\ref{fig1}. 
(b) A plot of the maximum LCE vs. time for the orbit shown in (a).
(c) 
 The $S(c)$ spectrum of the orbit
shown in (a) and (d) 
The $P(f)$ indicator for the orbit shown in (a).\label{fig3}}

\vs  \centering
\resizebox{0.8\hsize}{!}{\rotatebox{0}{\includegraphics*{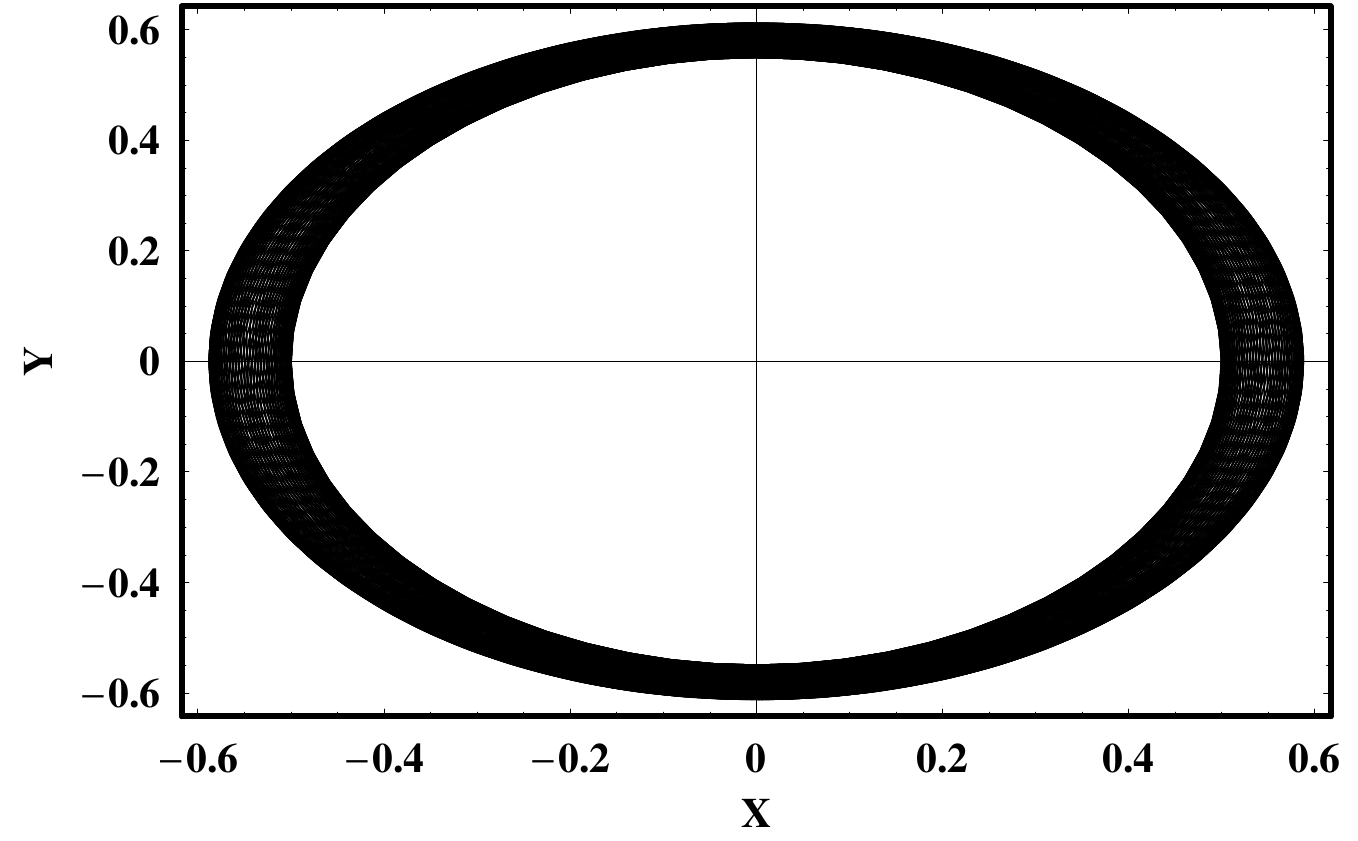}}
                         \rotatebox{0}{\includegraphics*{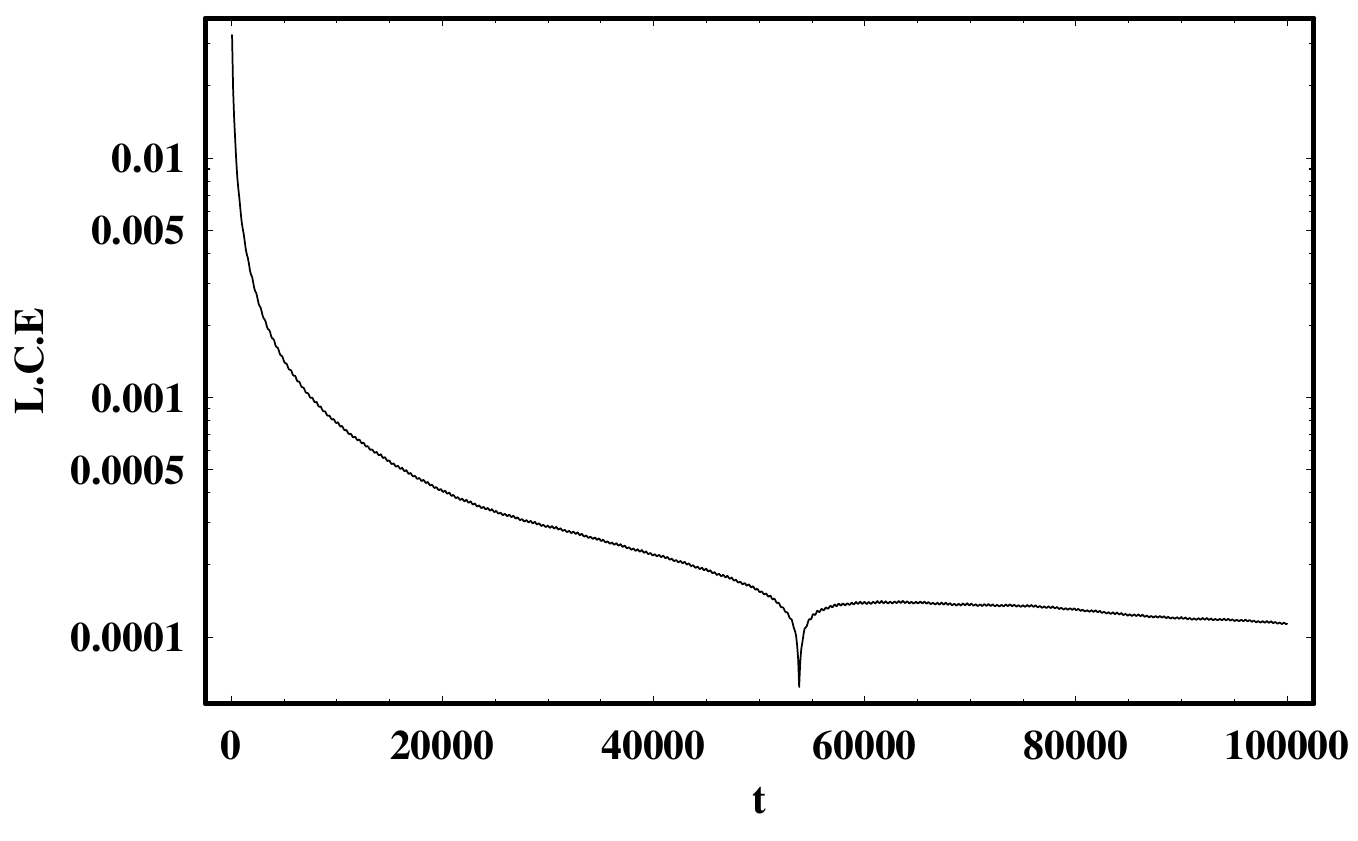}}}

\begin{minipage}{30mm}

\hspace{15mm}
 {\fns(a)}
\end{minipage}\hspace{25mm}
\begin{minipage}{30mm}
\hspace{16mm}{\fns(b)}\hs\hs\end{minipage}

\vs
\resizebox{0.8\hsize}{!}{\rotatebox{0}{\includegraphics*{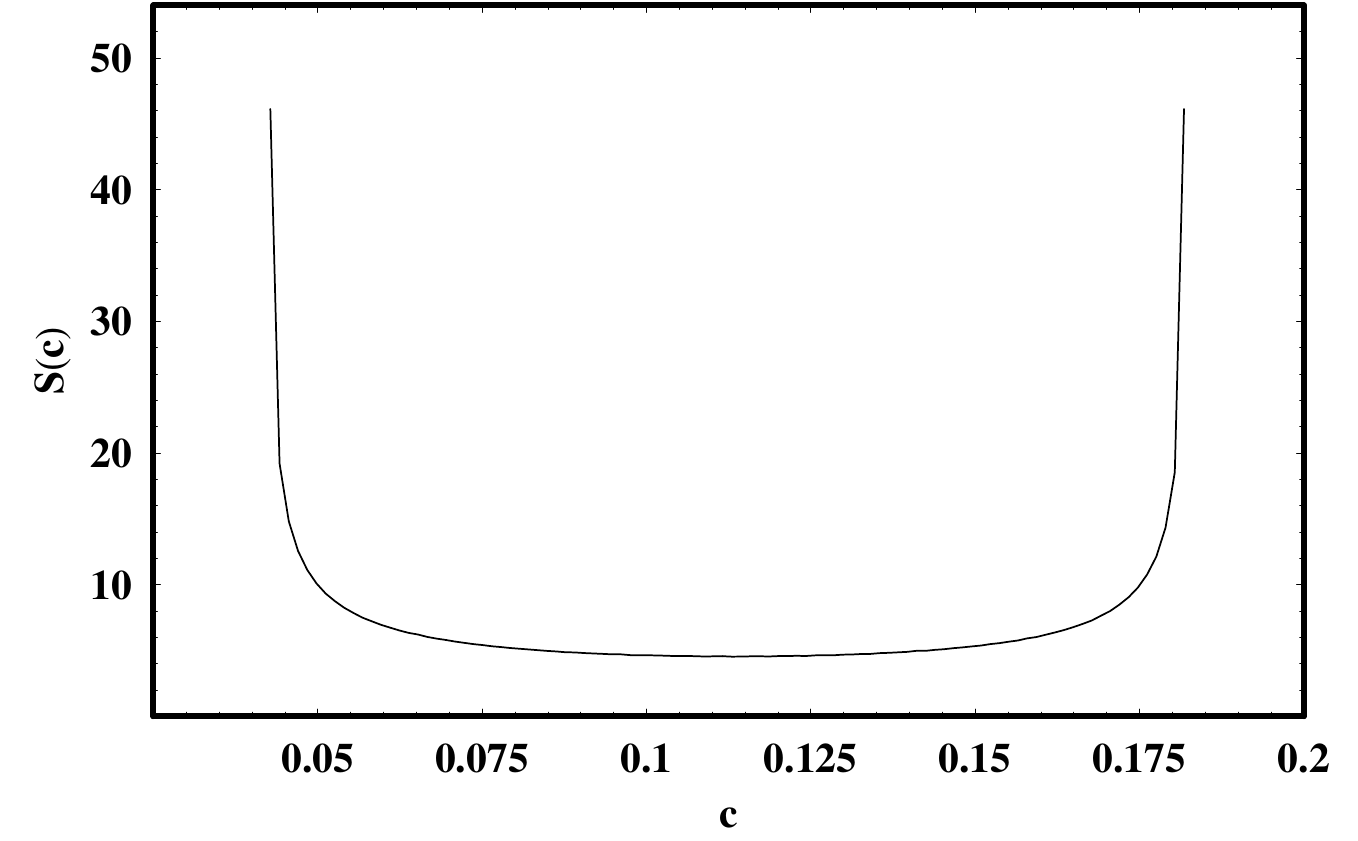}}
                         \rotatebox{0}{\includegraphics*{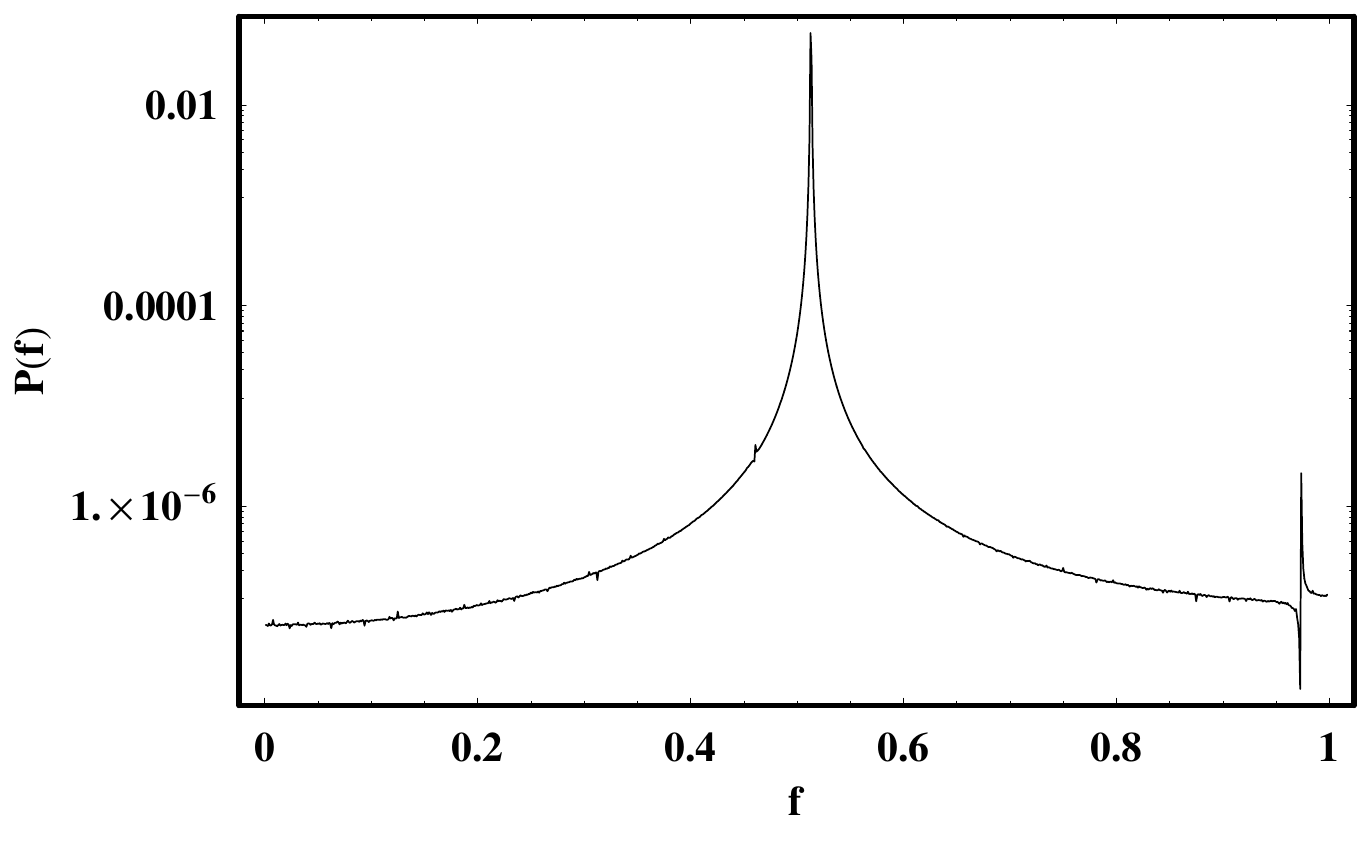}}}

\begin{minipage}{30mm}

\hspace{14mm}
 {\fns(c)}
\end{minipage}\hspace{25mm}
\begin{minipage}{30mm}
\hspace{18mm}{\fns(d)}\hs\hs\end{minipage}

\vspace{-2mm}

\caption{\baselineskip 3.6mm (a)--(d): Similar {to}
Fig.~\ref{fig3}(a)--(d) for the potential $V_{\rm tl}$. The values
of all other parameters and energy are {the same }as in
Fig.~\ref{fig2}.\label{fig4}}
\end{figure*}

Figure~\ref{fig3}(a)--(d) 
shows results for a regular orbit in potential $V_{\rm tg}$. The
orbit shown in Figure~\ref{fig3}(a) belongs to family (i) and has
initial conditions: $x_0=0.5, y_0=0, p_{x0}=0$, while in all cases,
$p_{y0}$ is found from the energy integral (10). The corresponding
values of energy and all the other parameters are as in
Figure~\ref{fig1}. Figure~\ref{fig3}(b) shows the maximum LCE of the
orbit, which vanishes indicating regular motion.
Figure~\ref{fig3}(c) shows the $S(c)$ spectrum of the orbit. Here,
we see a well defined $U$ type spectrum characteristic of the
regular motion. In Figure~\ref{fig3}(d), we see a plot of the $P(f)$
indicator, which displays only two peaks indicating regular motion.
In order to help the reader, we note that the orbit shown in
Figure~\ref{fig3}(a) was calculated for a time period of 100 time
units. The time scale for the $S(c)$ spectrum and the $P(f)$
indicator was $10^3 - 10^4$ time units.

Figure~\ref{fig4}(a)--(d) 
  shows results for an
orbit, with the same initial conditions and with the same time
scales for all calculations, but for the potential $V_{\rm tl}$.
All other parameters are the same as in Figure~\ref{fig2}.

\begin{figure*}

\vs \centering
\resizebox{0.82\hsize}{!}{\rotatebox{0}{\includegraphics*{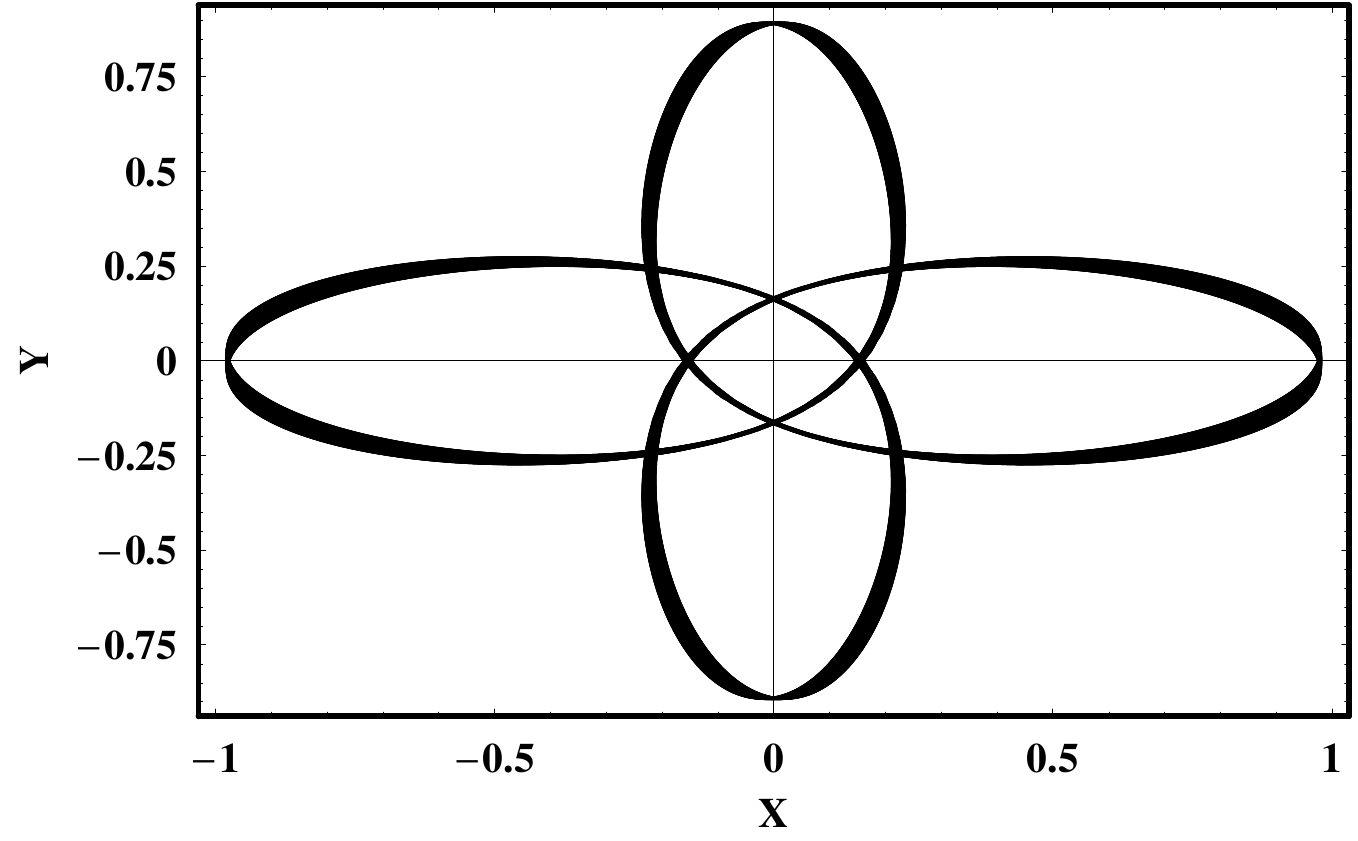}}
                          \rotatebox{0}{\includegraphics*{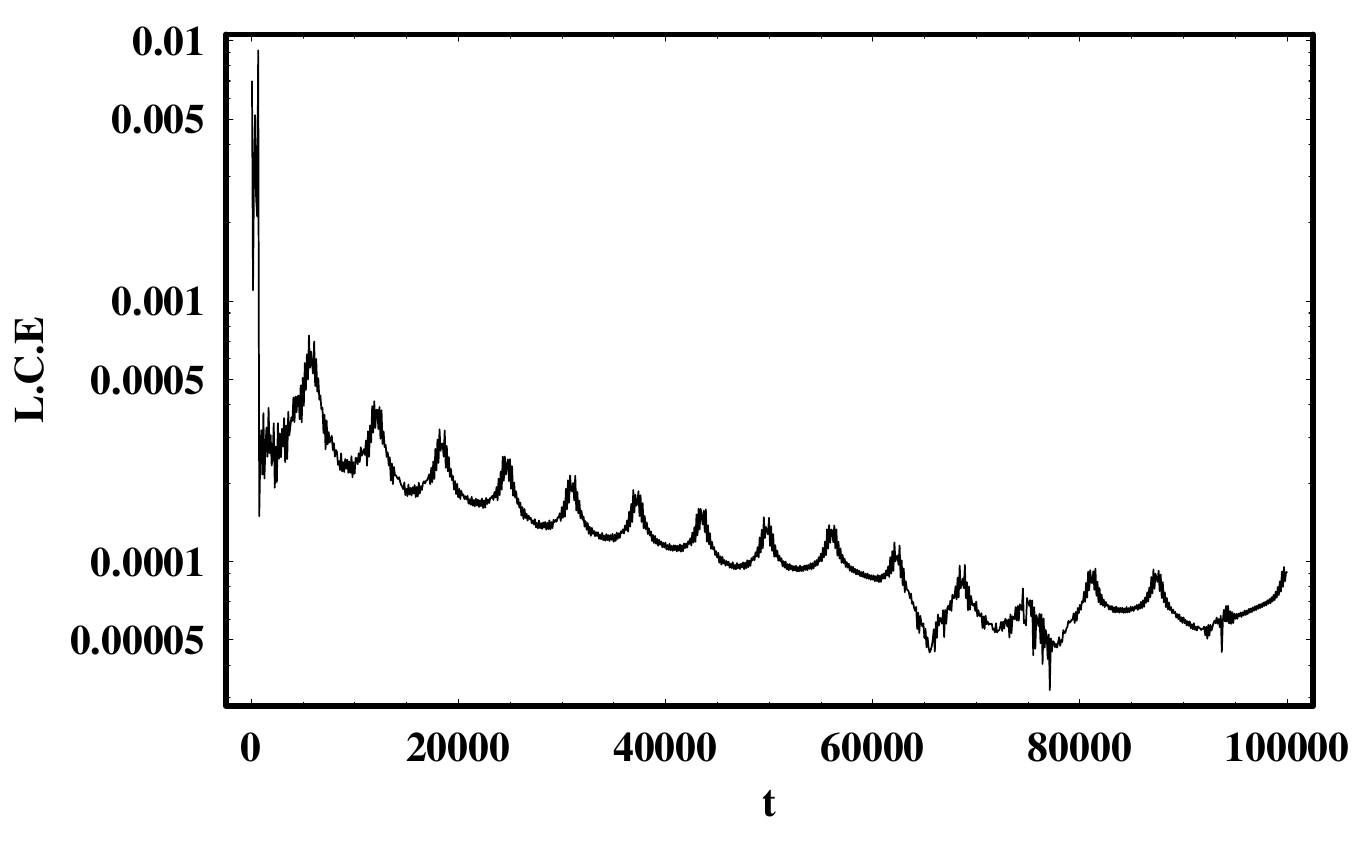}}}

\begin{minipage}{30mm}
\hspace{15mm}
 {\fns(a)}
\end{minipage}\hspace{25mm}
\begin{minipage}{30mm}
\hspace{18mm}{\fns(b)}\hs\hs\end{minipage}

\vs

\resizebox{0.80\hsize}{!}{\rotatebox{0}{\includegraphics*{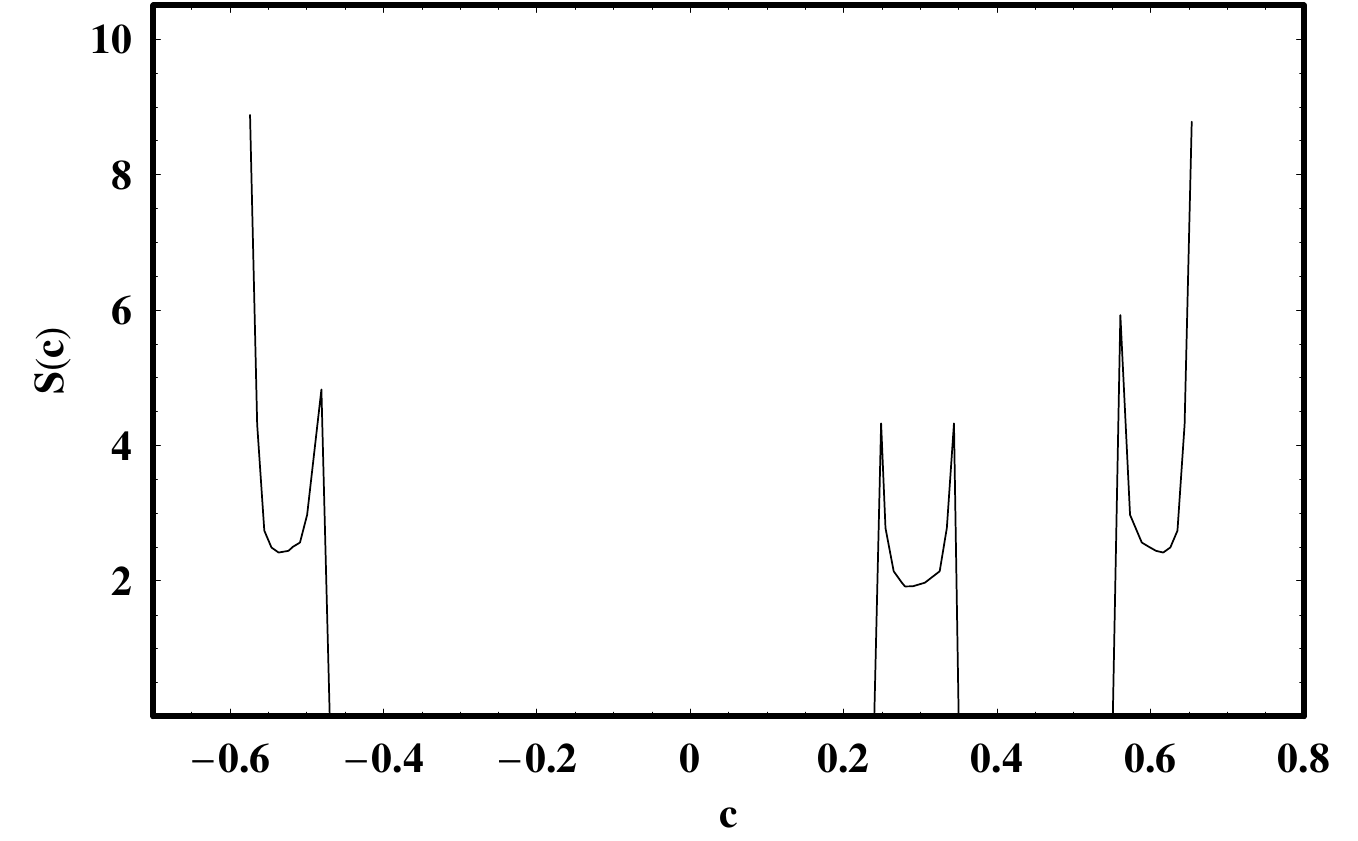}}
                          \rotatebox{0}{\includegraphics*{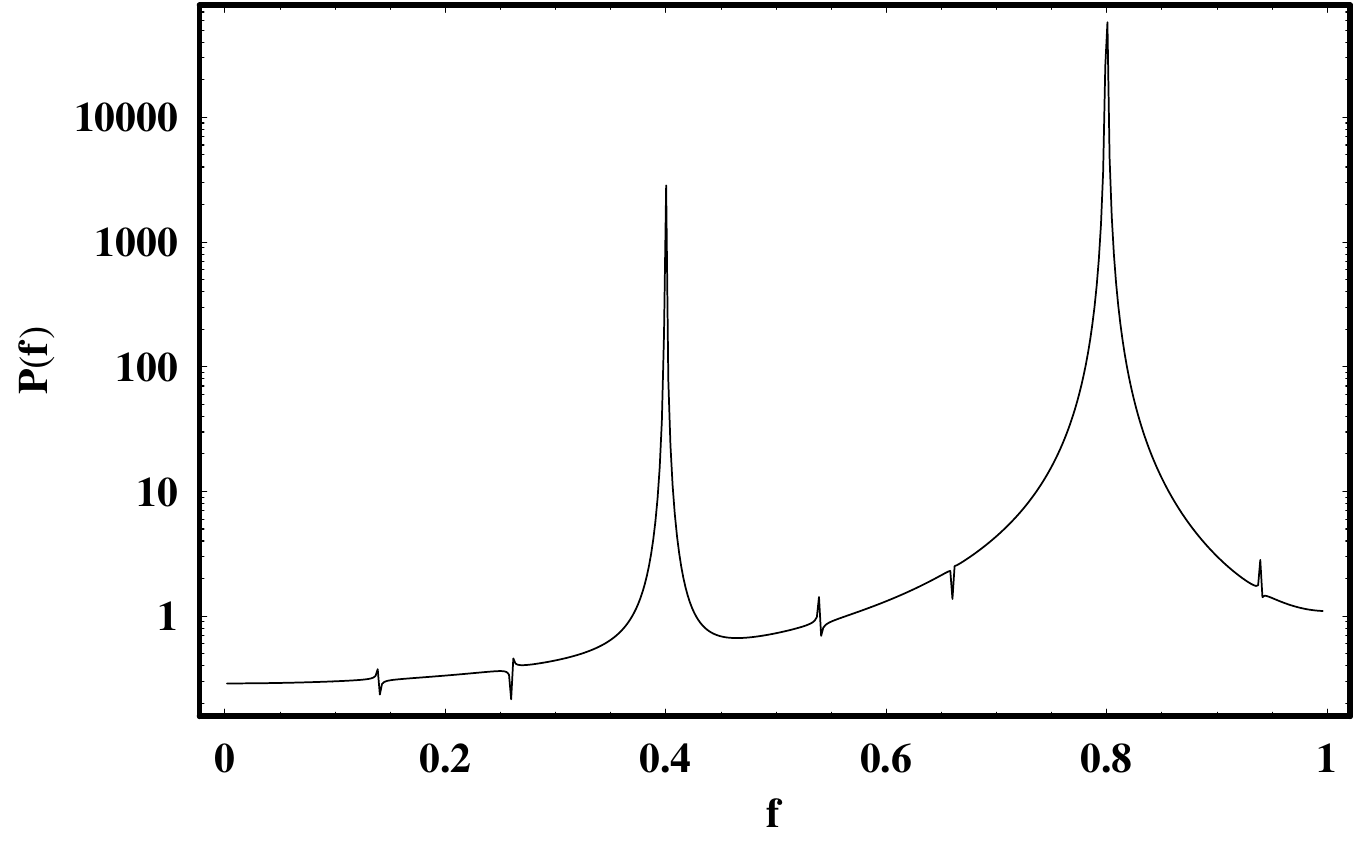}}}

\begin{minipage}{30mm}

\hspace{15mm}
 {\fns(c)}
\end{minipage}\hspace{25mm}
\begin{minipage}{30mm}
\hspace{18mm}{\fns(d)}\hs\hs\end{minipage}

\vspace{-2mm}
 \caption{\baselineskip 3.6mm \label{fig5}(a)--(d): Similar
{to} Fig.~\ref{fig3}(a)--(d) for a resonant orbit. Initial
conditions are: $x_0=0.15, y_0=0, p_{x0}=4.5$. The values of all
other parameters and energy are {the same }as in Fig.~\ref{fig1}.}
\end{figure*}
\begin{figure*}

\vs \centering
\resizebox{0.85\hsize}{!}{\rotatebox{0}{\includegraphics*{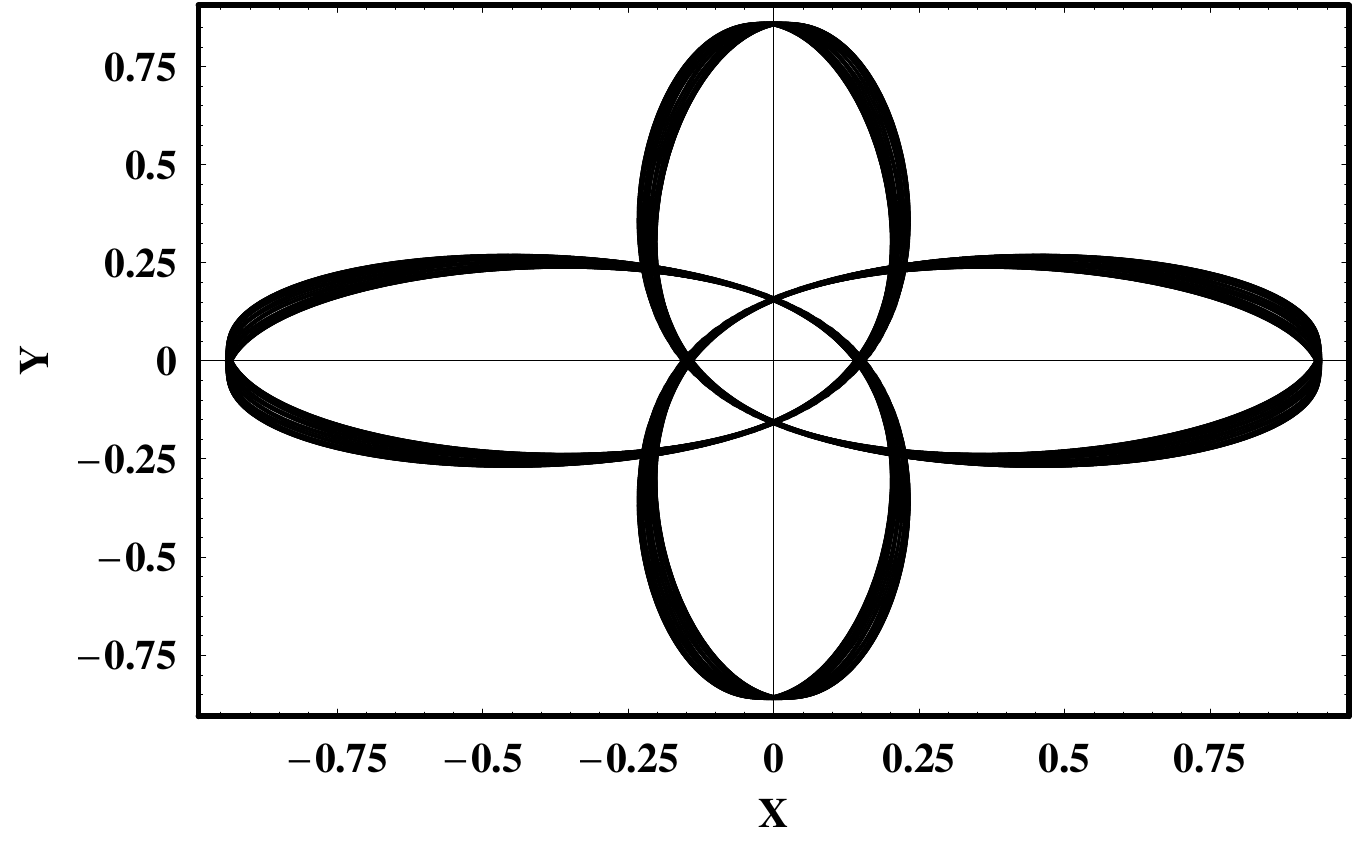}}
                          \rotatebox{0}{\includegraphics*{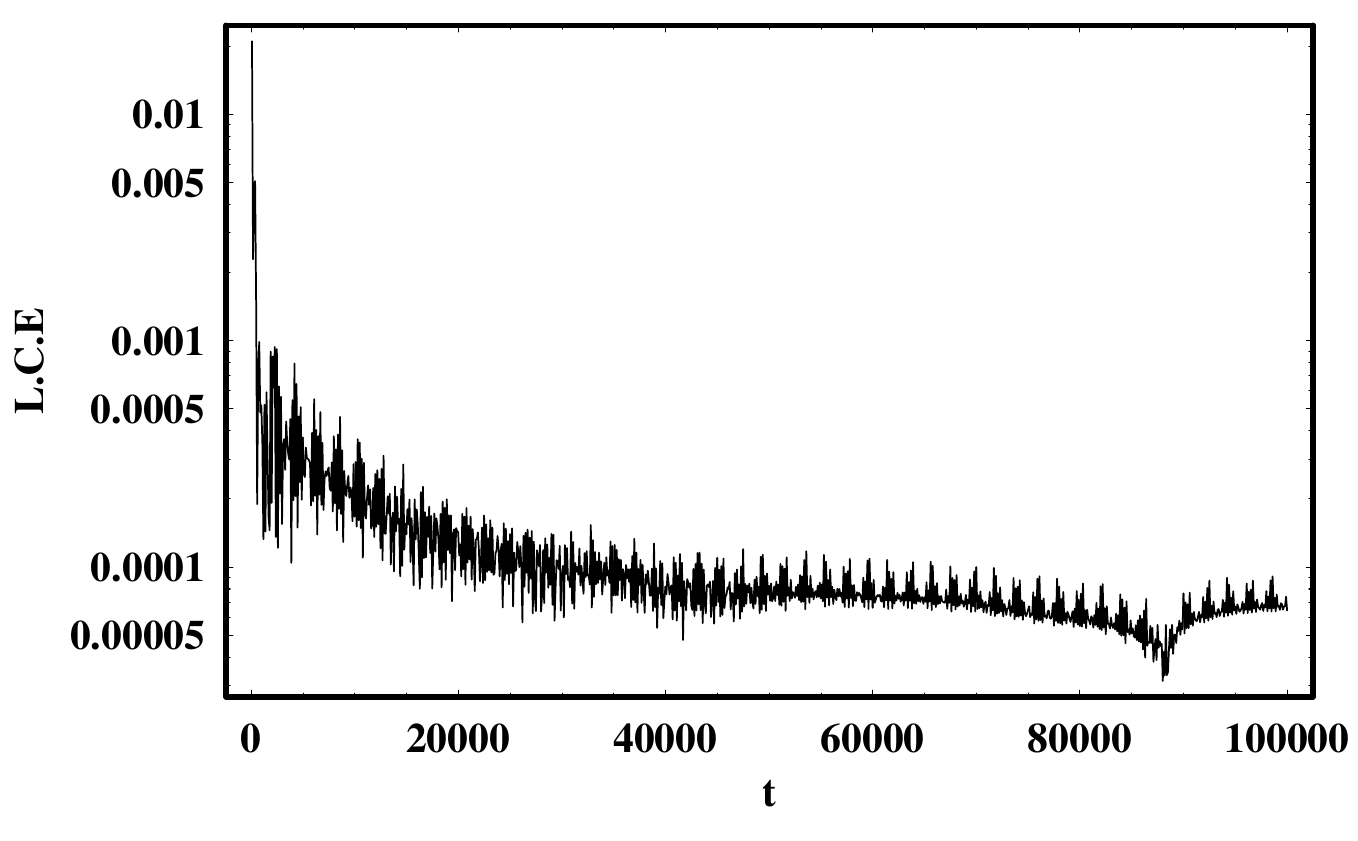}}}

\begin{minipage}{30mm}
\hspace{14mm}
 {\fns(a)}
\end{minipage}\hspace{25mm}
\begin{minipage}{30mm}
\hspace{20mm}{\fns(b)}\hs\hs\end{minipage}

\vs
\resizebox{0.80\hsize}{!}{\rotatebox{0}{\includegraphics*{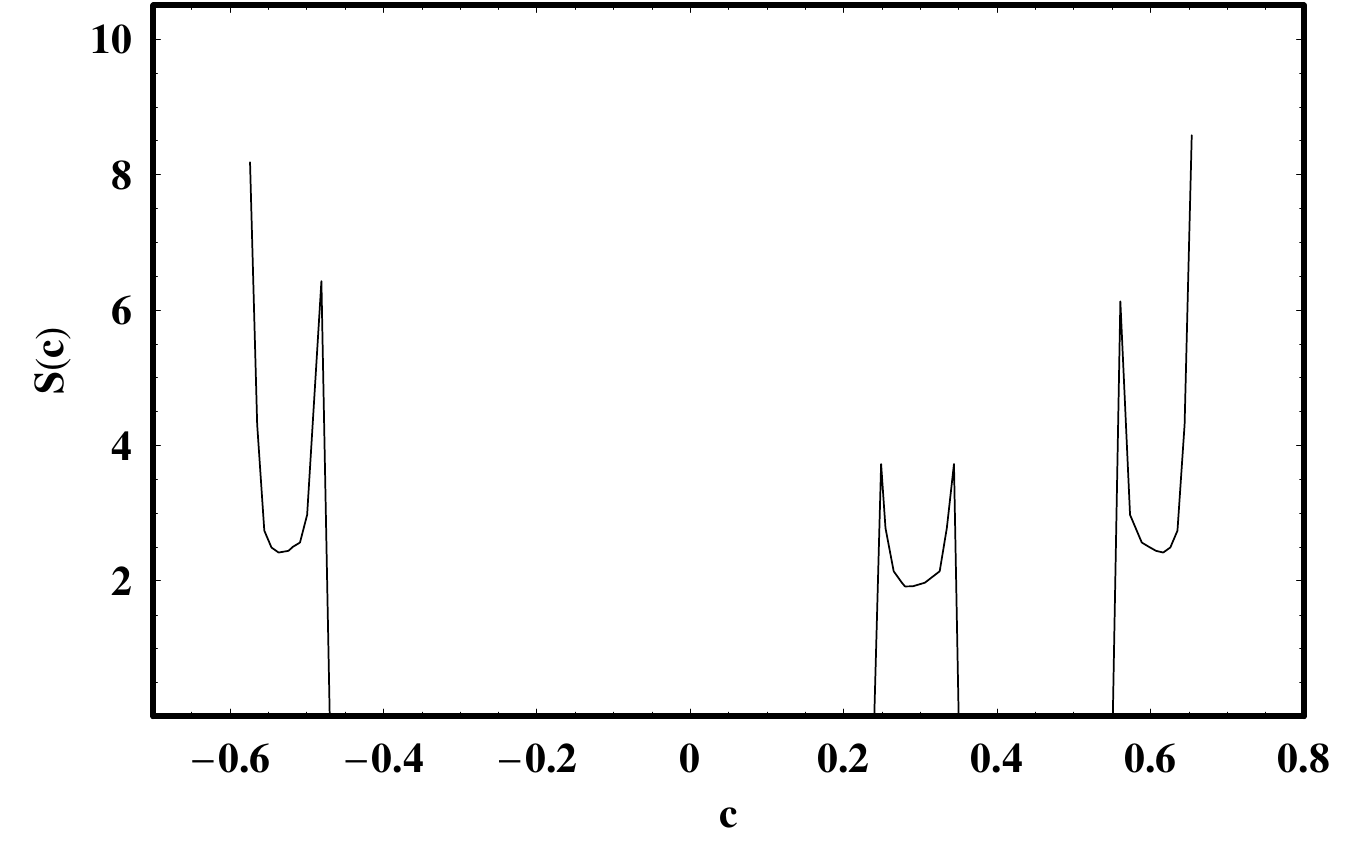}}
                          \rotatebox{0}{\includegraphics*{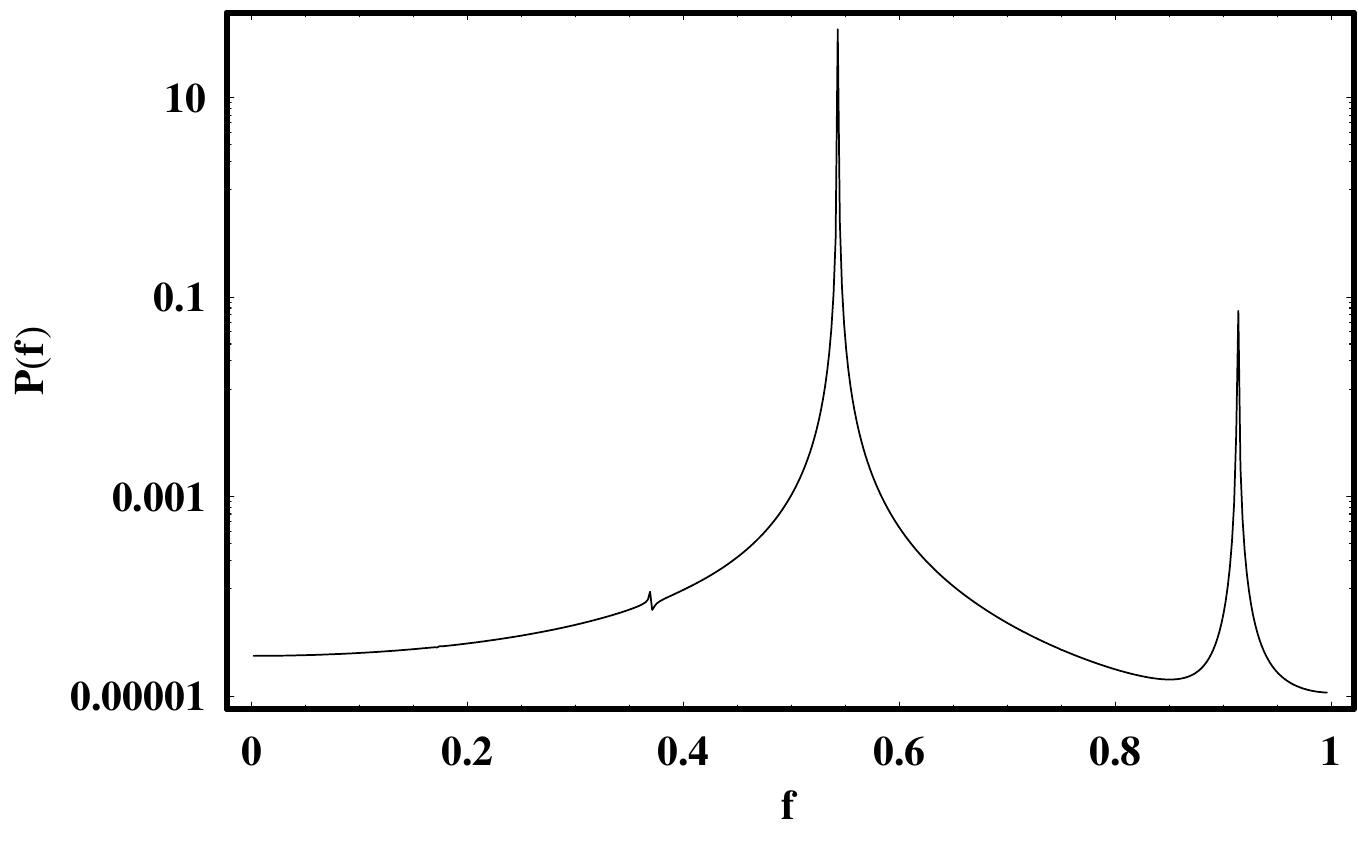}}}

\begin{minipage}{30mm}

\hspace{14mm}
 {\fns(c)}
\end{minipage}\hspace{25mm}
\begin{minipage}{30mm}
\hspace{19mm}{\fns(d)}\hs\hs\end{minipage}

\caption{\baselineskip 3.6mm \label{fig6}(a)--(d): Similar {to}
Fig.~\ref{fig5}(a)--(d) for the potential $V_{\rm tl}$. The values
of all other parameters and energy are {the same }as in
Fig.~\ref{fig2}.}
\end{figure*}
\begin{figure*}[!tH]

\vs \centering
\resizebox{0.8\hsize}{!}{\rotatebox{0}{\includegraphics*{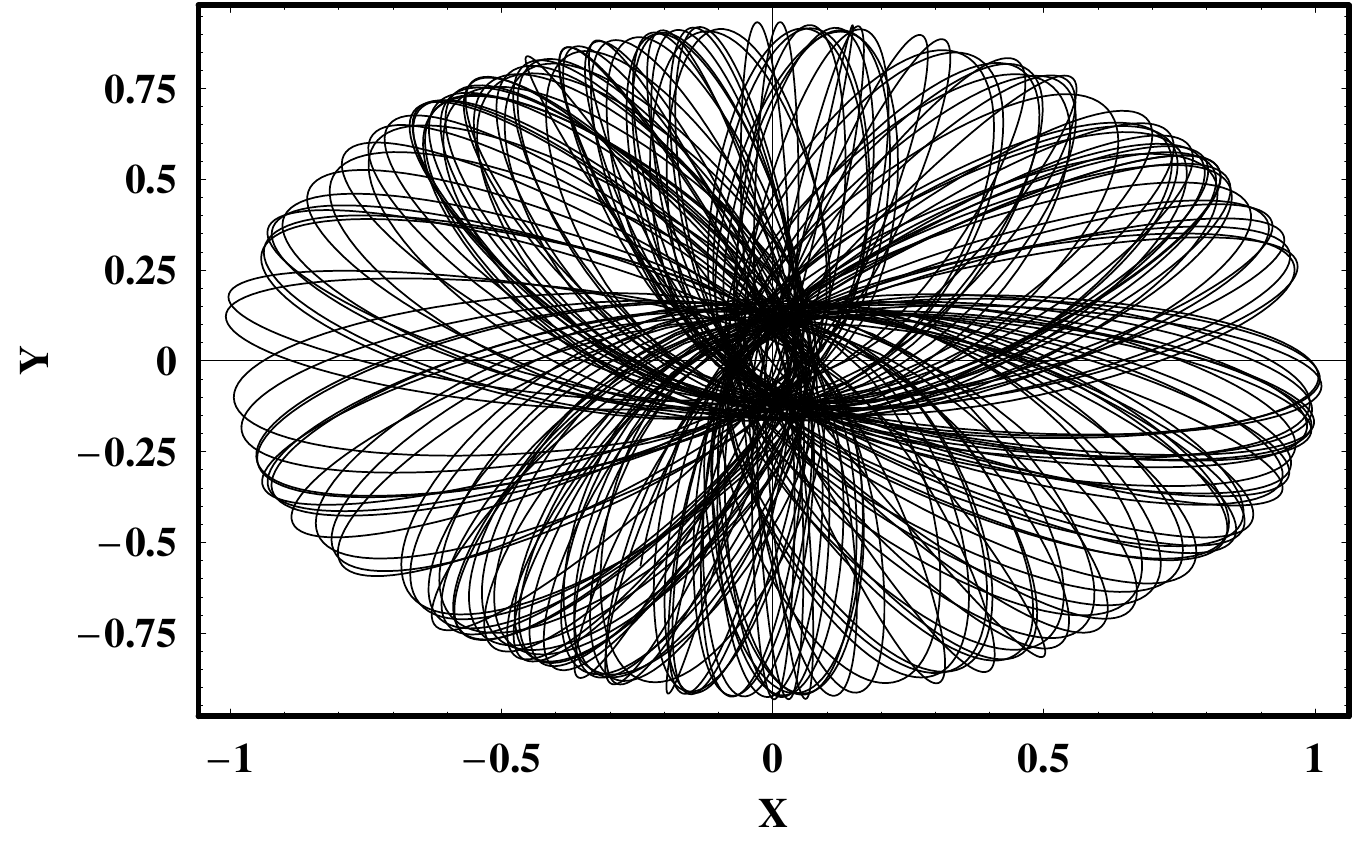}}
                         \rotatebox{0}{\includegraphics*{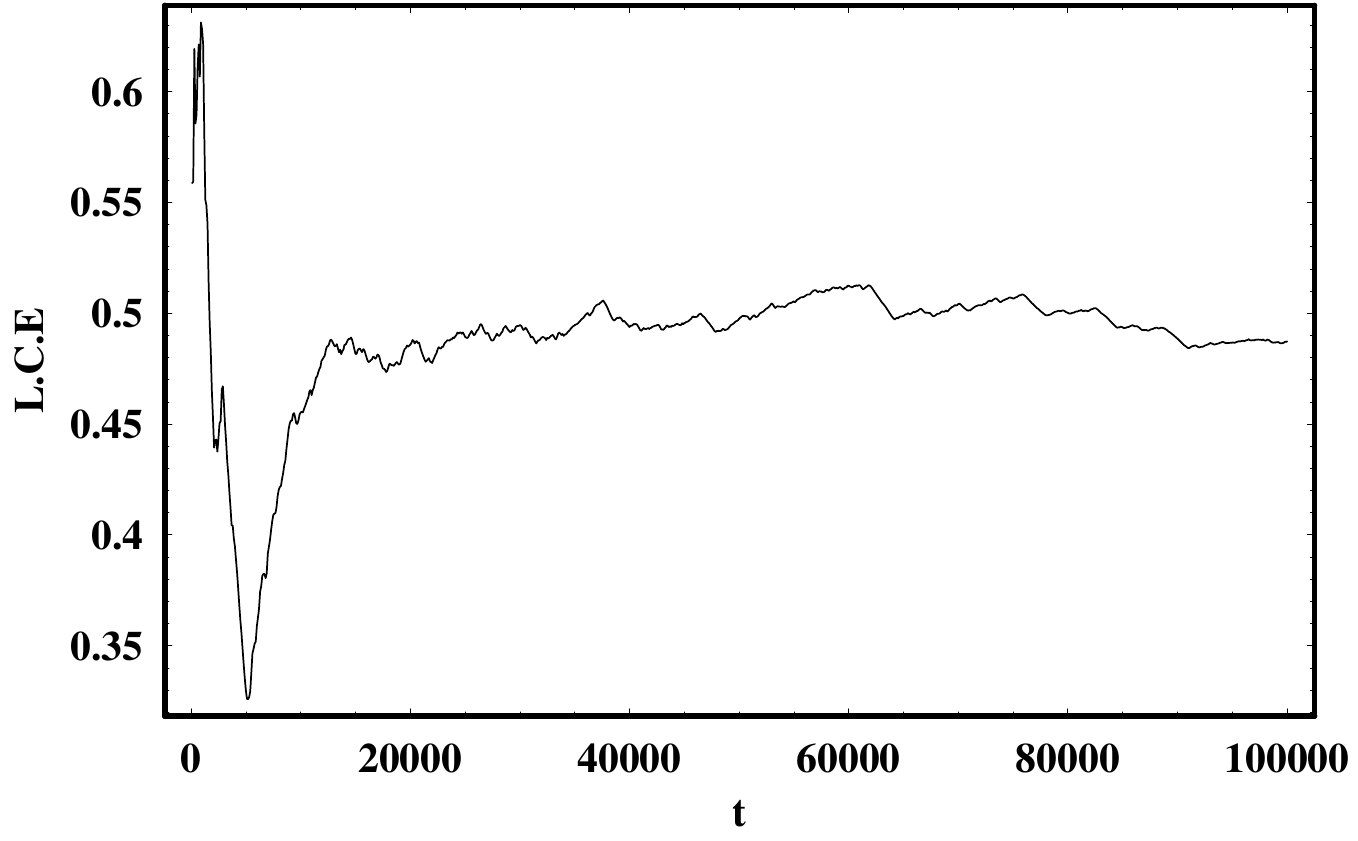}}}

\begin{minipage}{30mm}
\hspace{16mm}
 {\fns(a)}
\end{minipage}\hspace{25mm}
\begin{minipage}{30mm}
\hspace{16mm}{\fns(b)}\hs\hs\end{minipage}

\vs
\resizebox{0.8\hsize}{!}{\rotatebox{0}{\includegraphics*{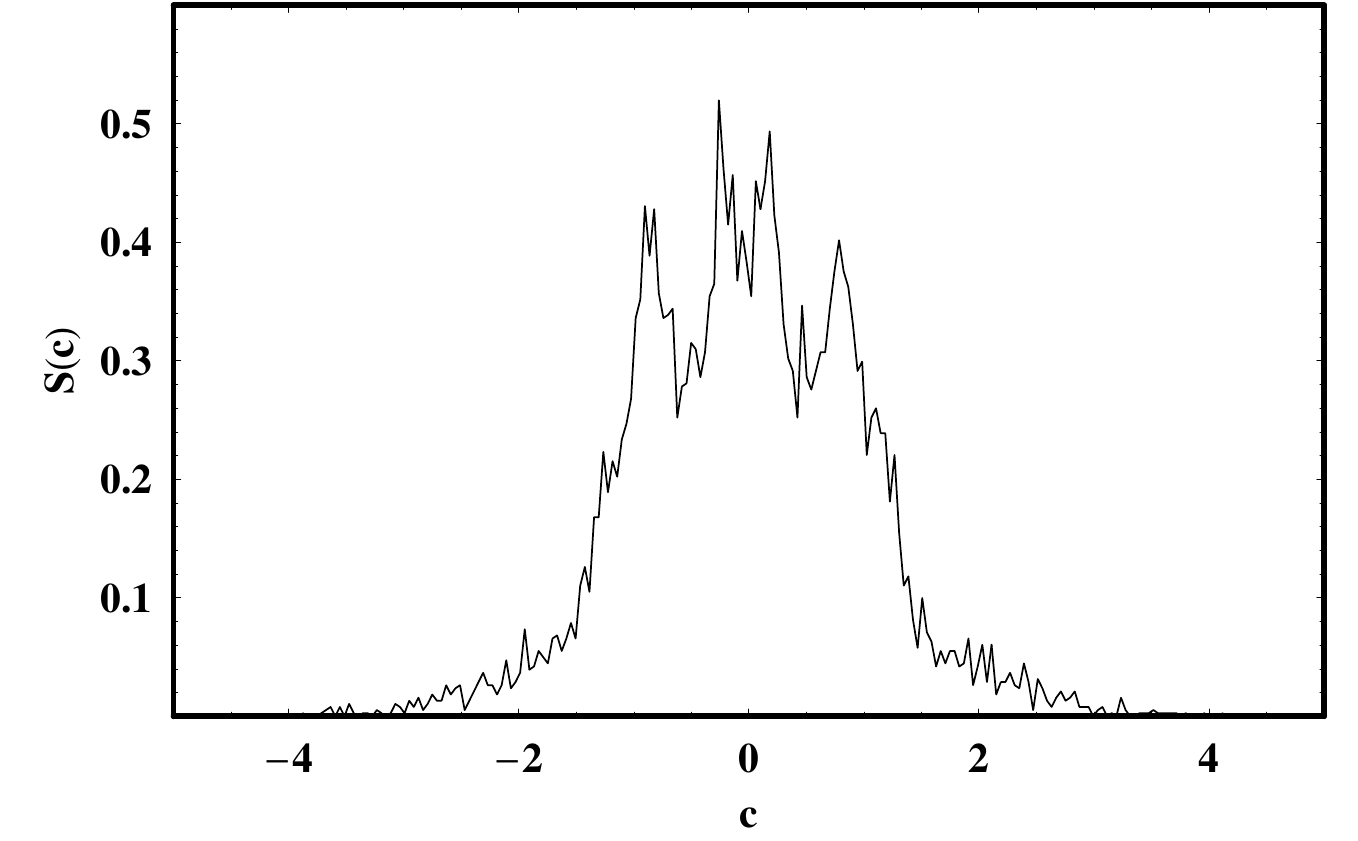}}
                         \rotatebox{0}{\includegraphics*{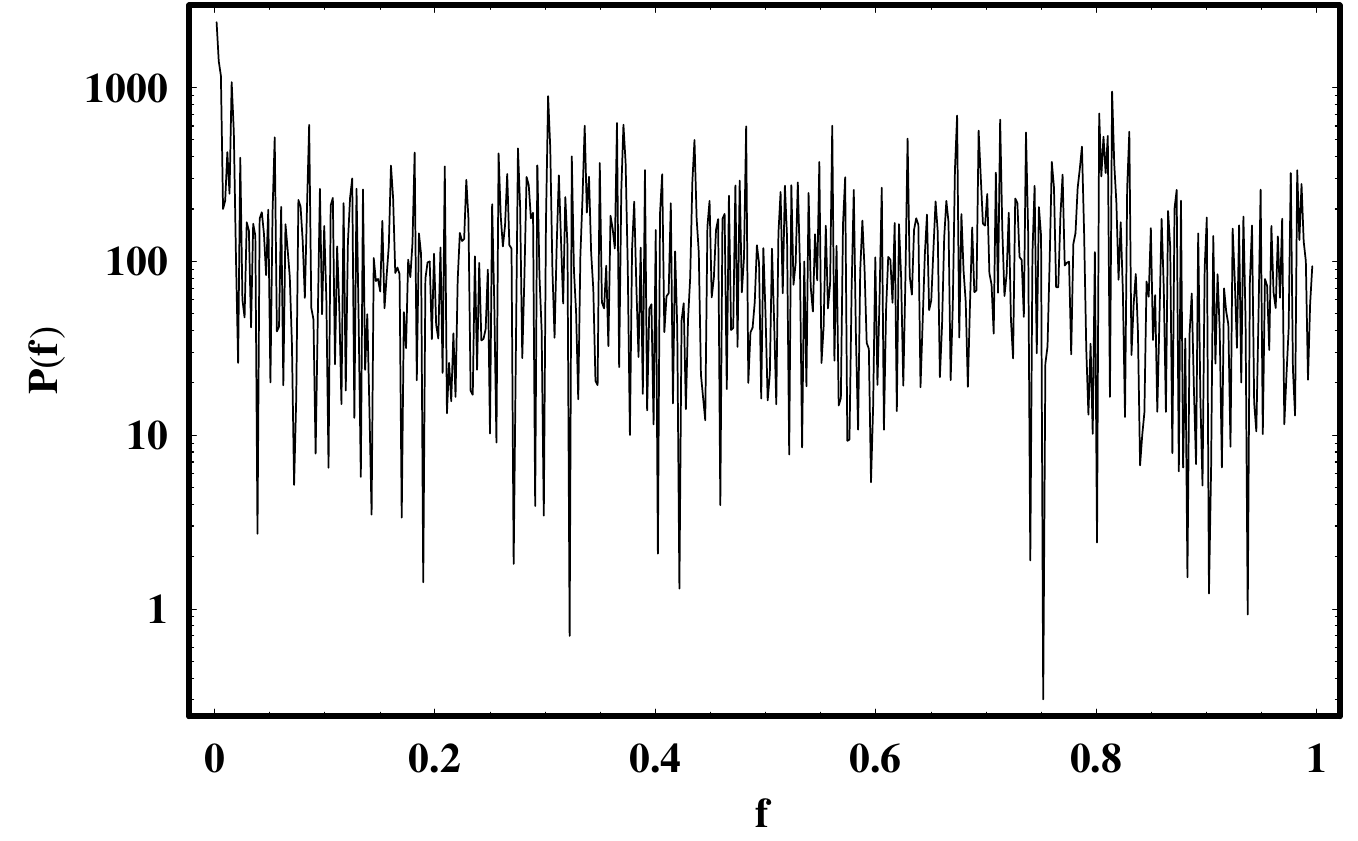}}}

\begin{minipage}{30mm}

\hspace{15mm}
 {\fns(c)}
\end{minipage}\hspace{25mm}
\begin{minipage}{30mm}
\hspace{17mm}{\fns(d)}\hs\hs\end{minipage}

\caption{\baselineskip 3.6mm \label{fig7}(a)--(d): Similar {to}
Fig.~\ref{fig3}(a)--(d) for a chaotic orbit. Initial conditions are:
$x_0=0.02, y_0=0, p_{x0}=2.5$. The values of all other parameters
and energy are {the same }as in Fig.~\ref{fig1}.}
\end{figure*}
\begin{figure*}

\vs \centering
\resizebox{0.85\hsize}{!}{\rotatebox{0}{\includegraphics*{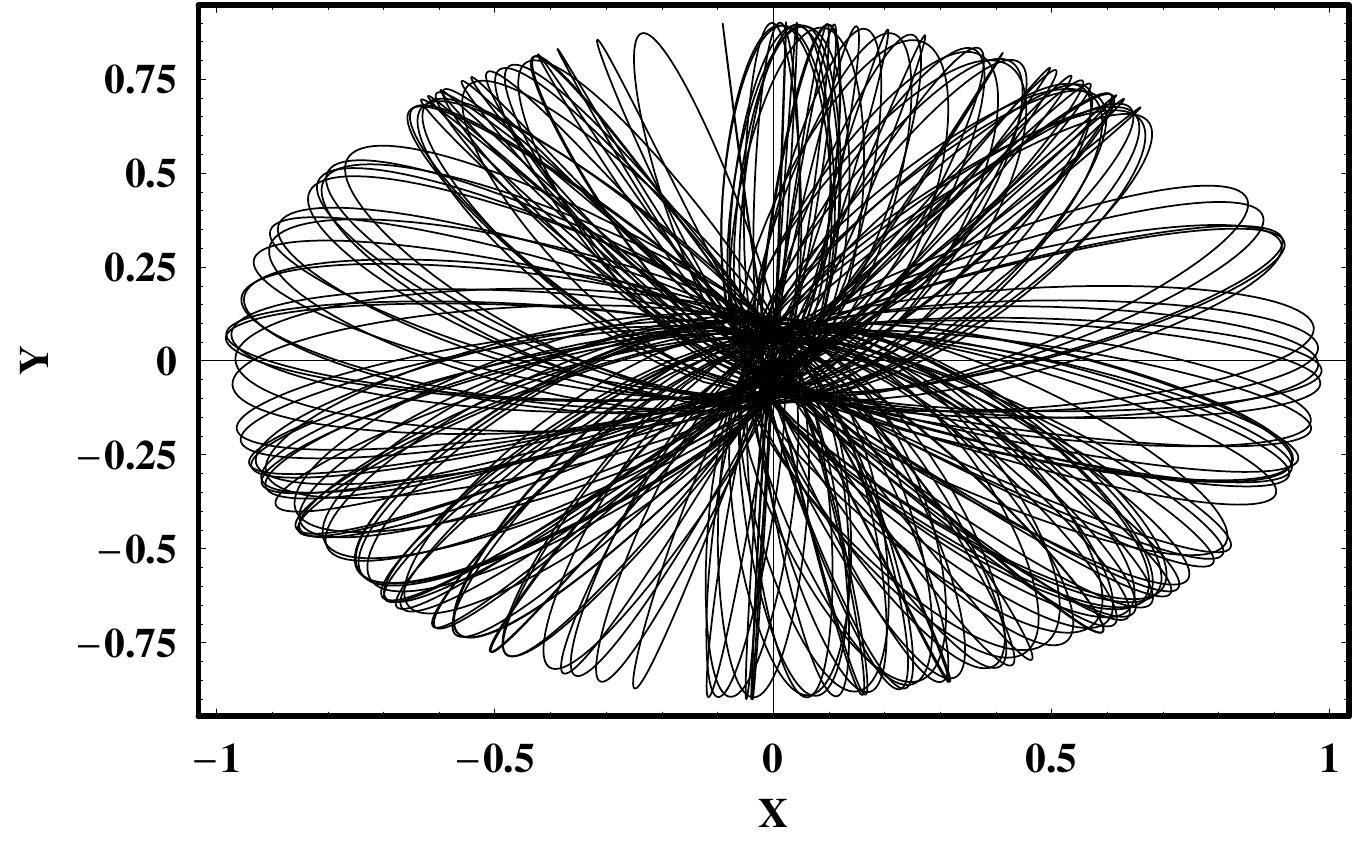}}
                          \rotatebox{0}{\includegraphics*{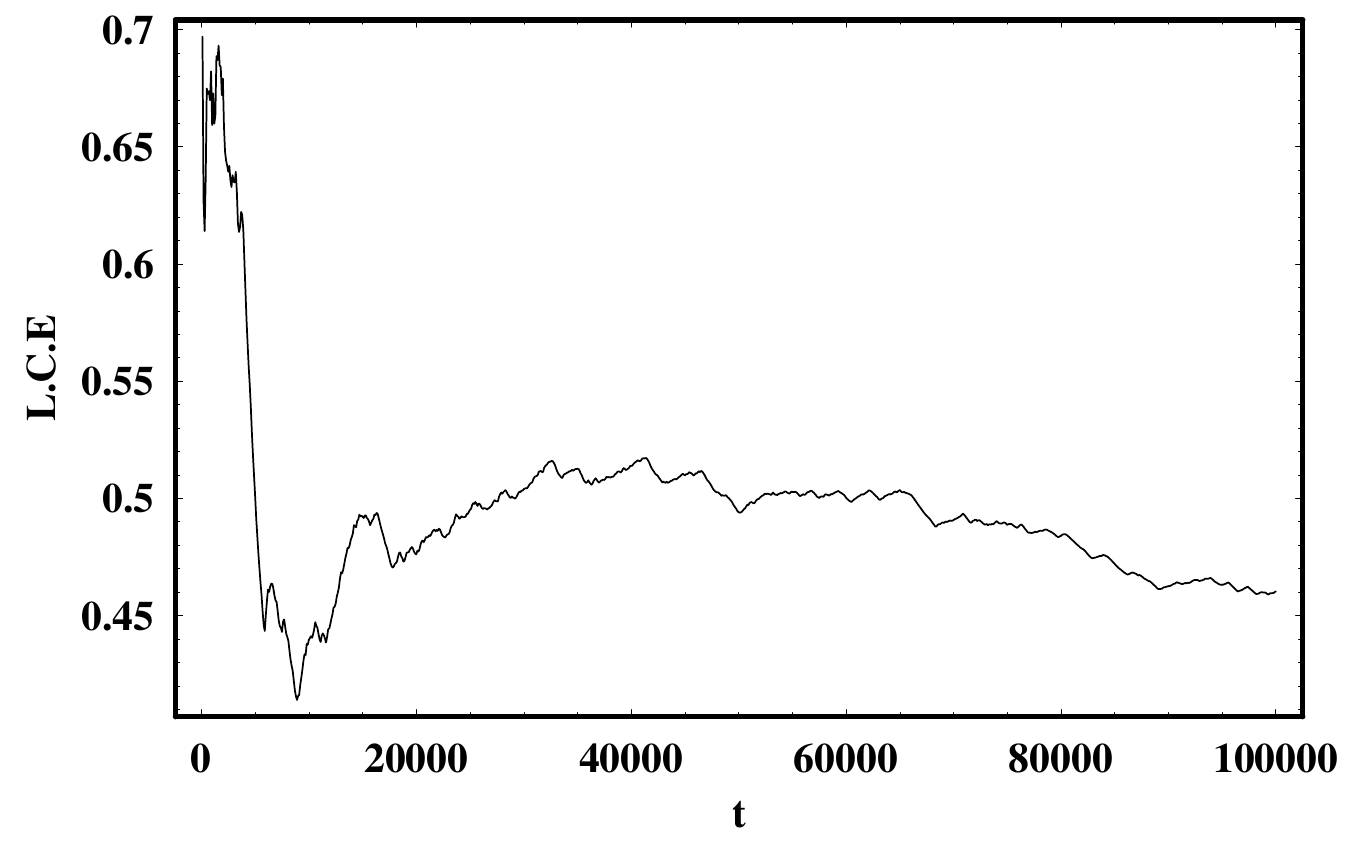}}}

\begin{minipage}{30mm}
\hspace{15mm}
 {\fns(a)}
\end{minipage}\hspace{25mm}
\begin{minipage}{30mm}
\hspace{18mm}{\fns(b)}\hs\hs\end{minipage}

\vs
\resizebox{0.9\hsize}{!}{\rotatebox{0}{\includegraphics*{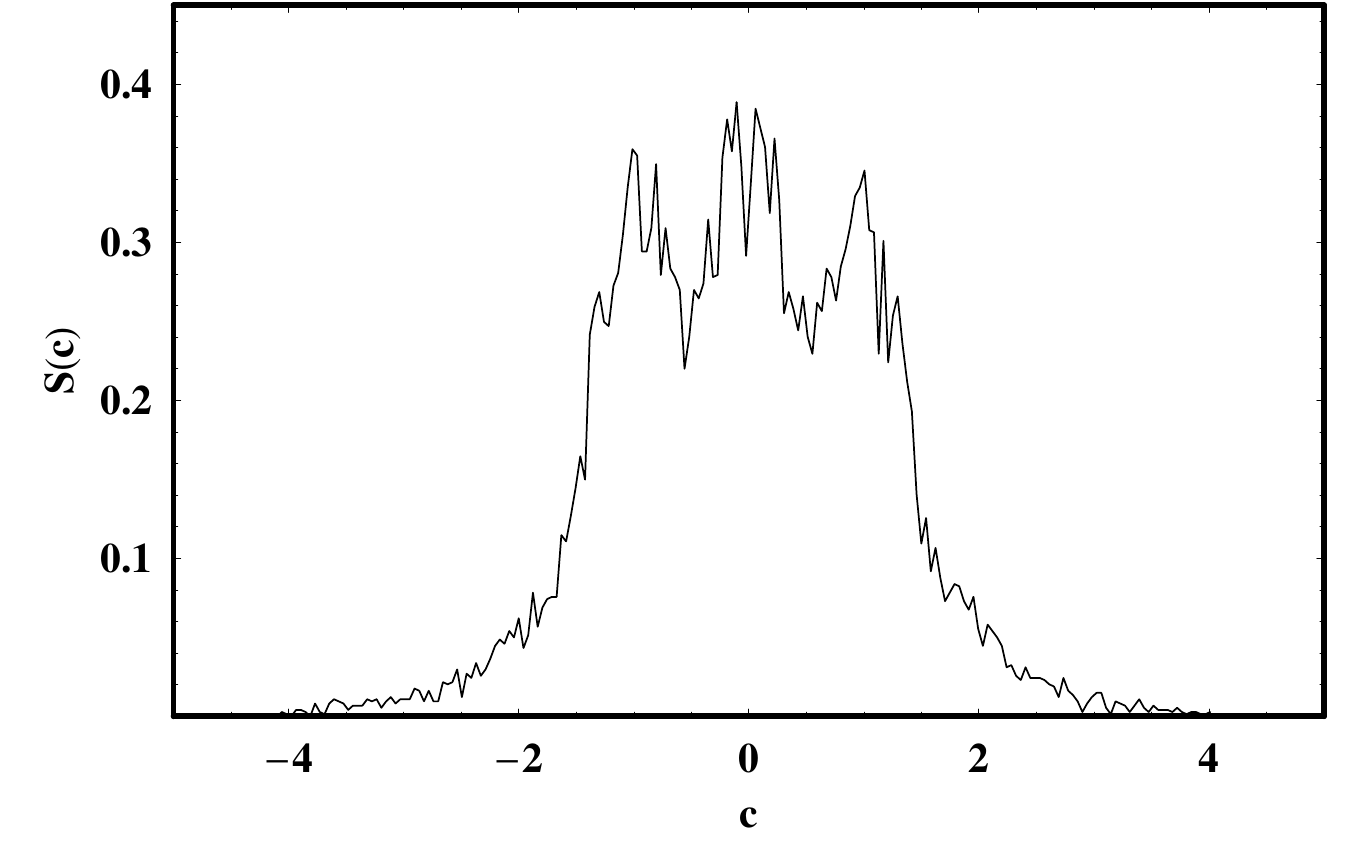}}
                         \rotatebox{0}{\includegraphics*{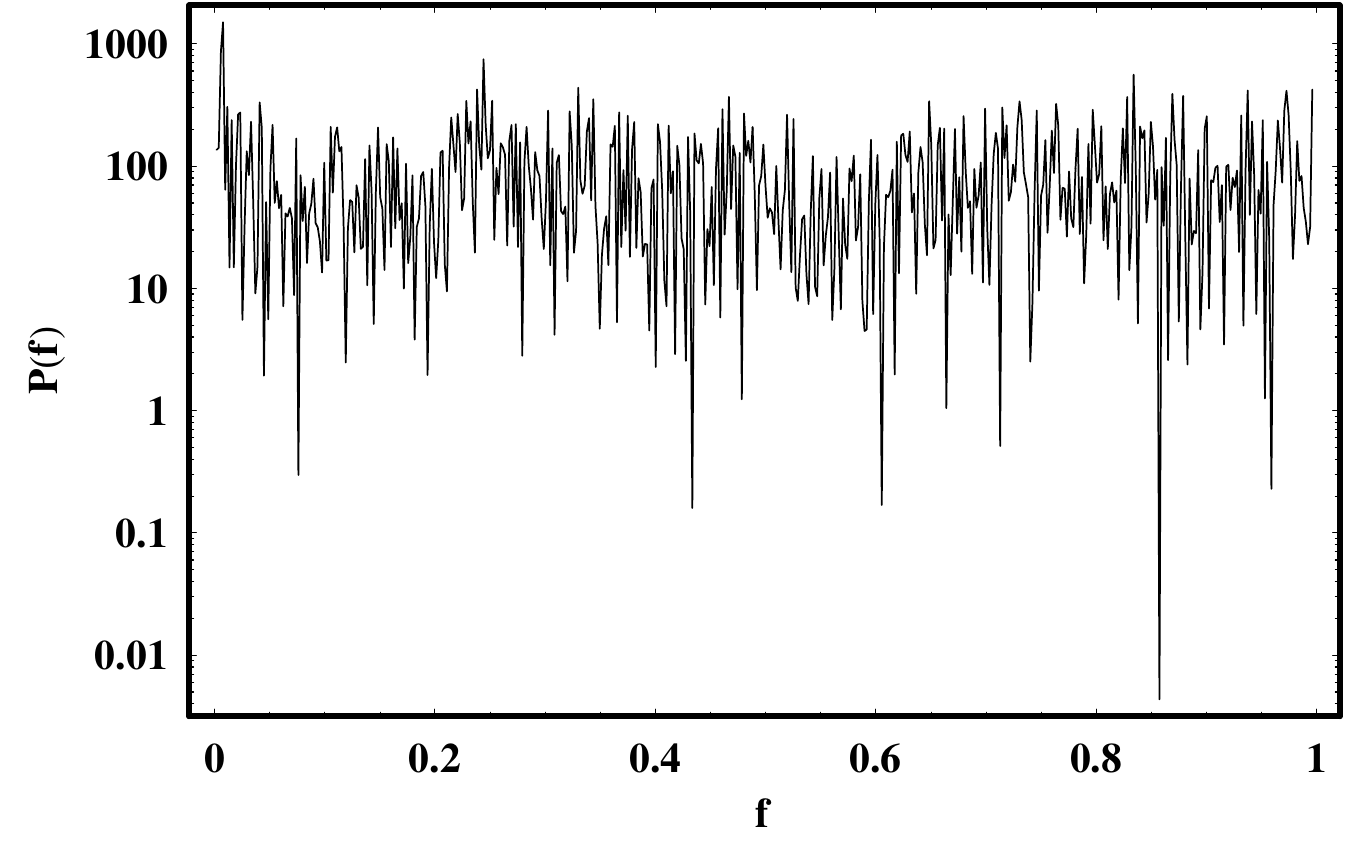}}}

\begin{minipage}{30mm}

\hspace{12mm}
 {\fns(c)}
\end{minipage}\hspace{25mm}
\begin{minipage}{30mm}
\hspace{21mm}{\fns(d)}\hs\hs\end{minipage}

\caption{\baselineskip 3.6mm \label{fig8}(a)--(d): Similar {to}
Fig.~\ref{fig7}(a)--(d) for the potential $V_{\rm tl}$. The values
of all other parameters and energy are {the same }as in
Fig.~\ref{fig2}.}
\end{figure*}

In order to have a better picture of the properties of motion in
the two potentials $V_{\rm tg}$ and $V_{\rm tl}$, we present
results for two more orbits.

Figures~\ref{fig5}(a)--(d) and \ref{fig6}(a)--(d) 
   are
similar to \ref{fig3}(a)--(d) and \ref{fig4}(a)--(d), for an orbit
with initial conditions: $x_0=0.15, y_0=0, p_{x0}=4.5$. This orbit
belongs to family (ii) and it is characteristic of the 3:3
resonance. As one can see, the outcomes presented in the two
{f}igures are very similar. The results presented in
Figures~\ref{fig7}(a)--(d) and \ref{fig8}(a)--(d)
  are similar to those of
Figures~\ref{fig3}(a)--(d) and \ref{fig4}(a)--(d) but for a chaotic
orbit. Initial conditions are: $x_0=0.02, y_0=0, p_{x0}=2.5$. Here
the maximum LCE has a positive value indicating chaotic motion.
Moreover, the $S(c)$ spectrum shows a number of large and small
peaks which is characteristic of chaotic motion. Finally, the $P(f)$
indicator is highly asymmetric with a large number of peaks, also
indicating chaotic motion. Comparing the results given in
Figures~\ref{fig7}(a)--(d) and \ref{fig8}({a})--(d), we can say that
they are very similar.

\begin{figure}

\vs \centering

\includegraphics[width=43mm]{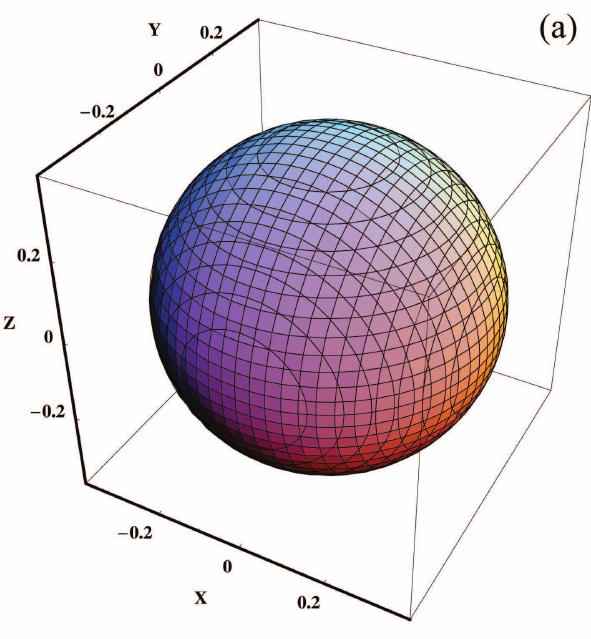}~~~~~~~~
\includegraphics[width=43mm]{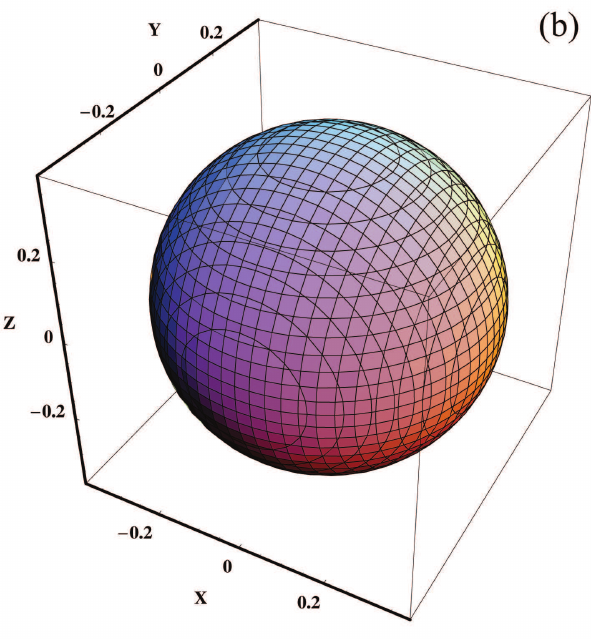}

\vs
\includegraphics[width=43mm]{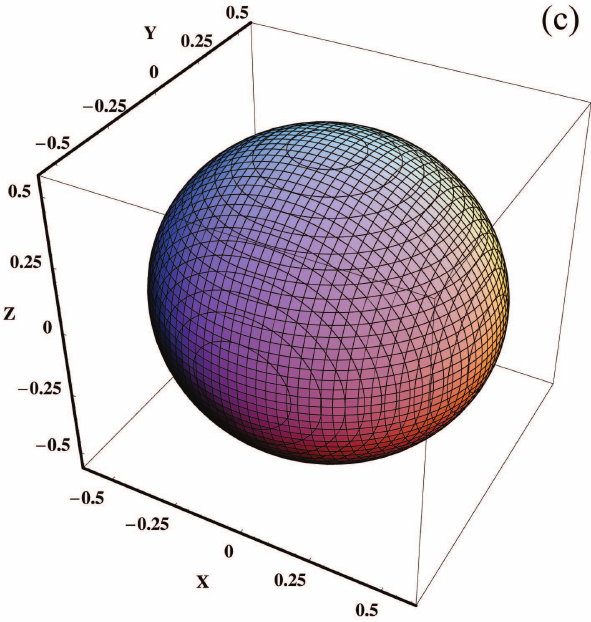}~~~~~~~~
\includegraphics[width=43mm]{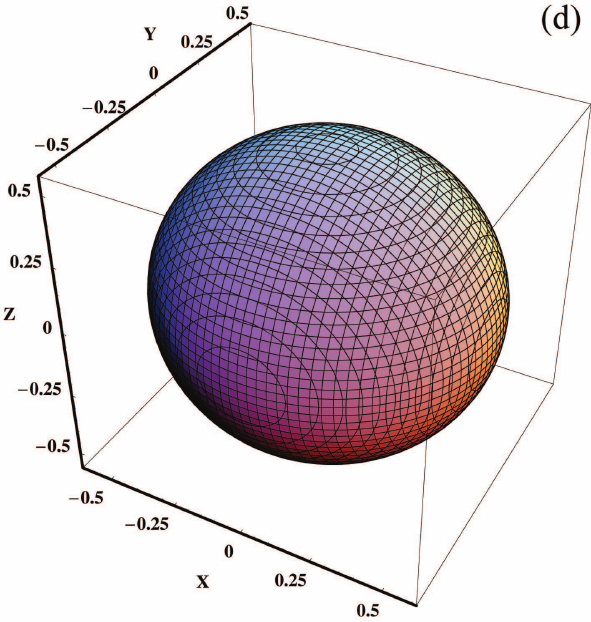}

\vs
\includegraphics[width=43mm]{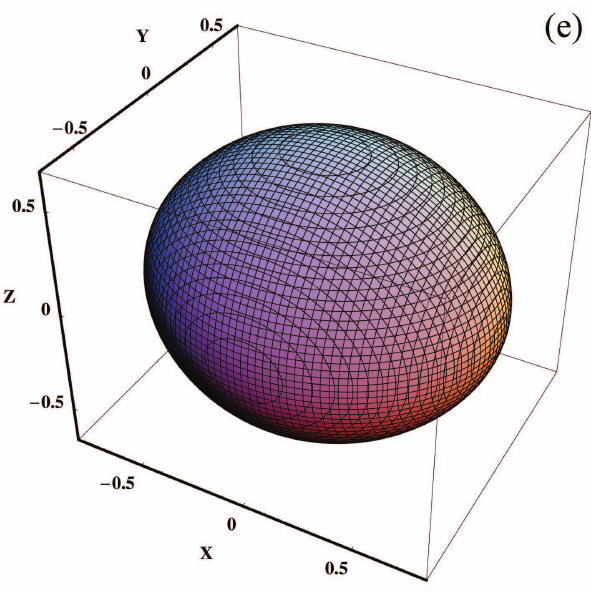}~~~~~~~~
\includegraphics[width=43mm]{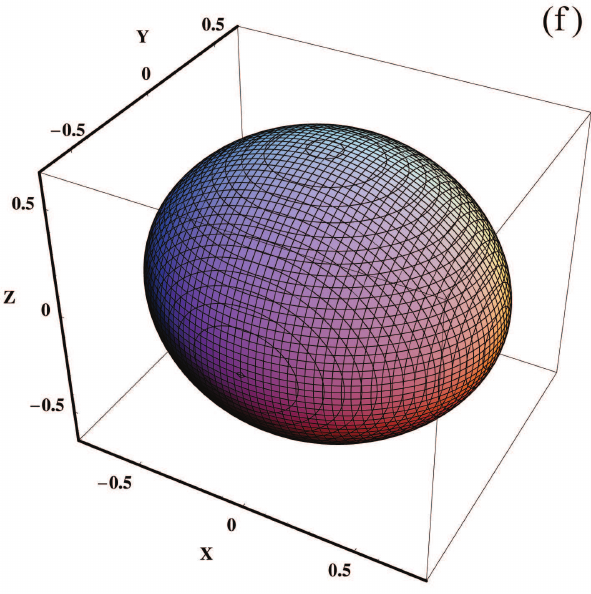}

\vs
\includegraphics[width=43mm]{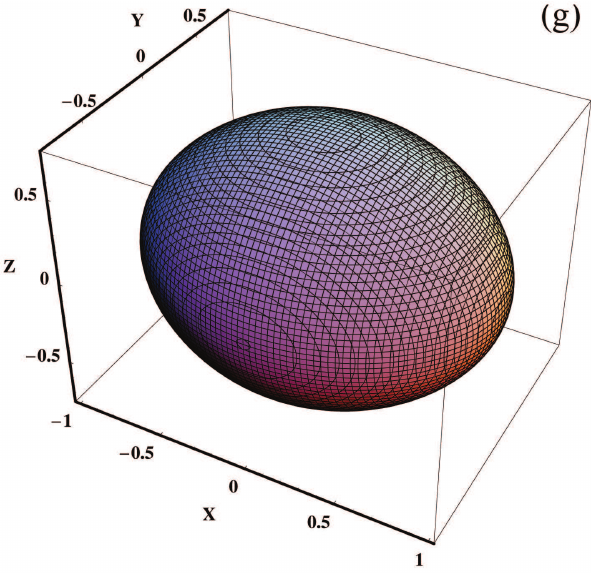}~~~~~~~~
\includegraphics[width=43mm]{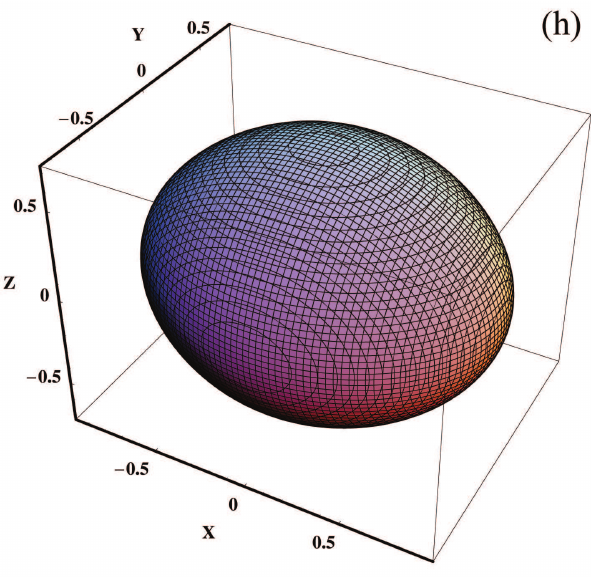}

\caption{\baselineskip 3.6mm \label{fig9}(a)--(h): Surfaces of
equal density for the 3D potentials. Left patterns correspond to
potential $V_{\rm tg}$, while right patterns to potential $V_{\rm
tl}$.}
\end{figure}

\begin{figure*}
\centering

\resizebox{0.7\hsize}{!}{\rotatebox{0}{\includegraphics*{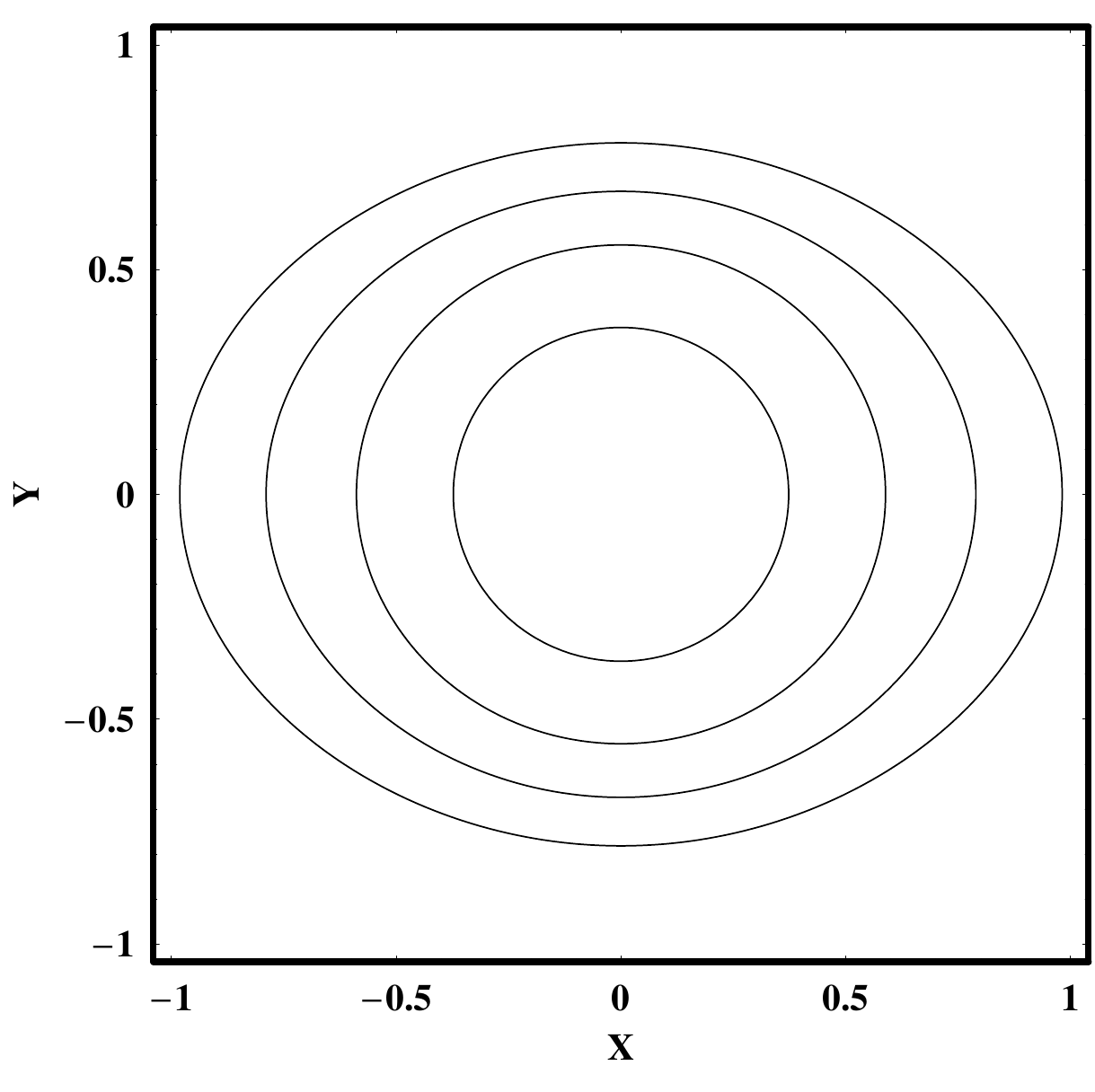}}
                         \rotatebox{0}{\includegraphics*{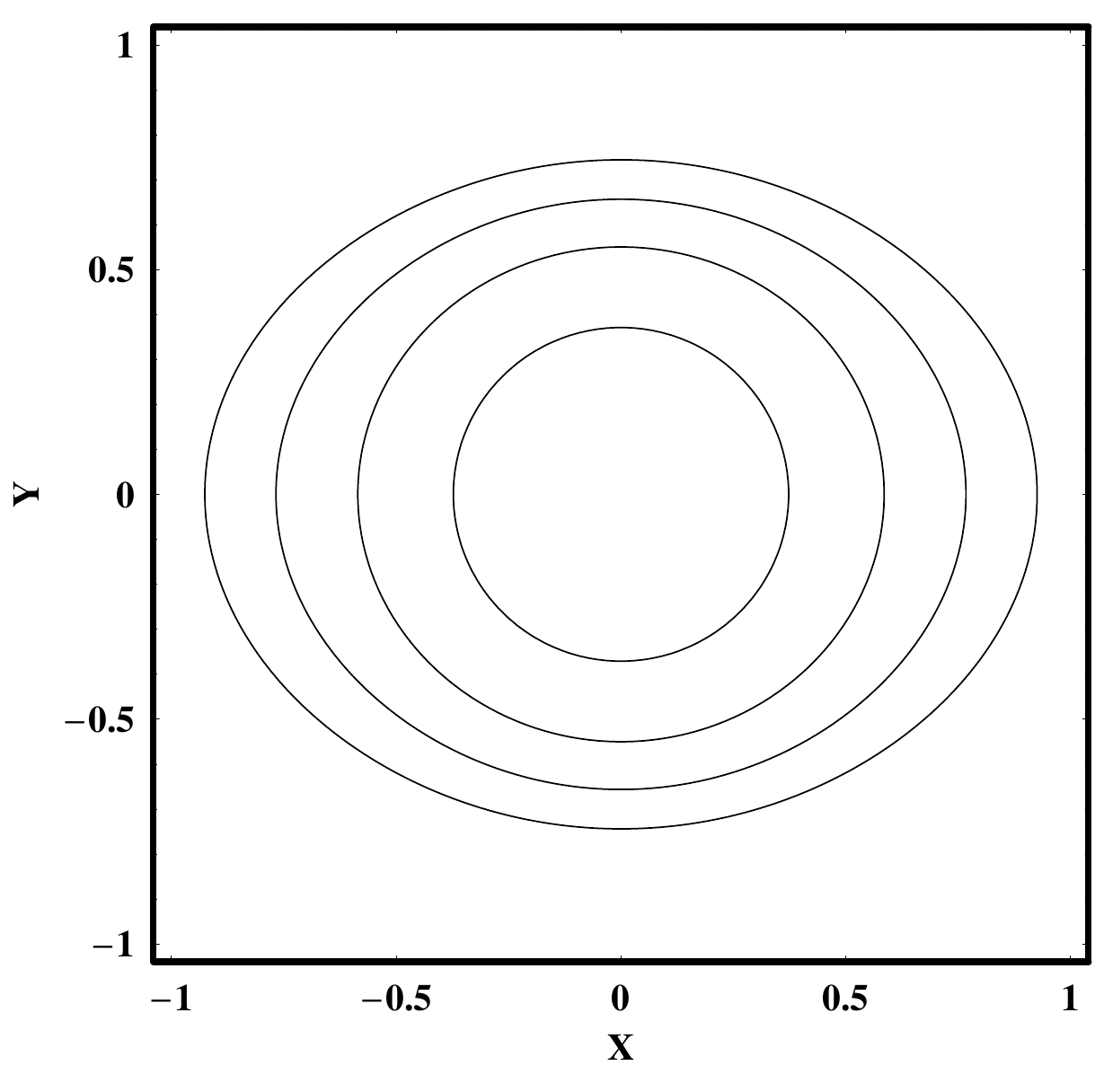}}}

\begin{minipage}{40mm}\centering

\hs\hs\hs {\fns(a)}
\end{minipage}\hspace{20mm}
\begin{minipage}{30mm}
\centering {\fns(b)}~~~~~~~~\end{minipage}

\resizebox{0.7\hsize}{!}{\rotatebox{0}{\includegraphics*{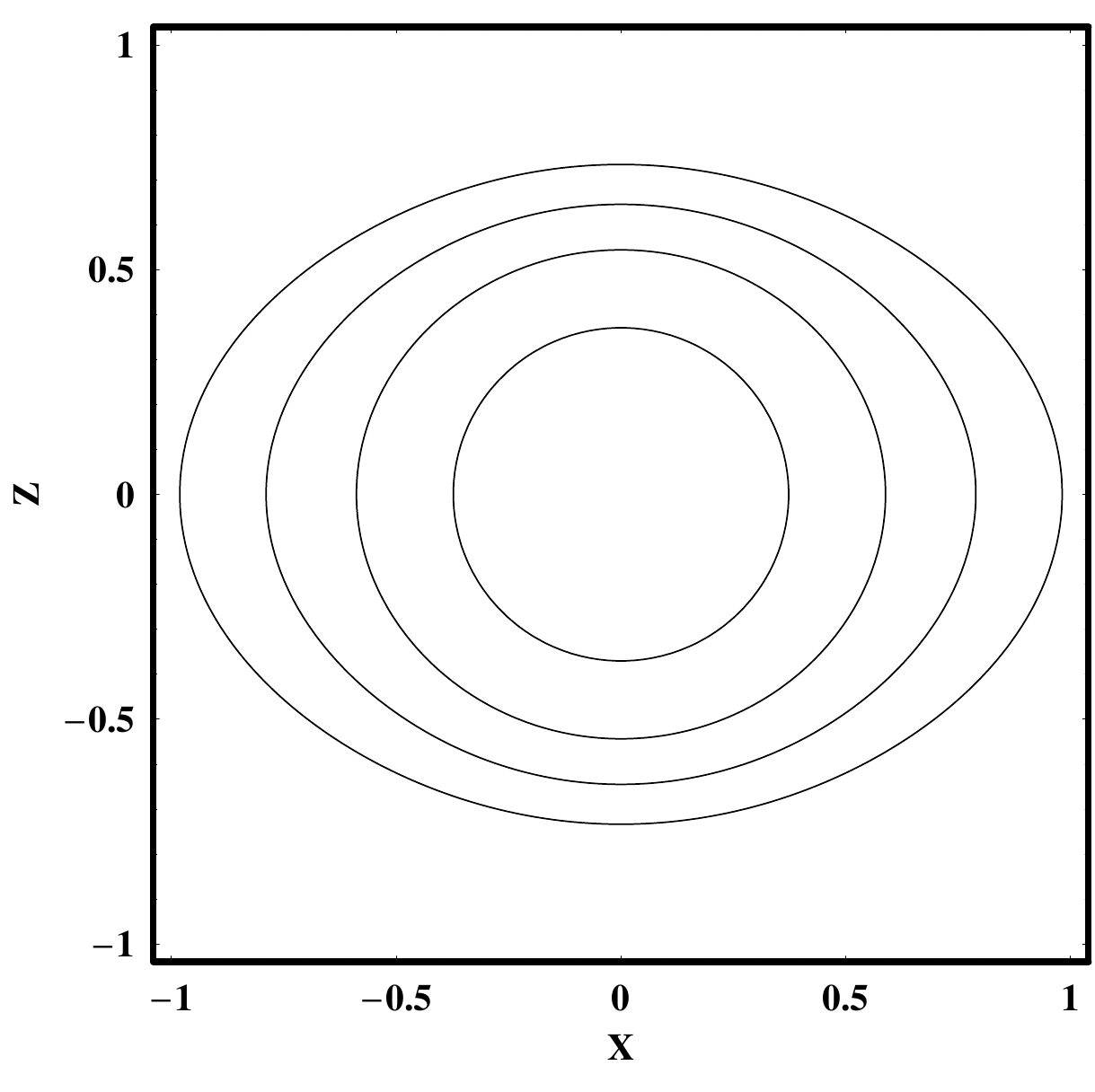}}
                         \rotatebox{0}{\includegraphics*{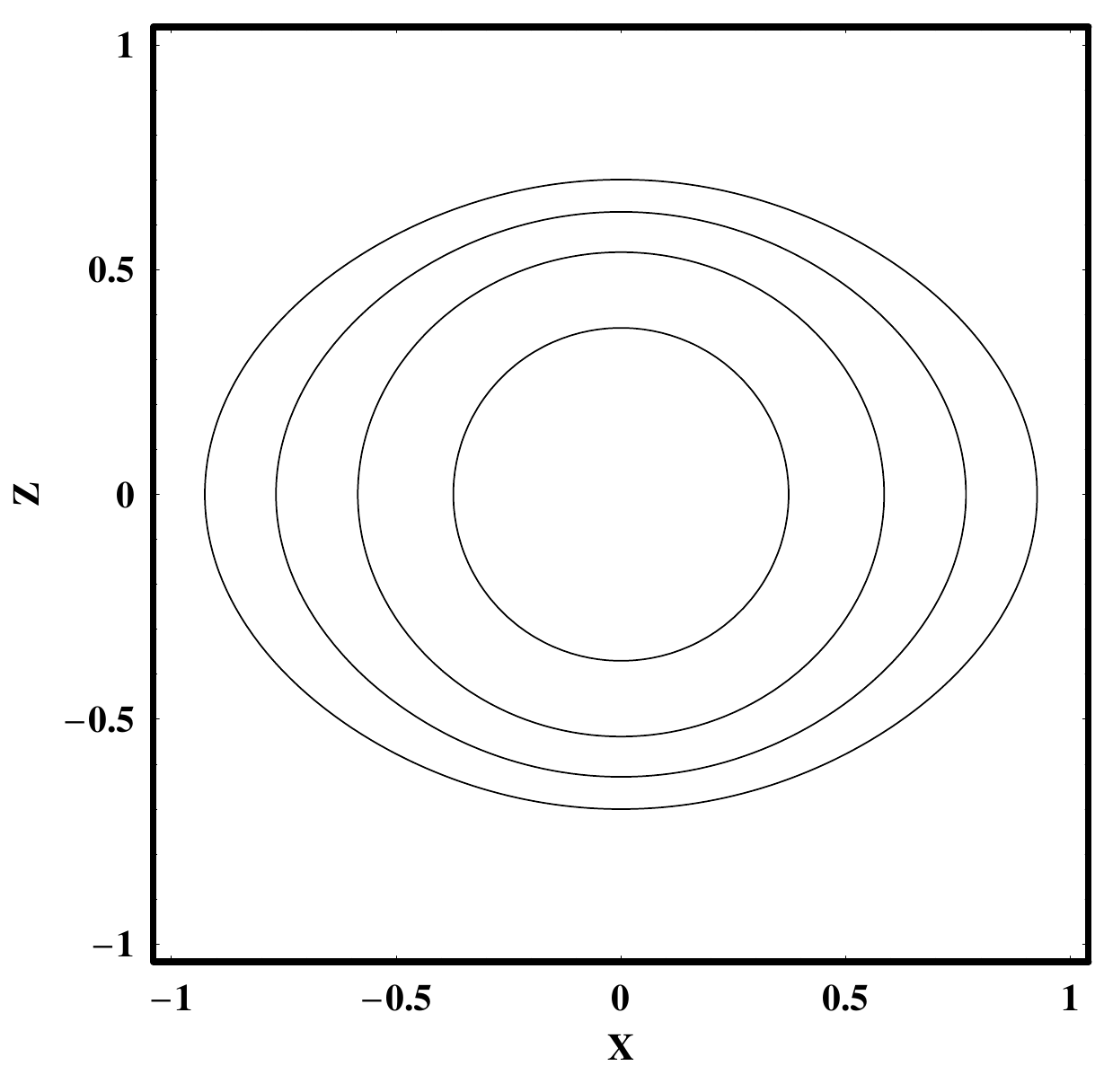}}}

\begin{minipage}{40mm}\centering

\hs\hs\hs {\fns(c)}
\end{minipage}\hspace{20mm}
\begin{minipage}{30mm}
\centering {\fns(d)}~~~~~~~~\end{minipage}

\resizebox{0.7\hsize}{!}{\rotatebox{0}{\includegraphics*{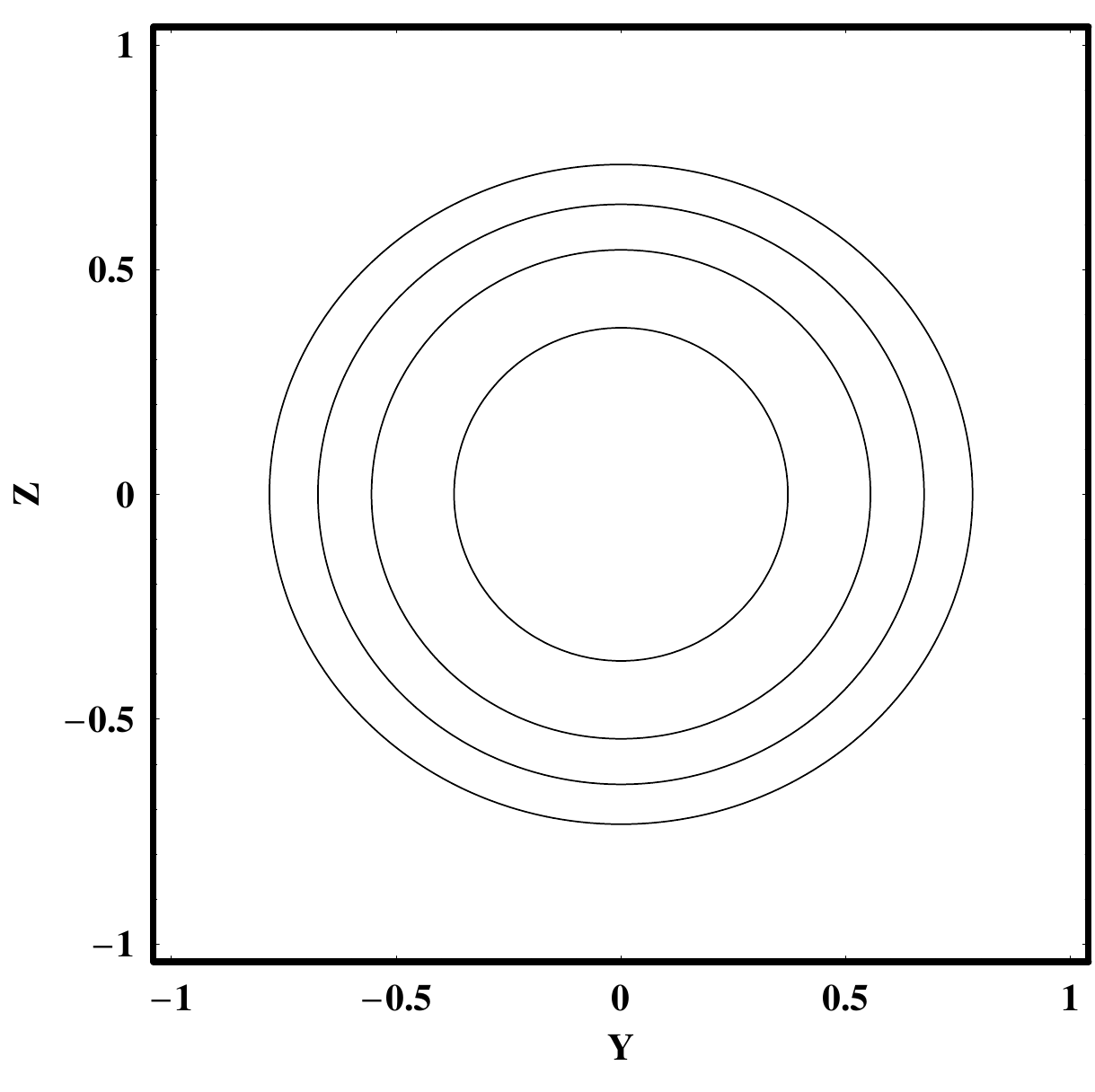}}
                         \rotatebox{0}{\includegraphics*{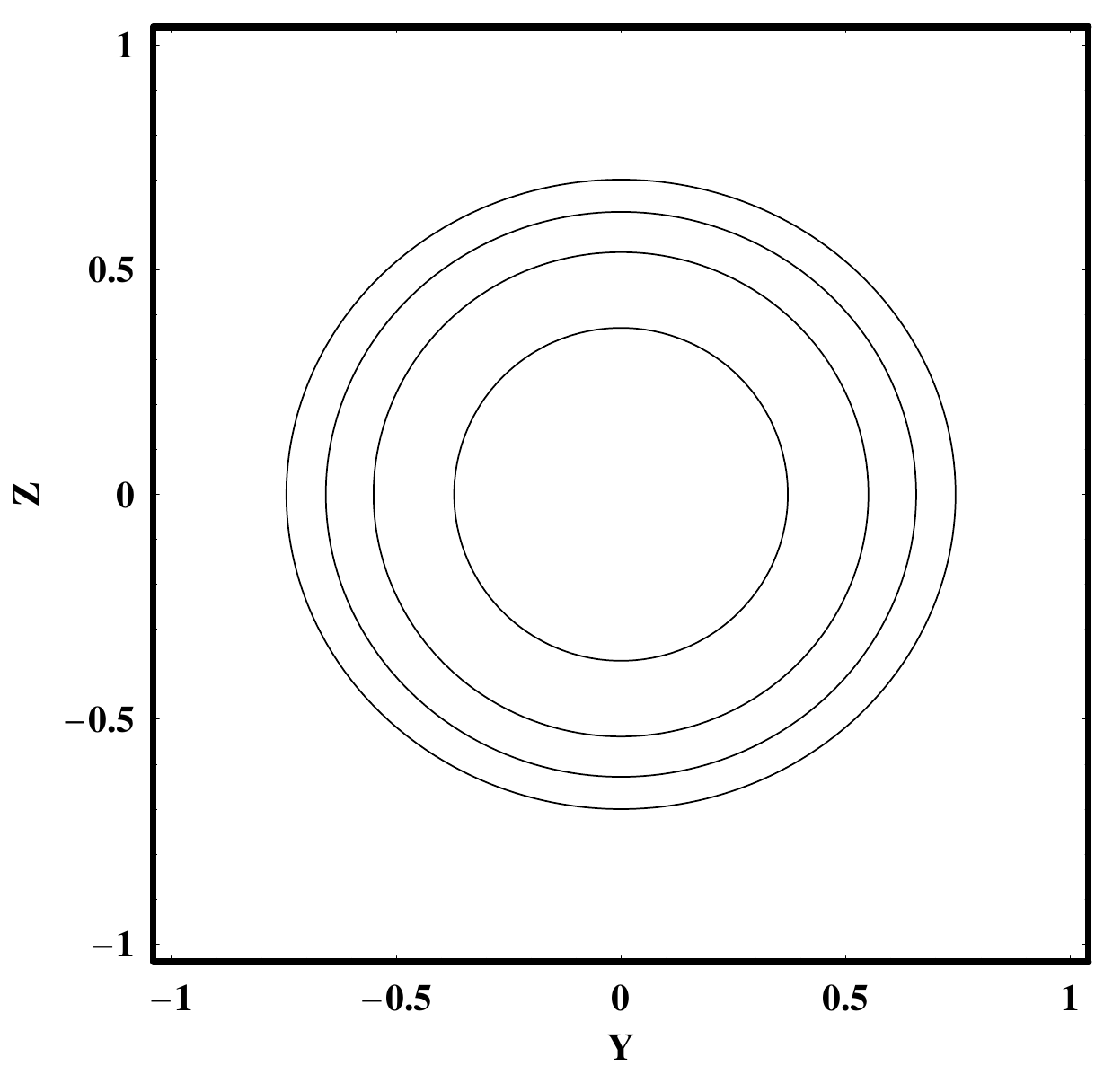}}}

\begin{minipage}{40mm}\centering

\hs\hs\hs {\fns(e)}
\end{minipage}\hspace{20mm}
\begin{minipage}{30mm}
\centering {\fns(f)}~~~~~~~~\end{minipage}

\vspace{-3mm} \caption{\baselineskip 3.6mm \label{fig10}(a)--(f):
Contours of equal density in the $xy, xz$ and $yz$ planes. Left
patterns correspond to potential $V_{\rm tg}$, while right patterns
to potential $V_{\rm tl}$.}
\end{figure*}

\begin{figure*}
\centering
\resizebox{0.7\hsize}{!}{\rotatebox{0}{\includegraphics*{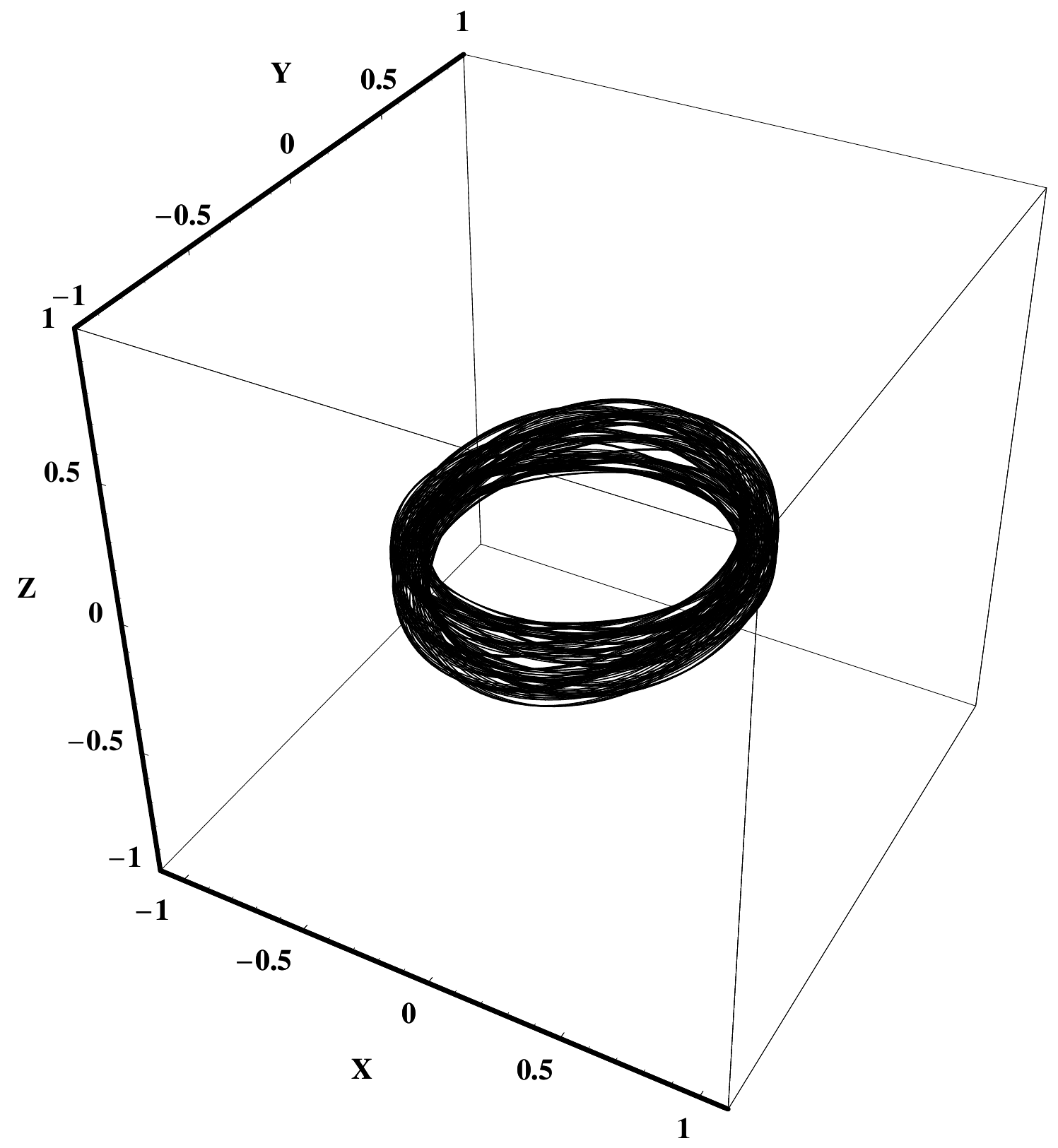}}
                         \rotatebox{0}{\includegraphics*{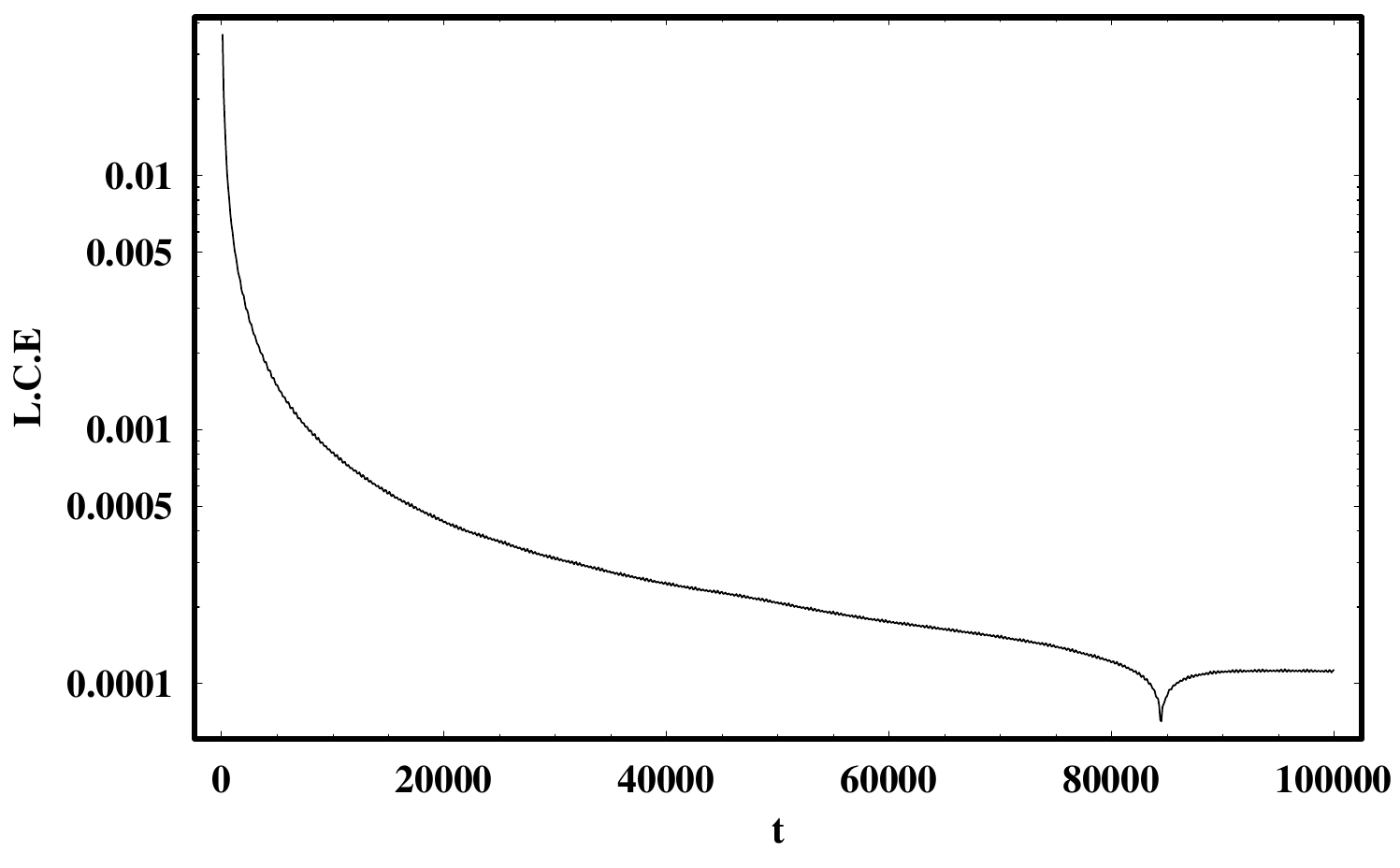}}}

\begin{minipage}{40mm}\centering

\hs\hs\hs {\fns(a)}
\end{minipage}\hspace{20mm}
\begin{minipage}{30mm}
\centering {\fns(b)}~~~~~~~~\end{minipage}

\vs
\resizebox{0.7\hsize}{!}{\rotatebox{0}{\includegraphics*{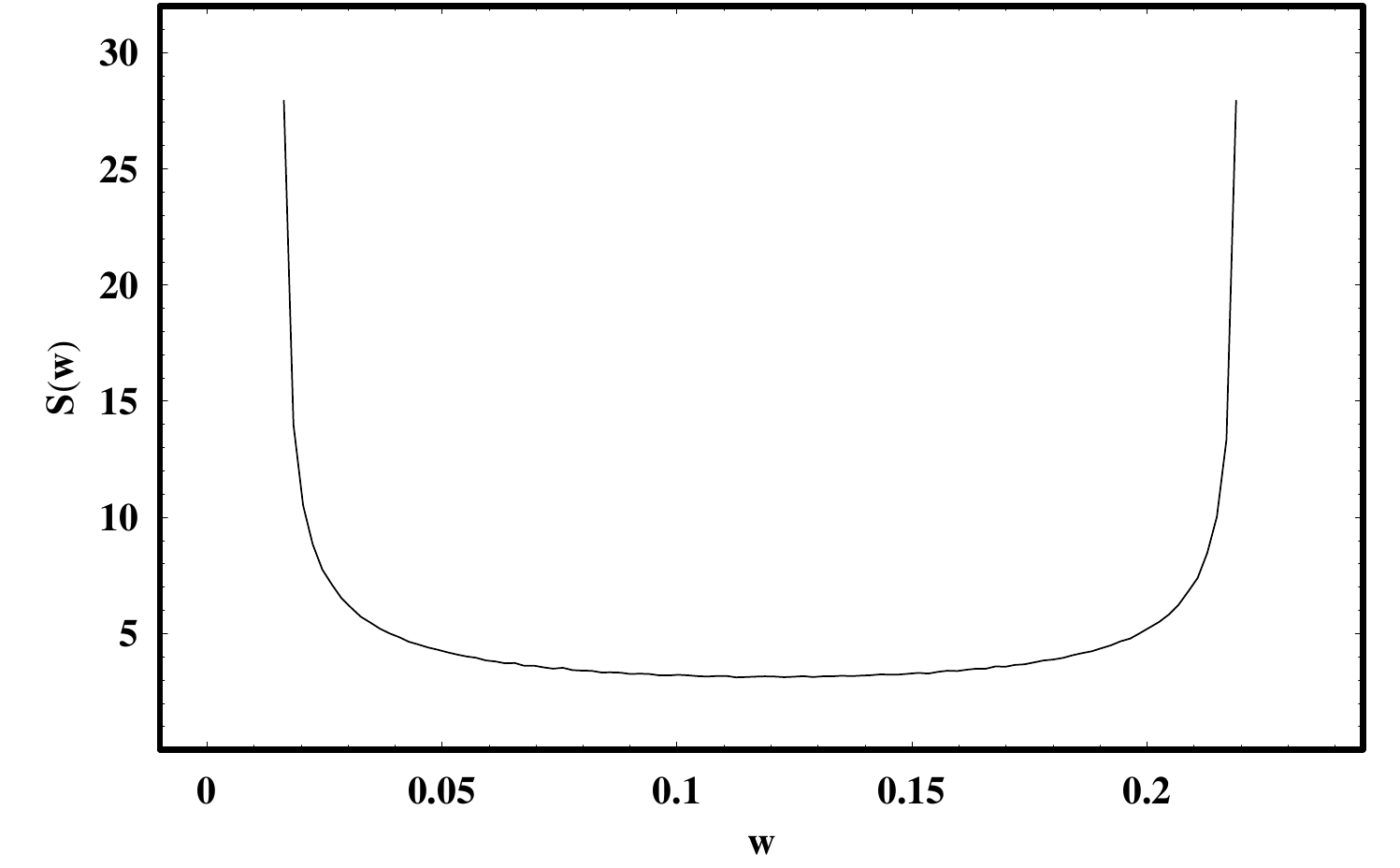}}
                         \rotatebox{0}{\includegraphics*{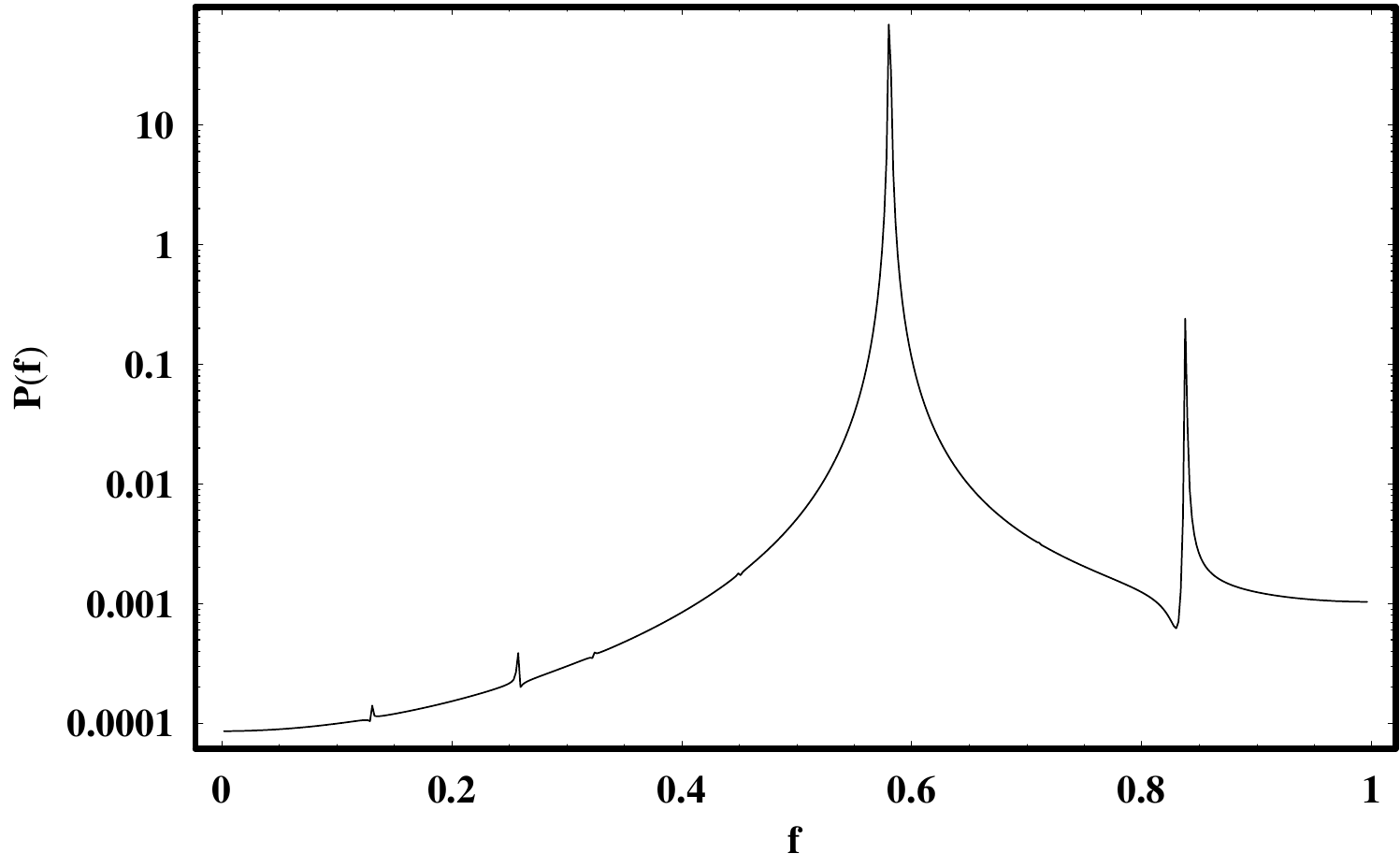}}}

\begin{minipage}{40mm}\centering

\hs\hs {\fns(c)}
\end{minipage}\hspace{20mm}
\begin{minipage}{30mm}
\centering {\fns(d)}~~~~~~~~\end{minipage}

\vspace{-3mm}

\caption{\baselineskip 3.6mm \label{fig11}(a)
A regular orbit in the 3D potential $V_{\rm tg}$. Initial
conditions are: $x_0=0.5, y_0=0, p_{x0}=0, z_0=0.1$, while
$p_{y0}$ is found from the energy integral. The values of all
other parameters and energy are as in Fig.~\ref{fig1}.
(b) 
A plot of the maximum LCE vs time for the orbit shown in (a). (c)
The $S(w)$ spectrum of the orbit shown in (a) and (d)
The $P(f)$ indicator for the orbit
shown in (a).}
\end{figure*}

\begin{figure*}
\centering
\resizebox{0.65\hsize}{!}{\rotatebox{0}{\includegraphics*{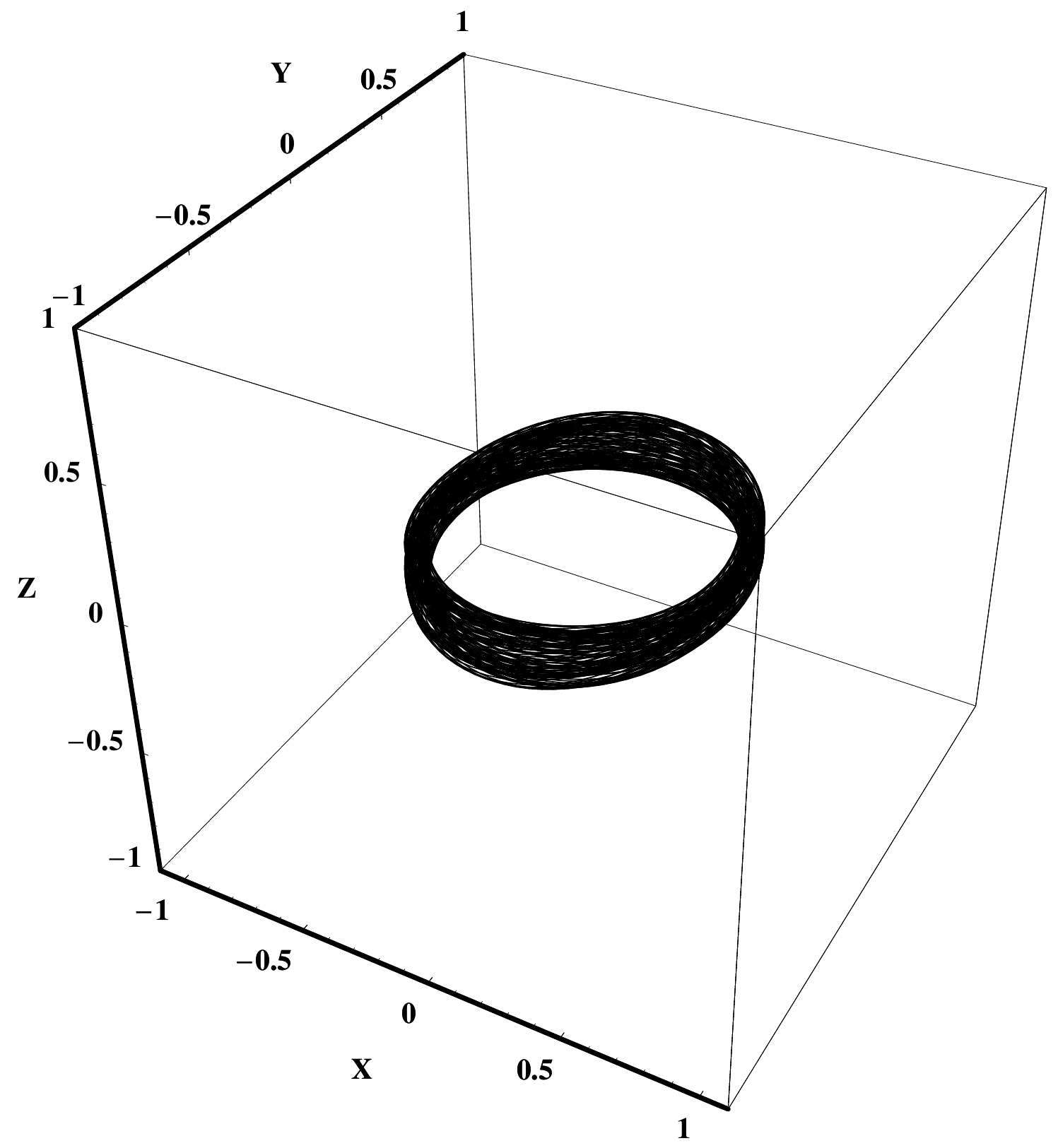}}~~~~~
                          \rotatebox{0}{\includegraphics*{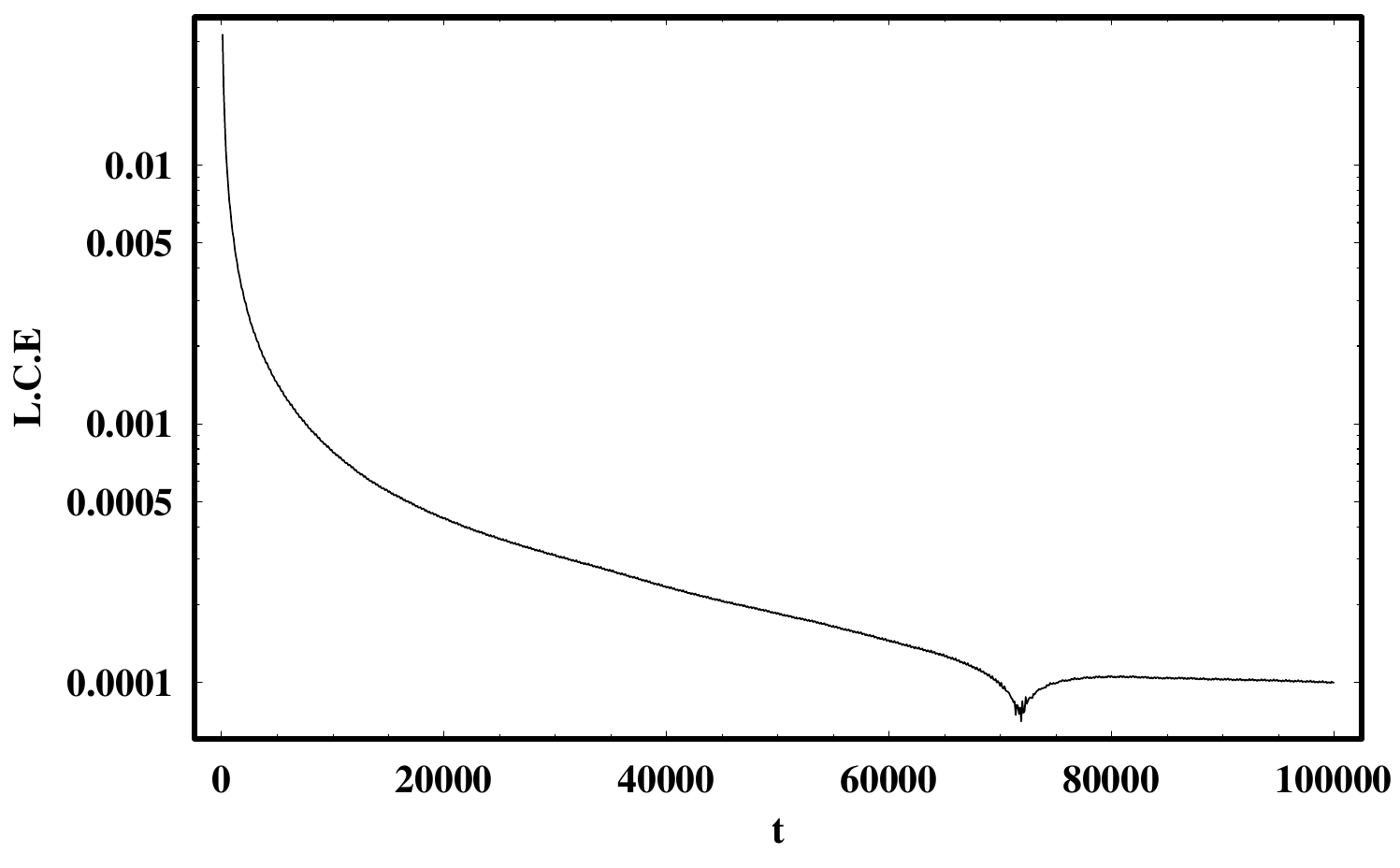}}}

\begin{minipage}{40mm}\centering

\hs\hs\hs {\fns(a)}
\end{minipage}\hspace{20mm}
\begin{minipage}{21mm}
\centering {\fns(b)}~~~~~~~~\end{minipage}

\vs

\resizebox{0.65\hsize}{!}{\rotatebox{0}{\includegraphics*{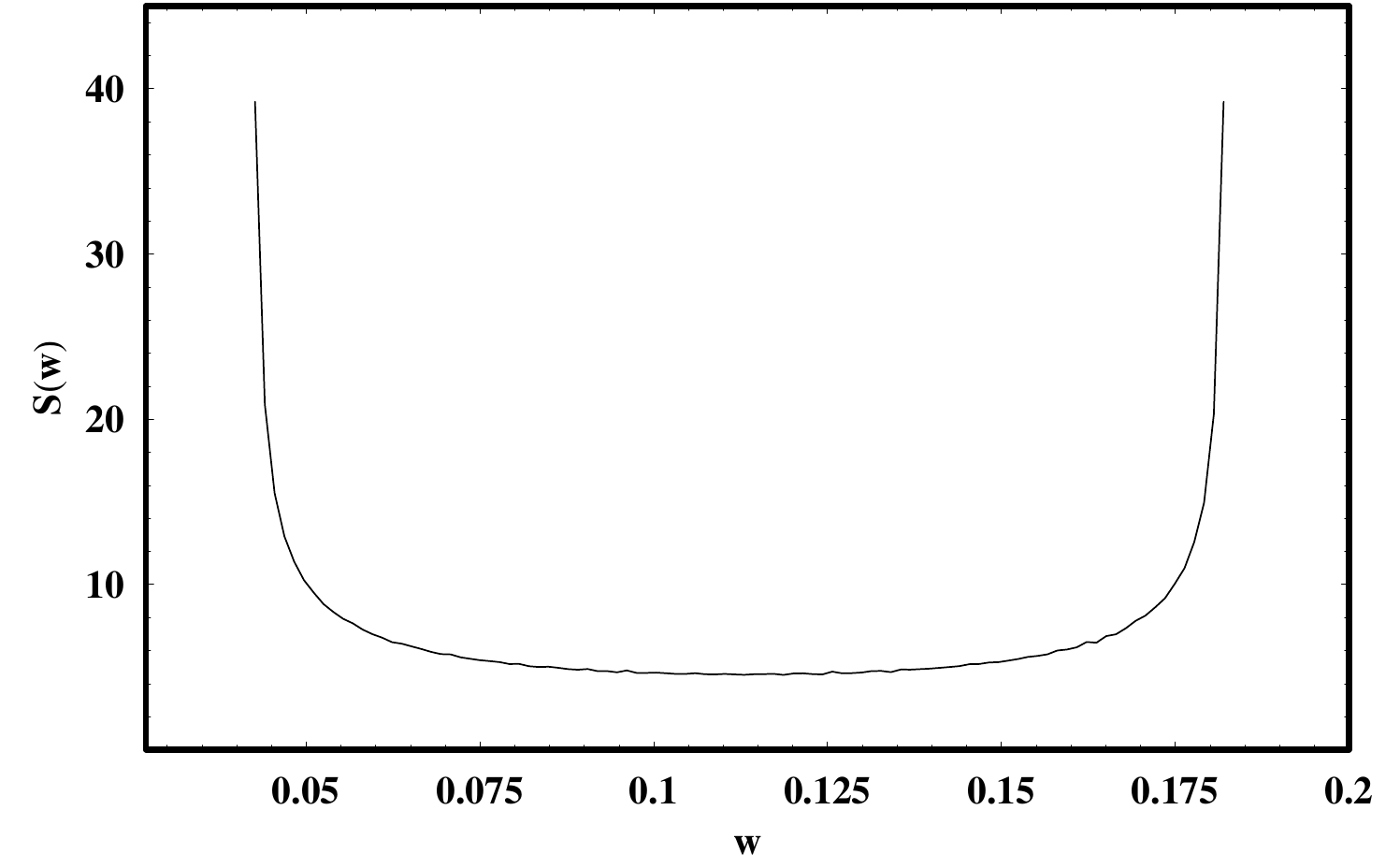}}~~~~~
                          \rotatebox{0}{\includegraphics*{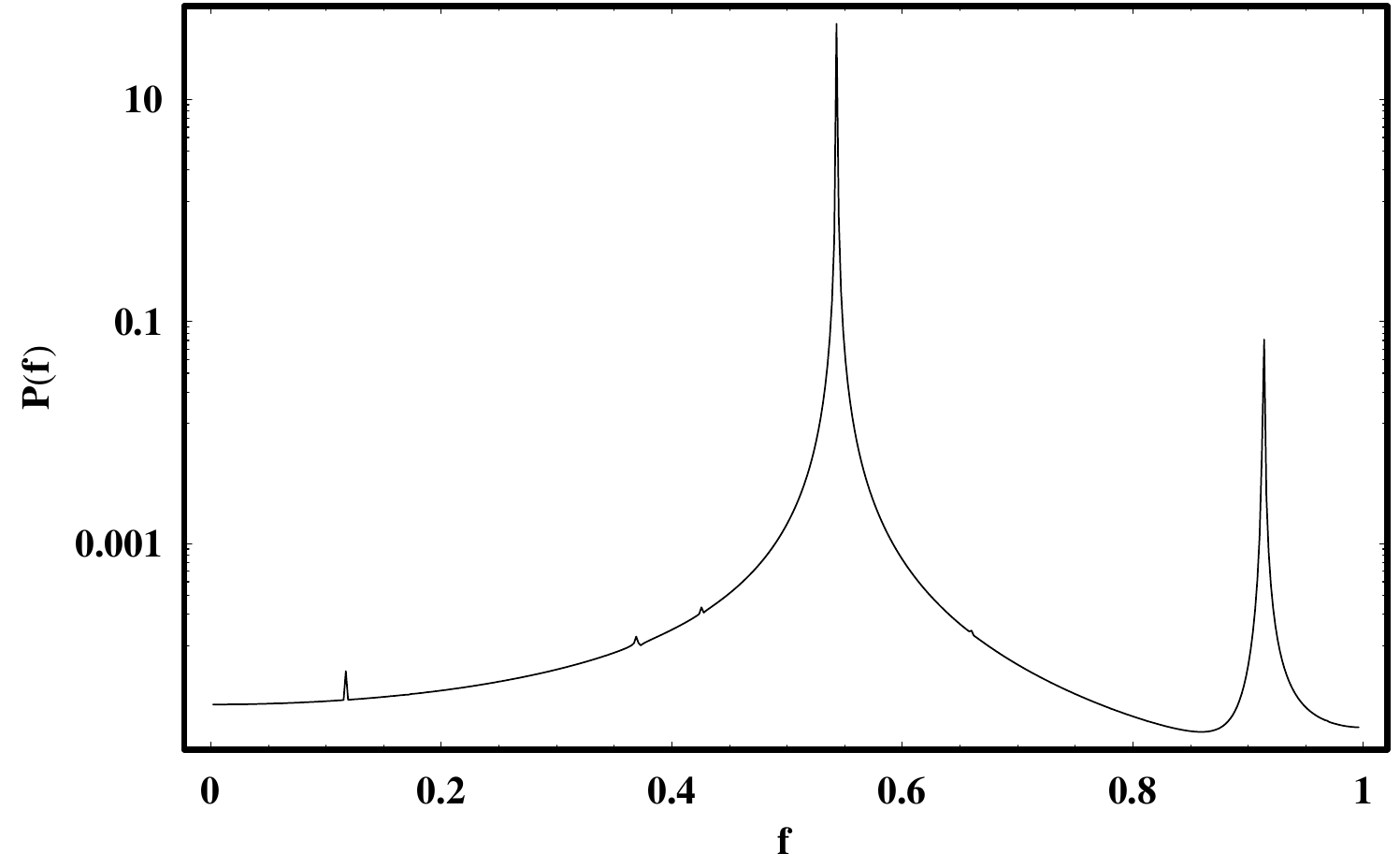}}}

\begin{minipage}{40mm}\centering

\hs\hs\hs {\fns(c)}
\end{minipage}\hspace{20mm}
\begin{minipage}{30mm}
\centering {\fns(d)}~~~~~~~~\end{minipage}

\vspace{-3mm} \caption{\baselineskip 3.6mm \label{fig12}(a)--(d):
Similar {to} Fig.~\ref{fig11}(a)--(d) for the potential $V_{\rm
tl}$. The values of all other parameters and energy are {the same
}as in Fig.~\ref{fig2}.}

\vs \centering
\resizebox{0.68\hsize}{!}{\rotatebox{0}{\includegraphics*{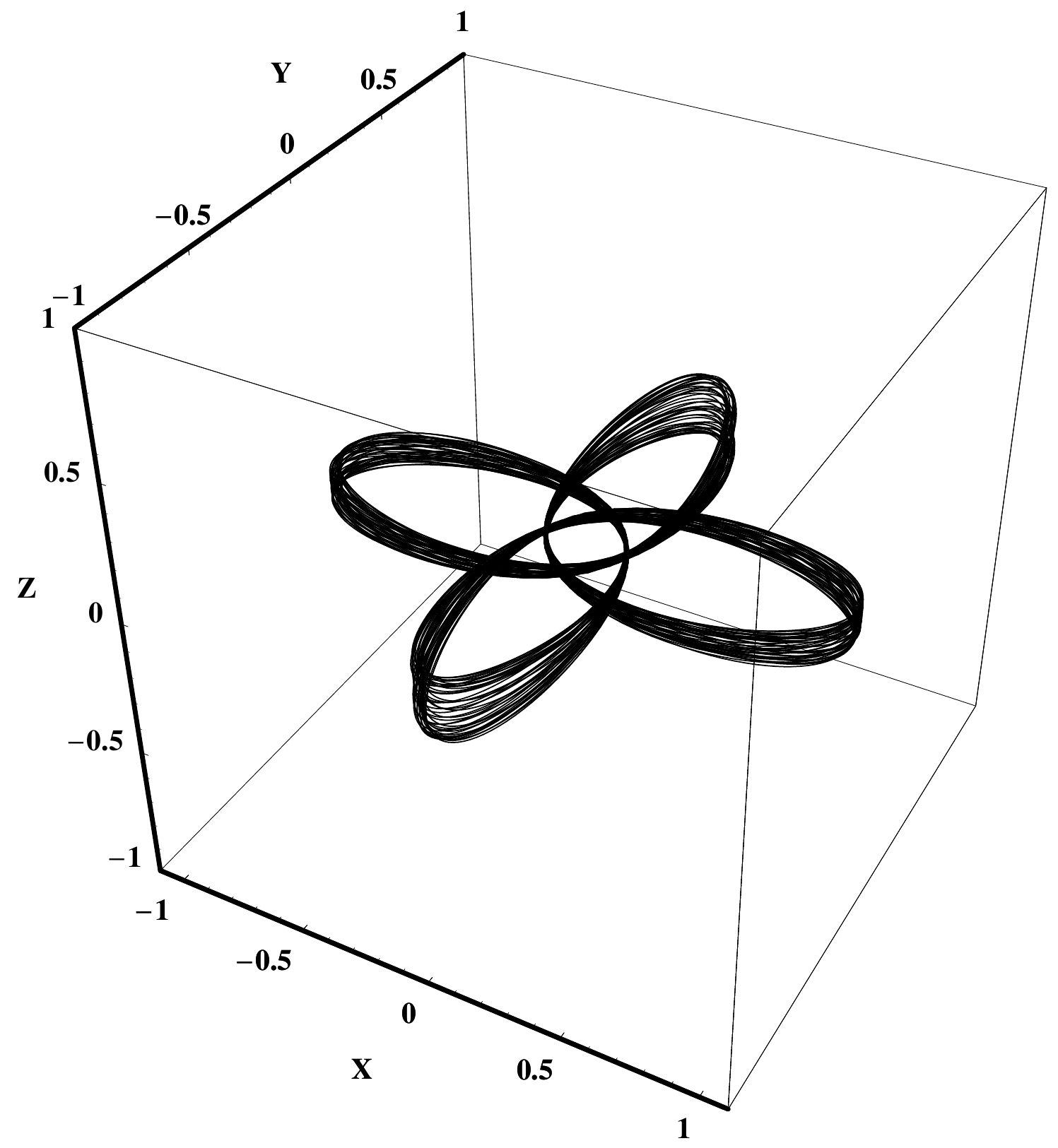}}~~~~~
                          \rotatebox{0}{\includegraphics*{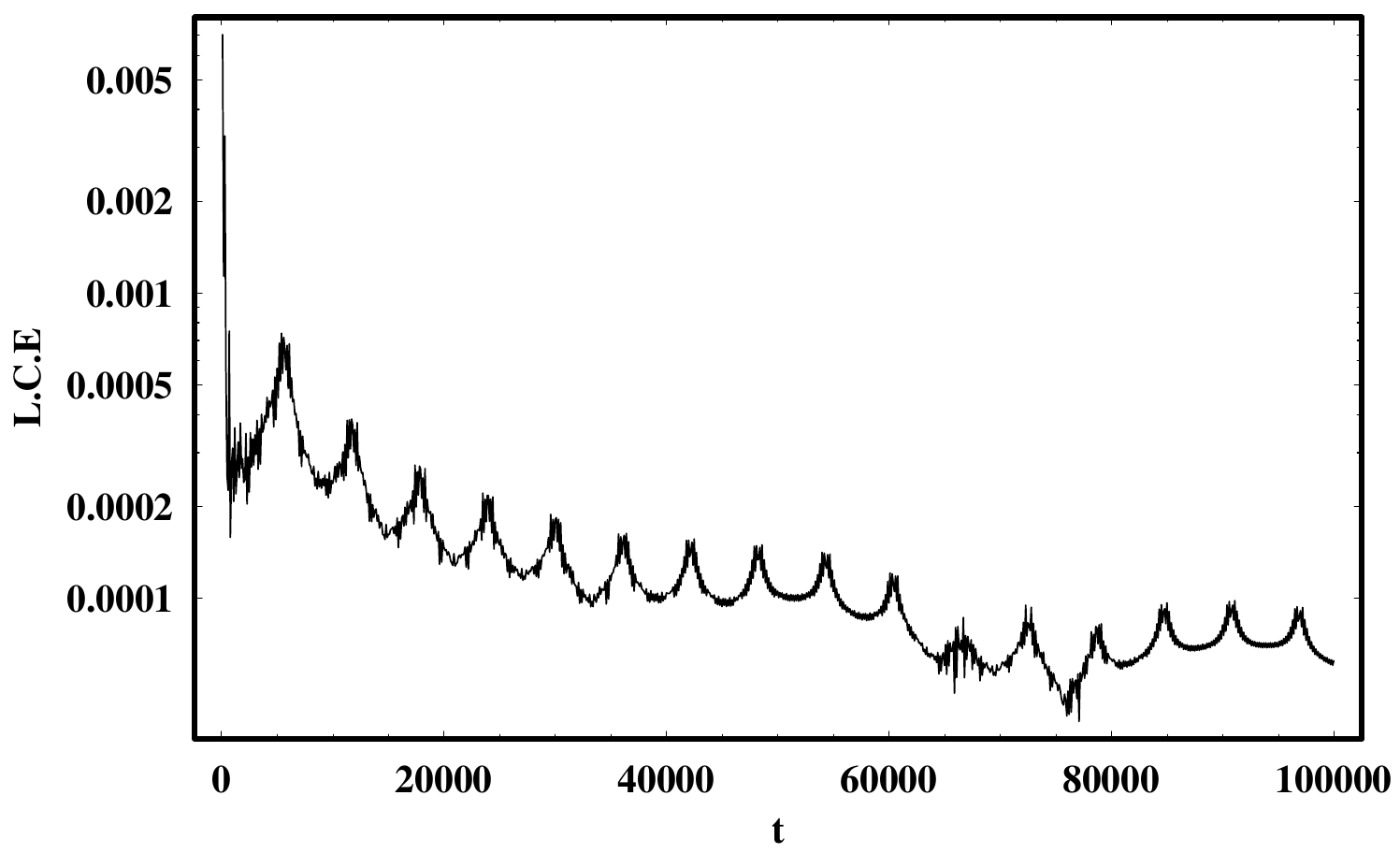}}}

\begin{minipage}{40mm}\centering

\hs\hs\hs {\fns(a)}
\end{minipage}\hspace{20mm}
\begin{minipage}{30mm}
\centering {\fns(b)}~~~~~~~~\end{minipage}

\vs

\resizebox{0.68\hsize}{!}{\rotatebox{0}{\includegraphics*{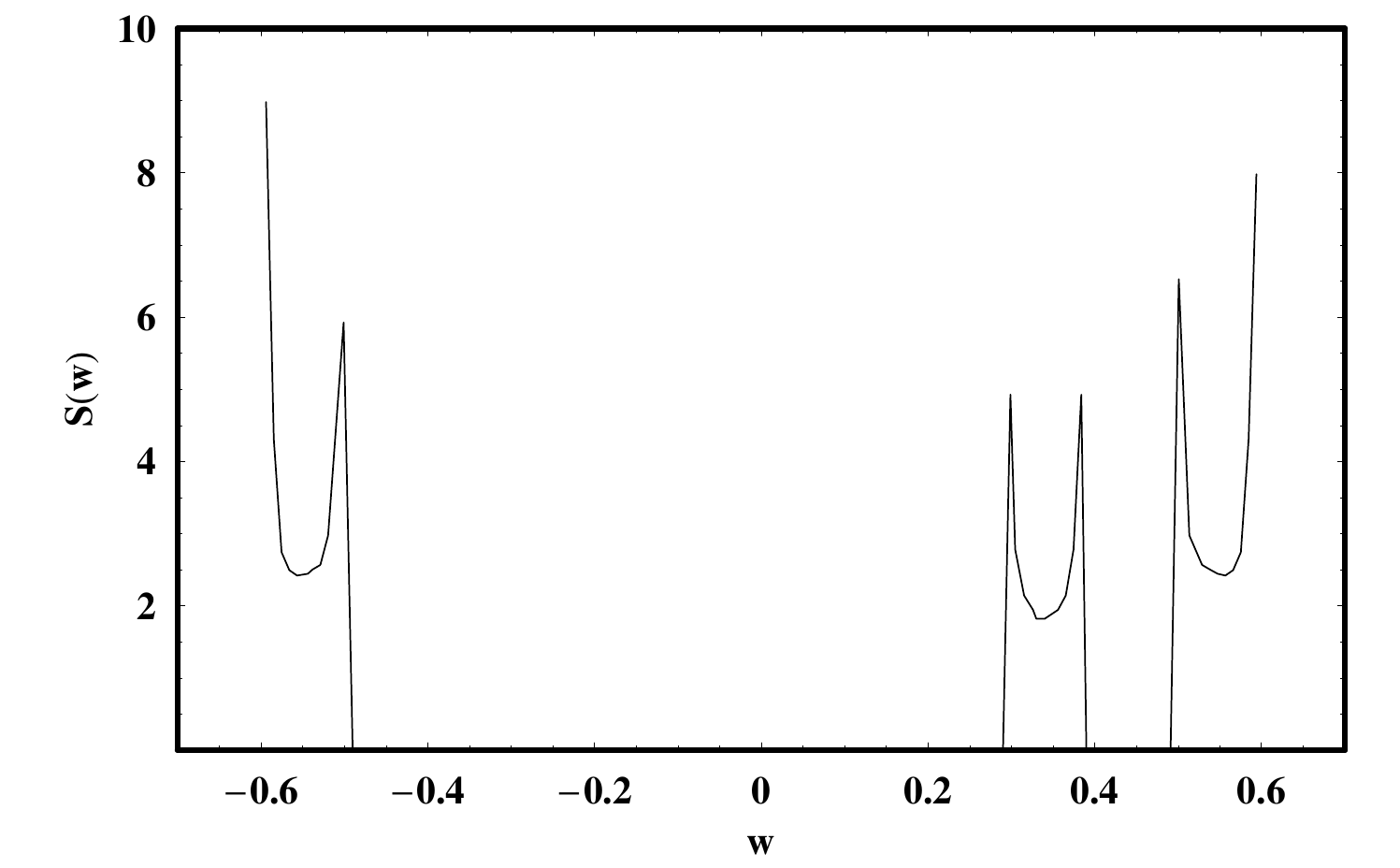}}~~~~~
                          \rotatebox{0}{\includegraphics*{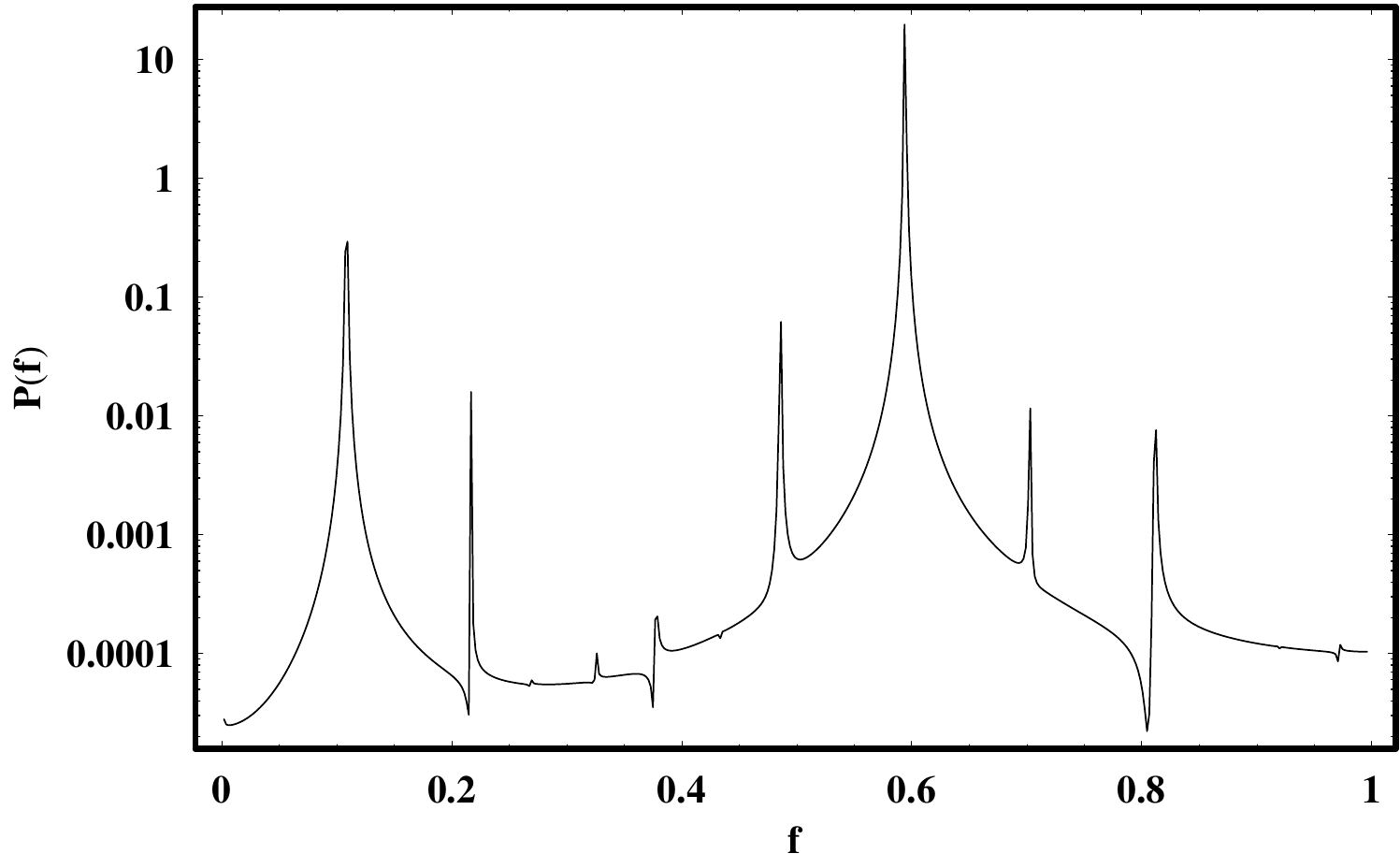}}}

\begin{minipage}{40mm}\centering

\hs\hs\hs {\fns(c)}
\end{minipage}\hspace{20mm}
\begin{minipage}{30mm}
\centering {\fns(d)}~~~~~~~~\end{minipage}

\vspace{-3mm} \caption{\baselineskip 3.6mm \label{fig13}(a)--(d):
Similar {to} Fig.~\ref{fig11}(a)--(d) for a resonant 3D orbit.
Initial conditions are: $x_0=0.15, y_0=0, p_{x0}=4.5, z_0=0.01$. The
values of all other parameters and energy are {the same }as in
Fig.~\ref{fig1}.}
\end{figure*}

\begin{figure*}
\centering
\resizebox{0.68\hsize}{!}{\rotatebox{0}{\includegraphics*{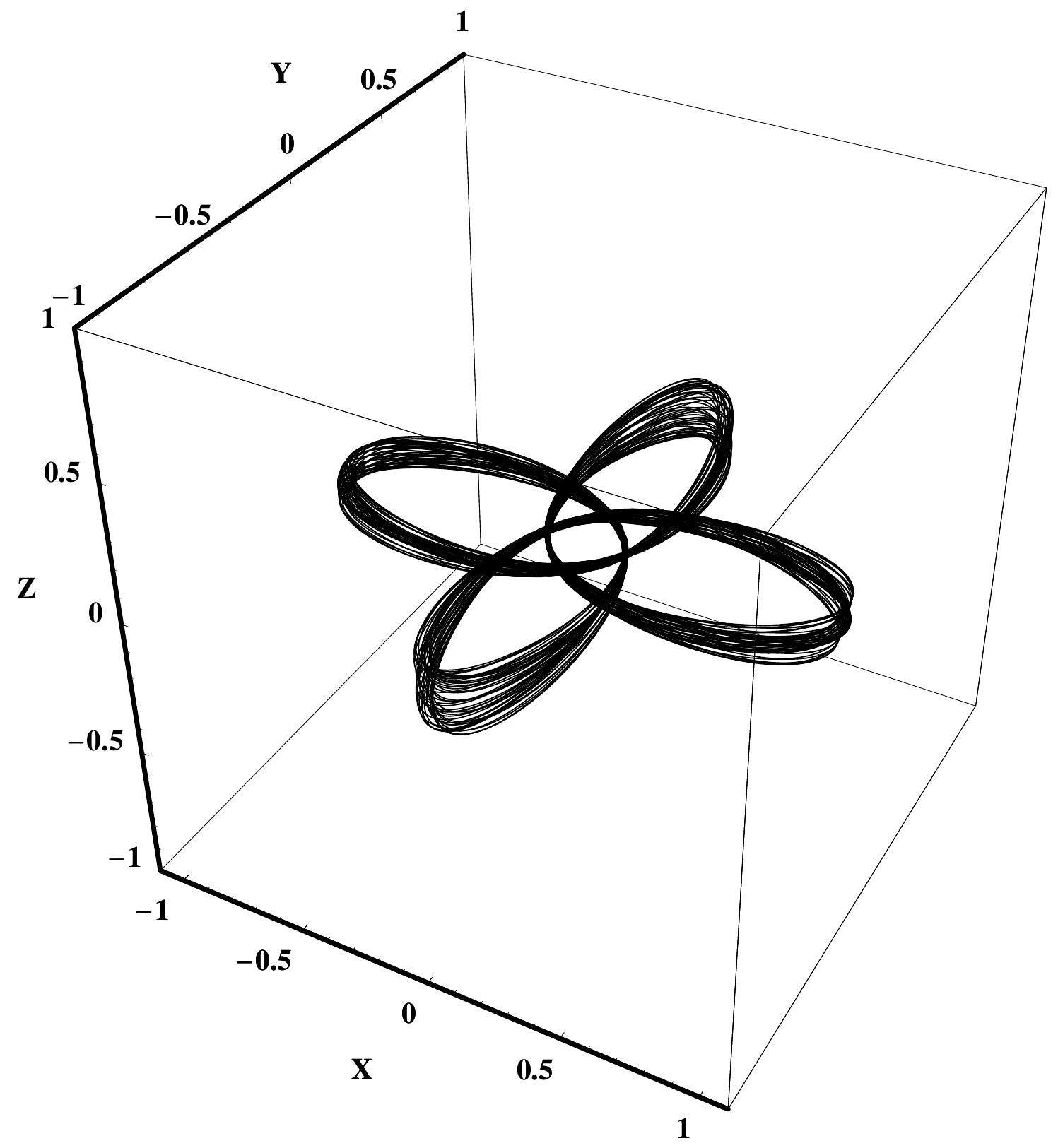}}~~~~~
                          \rotatebox{0}{\includegraphics*{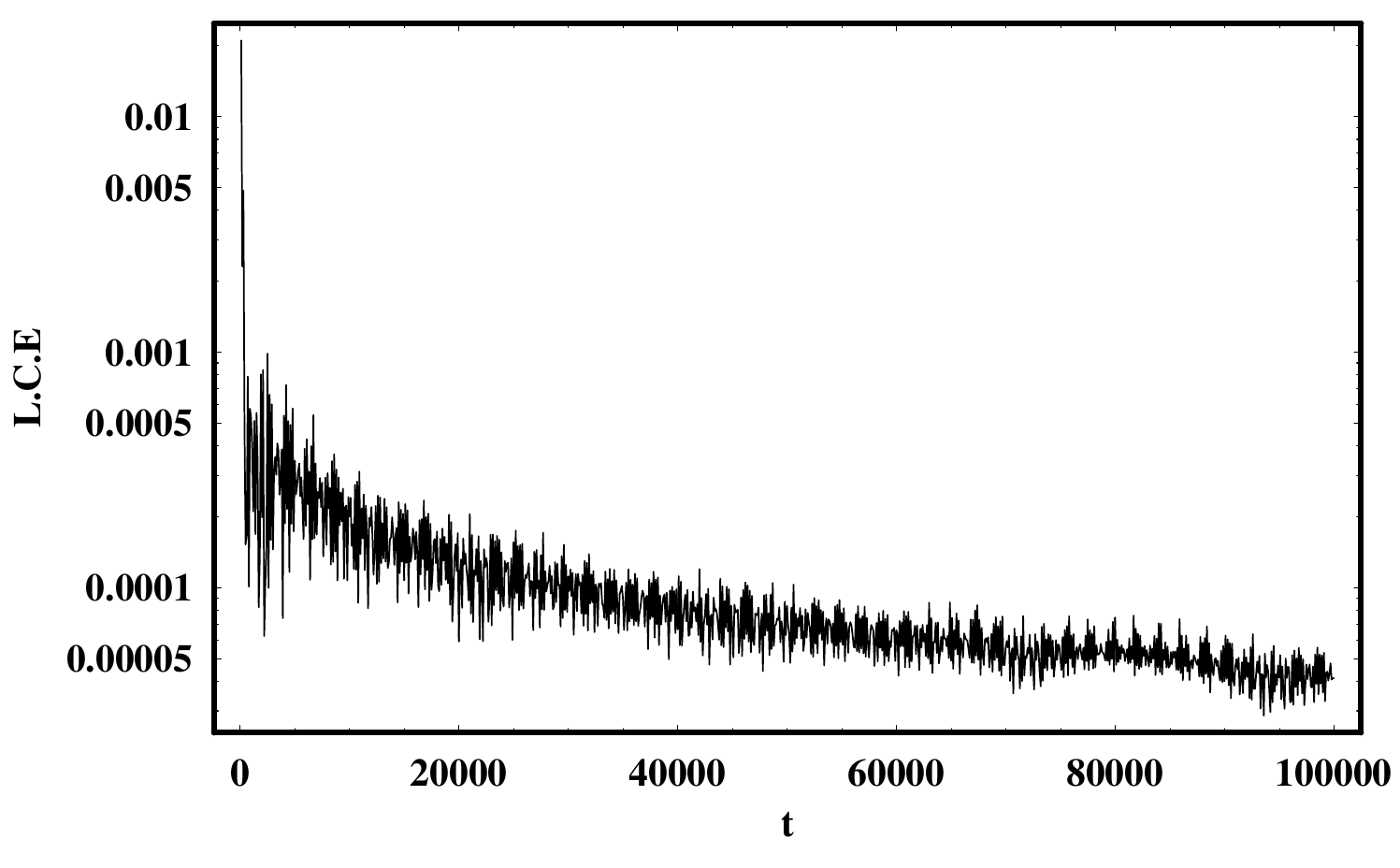}}}

\begin{minipage}{40mm}\centering

\hs\hs\hs {\fns(a)}
\end{minipage}\hspace{20mm}
\begin{minipage}{30mm}
\centering {\fns(b)}~~~~~~~~\end{minipage}

\vs

\resizebox{0.68\hsize}{!}{\rotatebox{0}{\includegraphics*{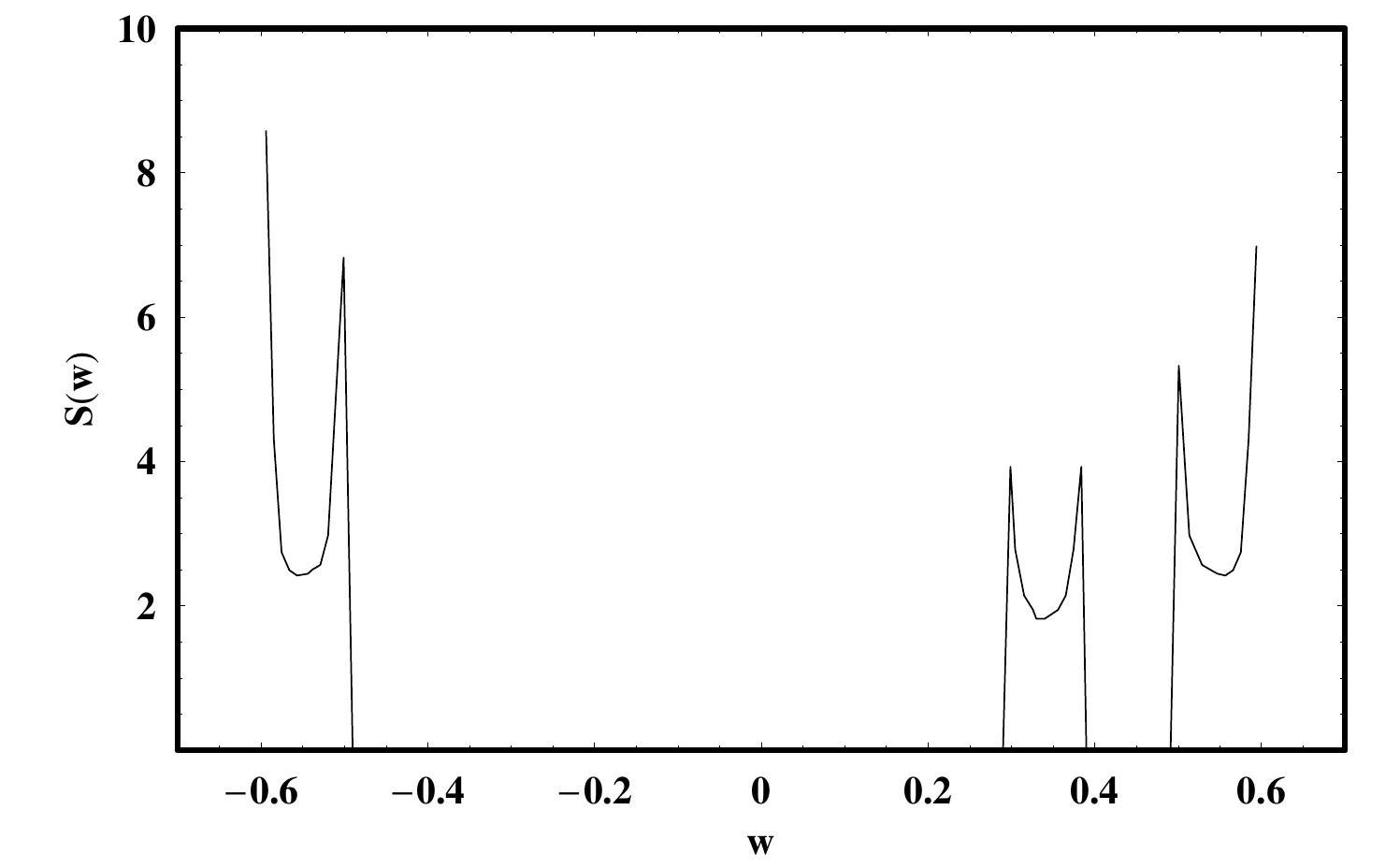}}~~~~~
                          \rotatebox{0}{\includegraphics*{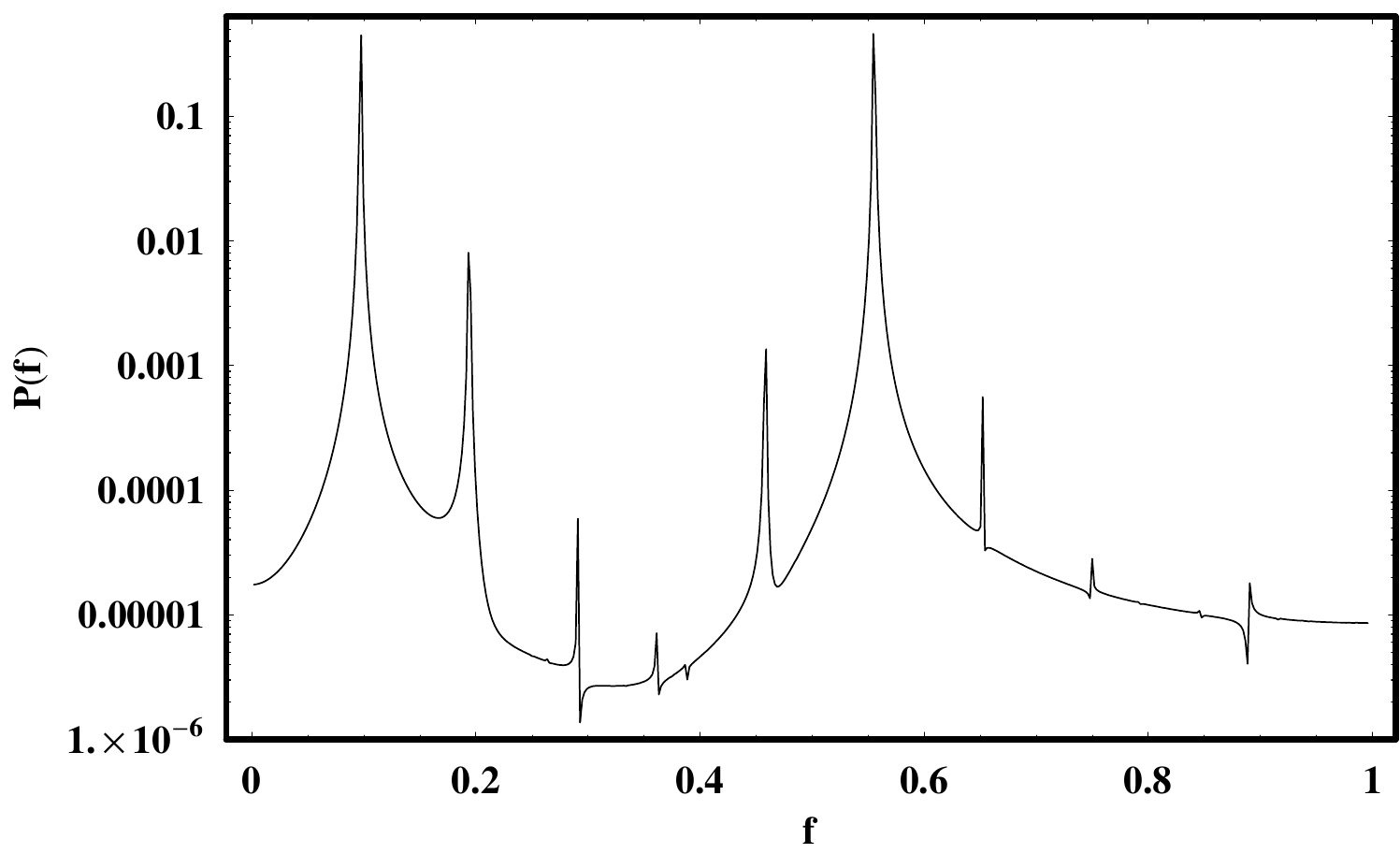}}}

\begin{minipage}{40mm}\centering

\hs\hs\hs {\fns(c)}
\end{minipage}\hspace{20mm}
\begin{minipage}{30mm}
\centering {\fns(d)}~~~~~~~~\end{minipage}

\vspace{-3mm} \caption{\baselineskip 3.6mm \label{fig14}(a)--(d):
Similar {to} Fig.~\ref{fig13}(a)--(d) for the potential $V_{\rm
tl}$. The values of all other parameters and energy are {the same
}as in Fig.~\ref{fig2}.}
\centering
\resizebox{0.68\hsize}{!}{\rotatebox{0}{\includegraphics*{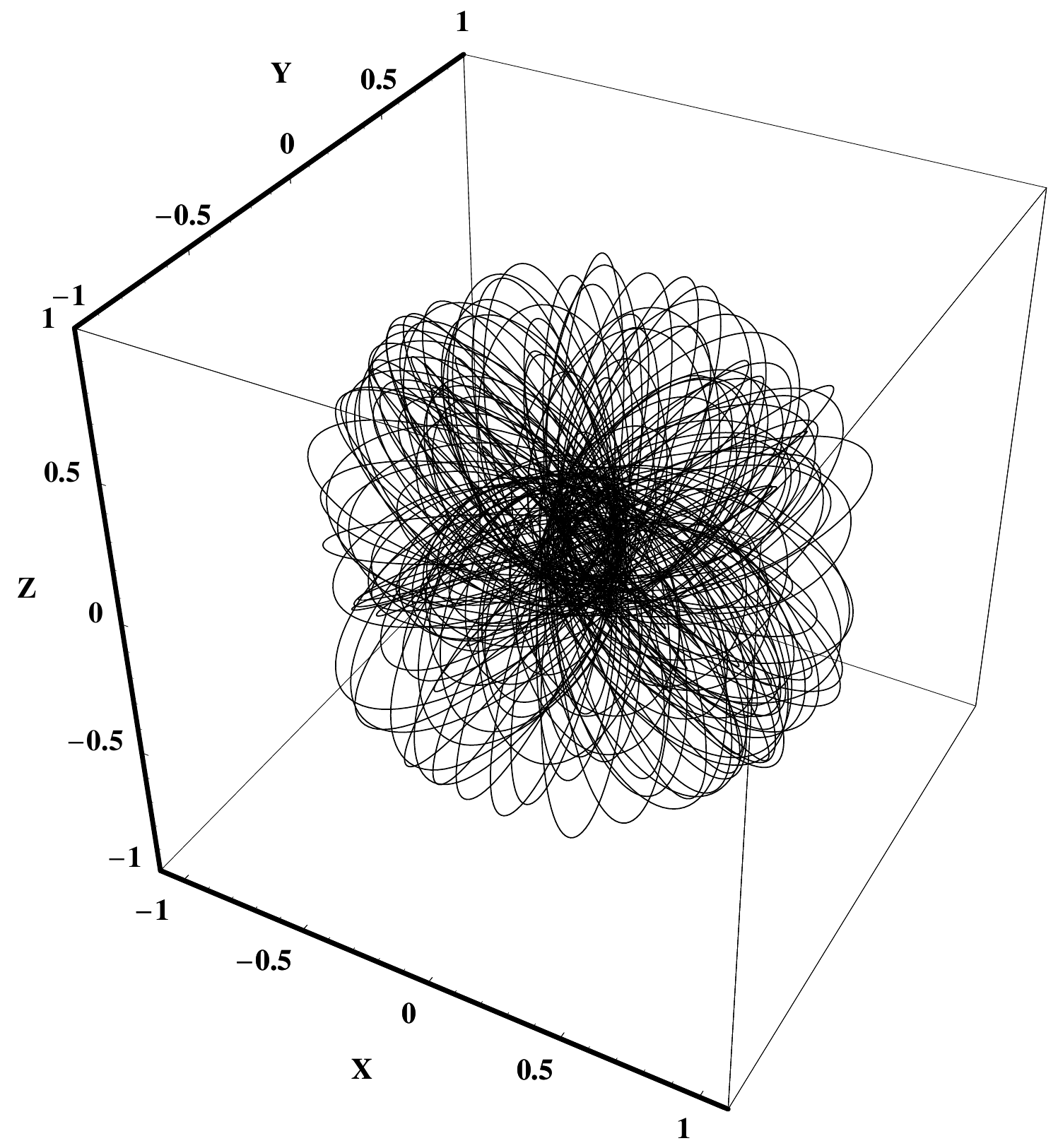}}
                          \rotatebox{0}{\includegraphics*{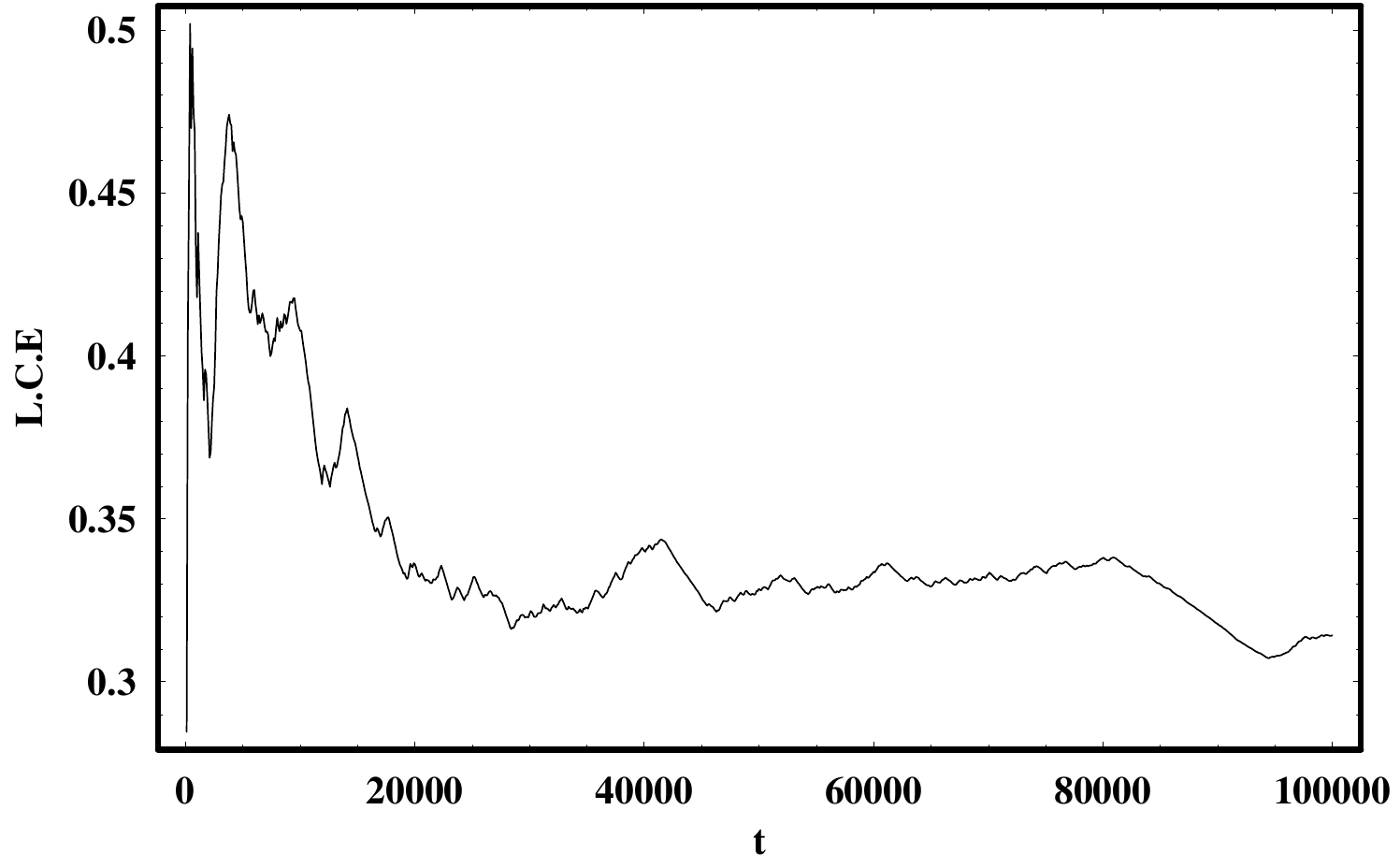}}}

\begin{minipage}{40mm}\centering

\hs\hs {\fns(a)}
\end{minipage}\hspace{20mm}
\begin{minipage}{24mm}
\centering {\fns(b)}~~~~~~~~\end{minipage}

\vs

\resizebox{0.68\hsize}{!}{\rotatebox{0}{\includegraphics*{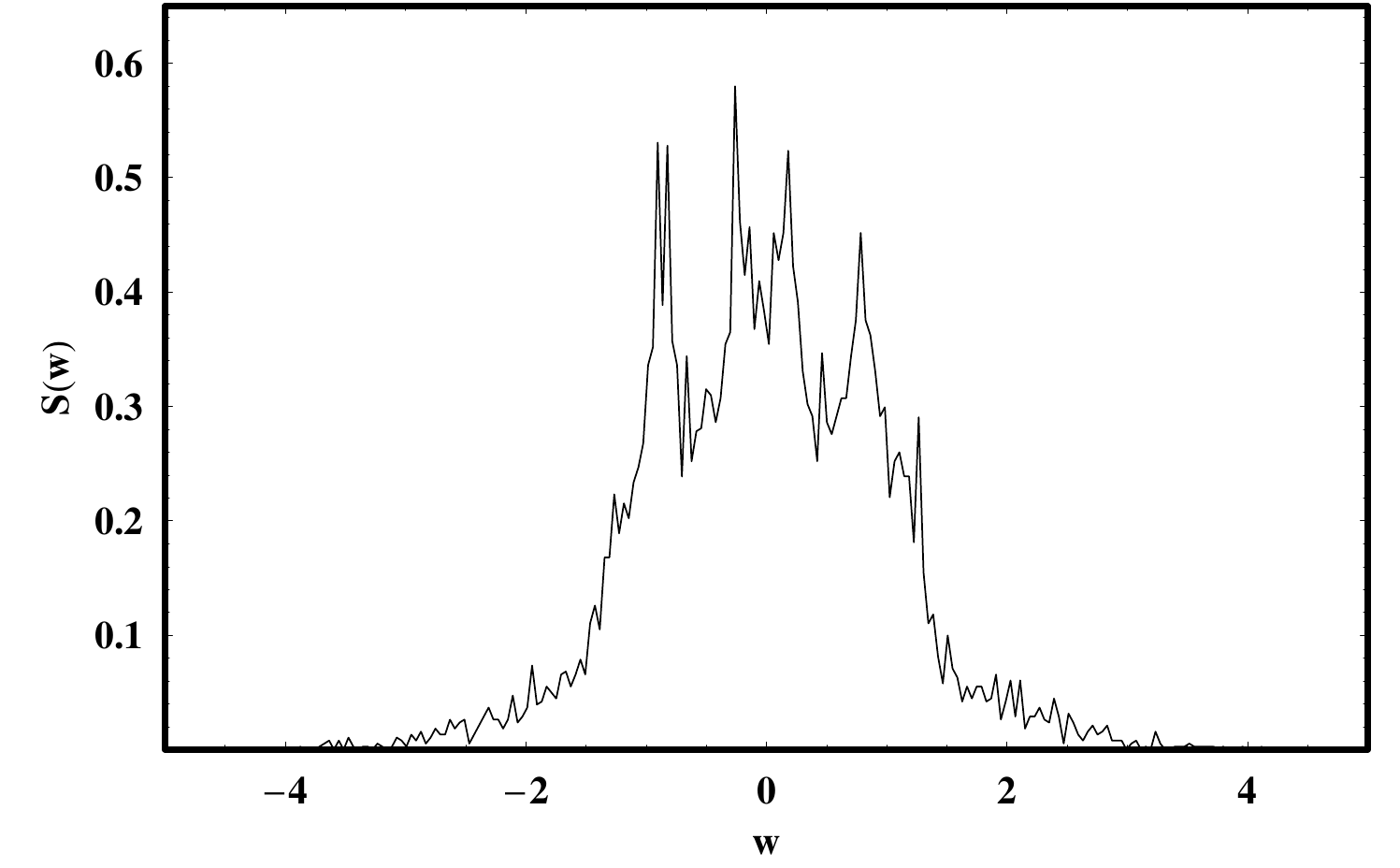}}
                          \rotatebox{0}{\includegraphics*{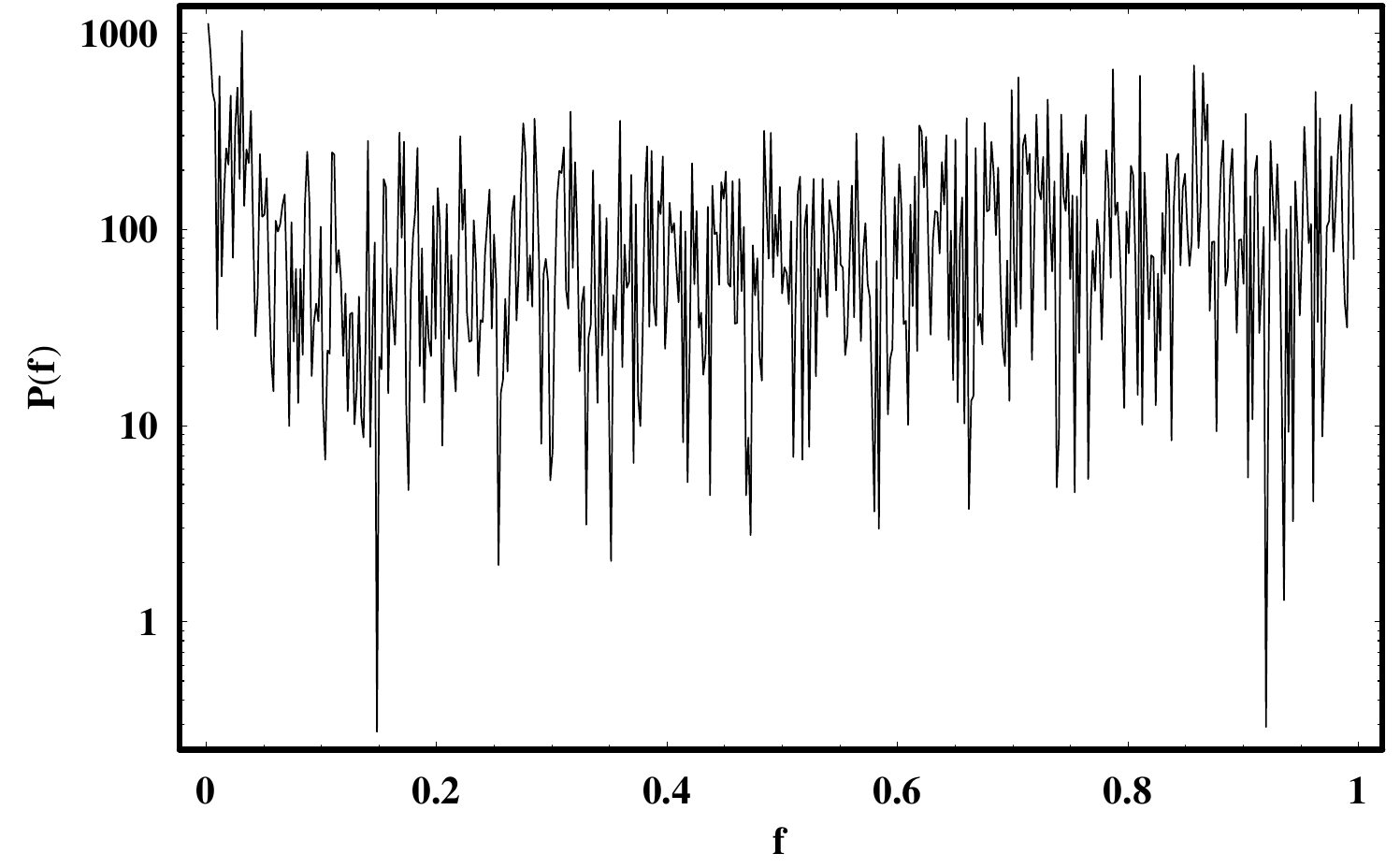}}}

\begin{minipage}{40mm}\centering

\hs\hs\hs {\fns(c)}
\end{minipage}\hspace{20mm}
\begin{minipage}{30mm}
\centering {\fns(d)}~~~~~~~~\end{minipage}

\vspace{-3mm} \caption{\baselineskip 3.6mm \label{fig15}(a)--(d):
Similar {to} Fig.~\ref{fig11}(a)--(d) for a chaotic 3D orbit.
Initial conditions are: $x_0=0.02, y_0=0, p_{x0}=2.5,z_0=0.1$. The
values of all other parameters and energy are {the same }as in
Fig.~\ref{fig1}.}
\end{figure*}

\begin{figure*}
\vs \centering
\resizebox{0.75\hsize}{!}{\rotatebox{0}{\includegraphics*{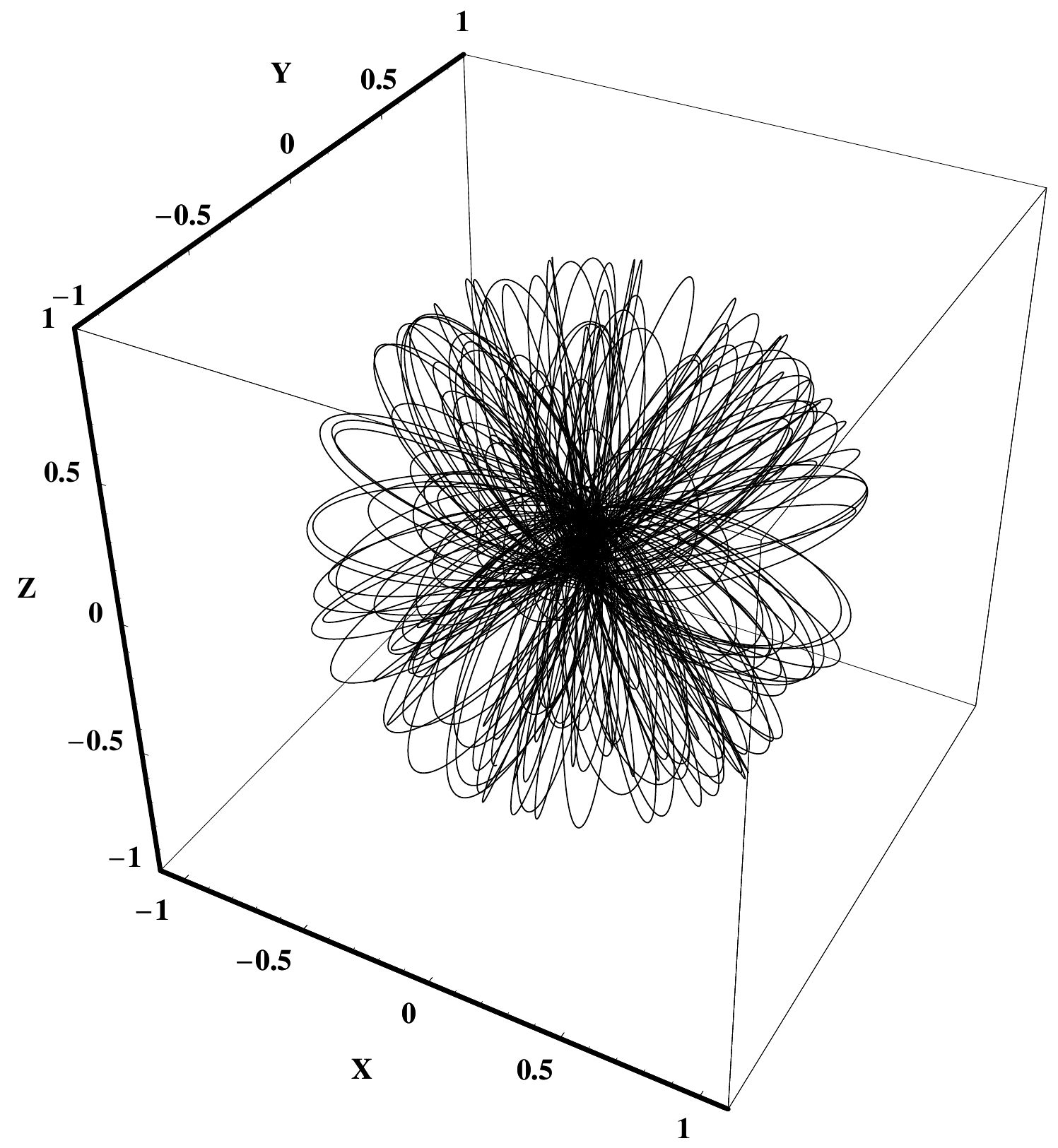}}~~~~~
                          \rotatebox{0}{\includegraphics*{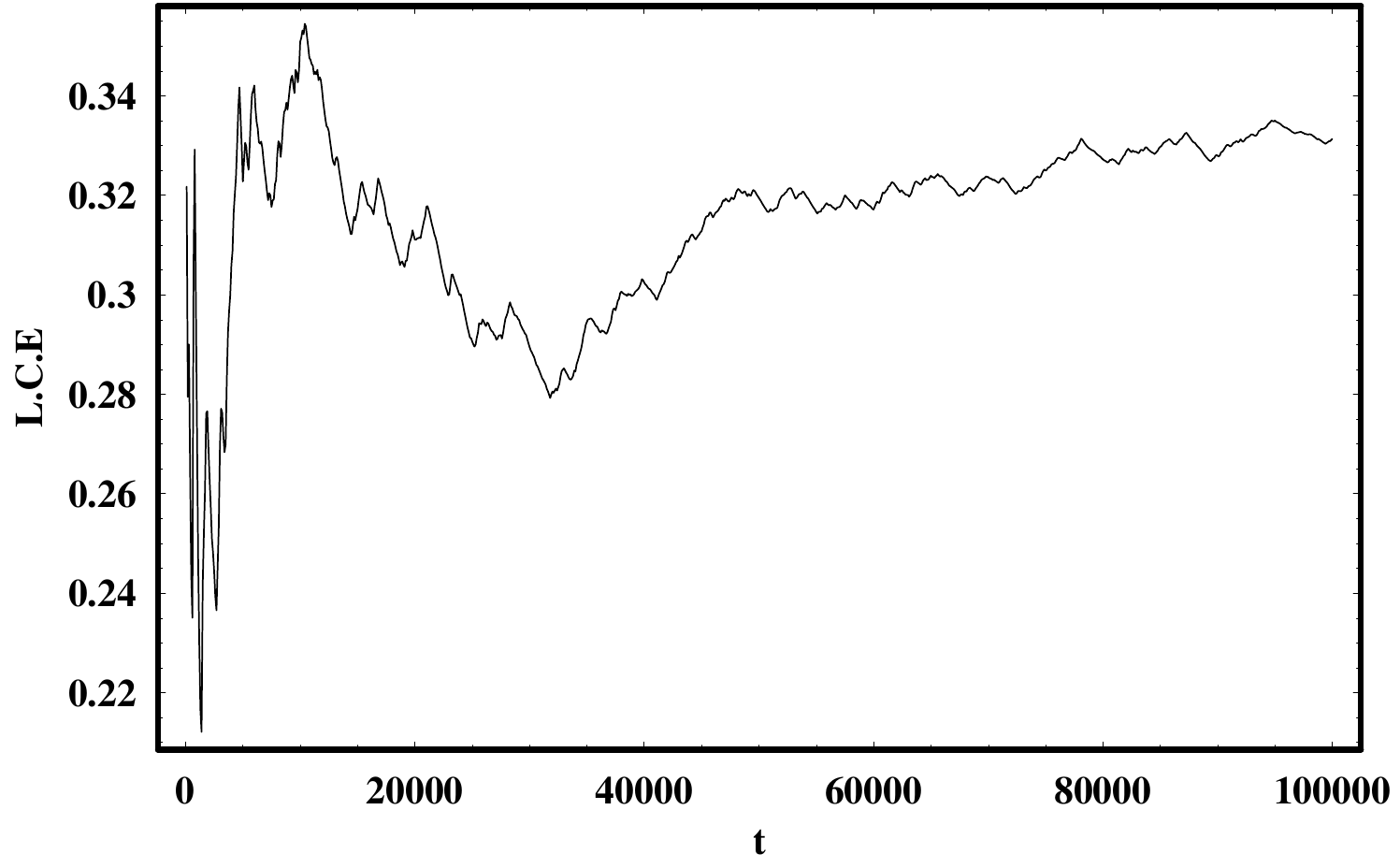}}}

\begin{minipage}{40mm}\centering

\hs\hs\hs {\fns(a)}
\end{minipage}\hspace{20mm}
\begin{minipage}{30mm}
\centering {\fns(b)}~~~~~~~~\end{minipage}

\vs
\resizebox{0.75\hsize}{!}{\rotatebox{0}{\includegraphics*{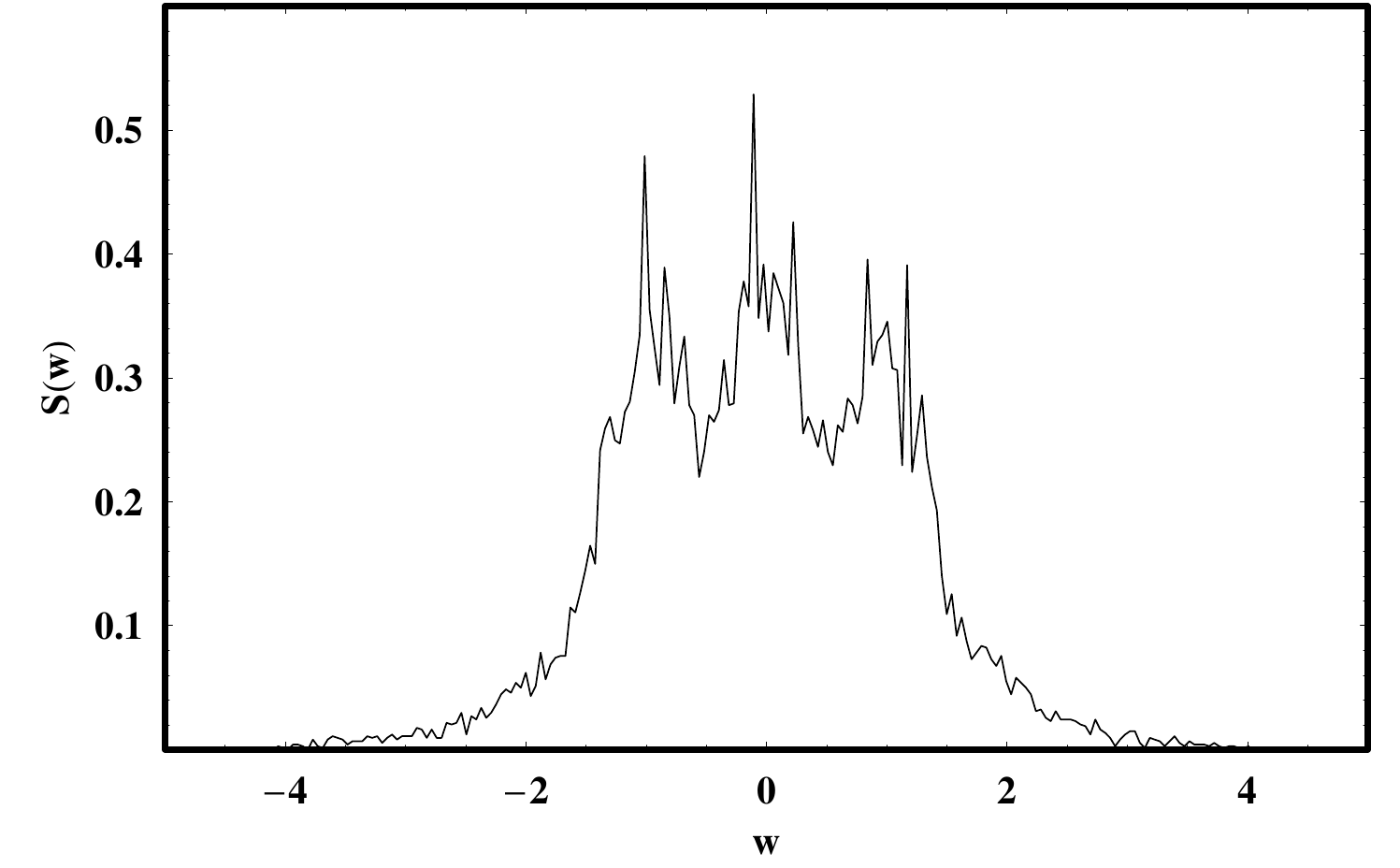}}~~~~~
                          \rotatebox{0}{\includegraphics*{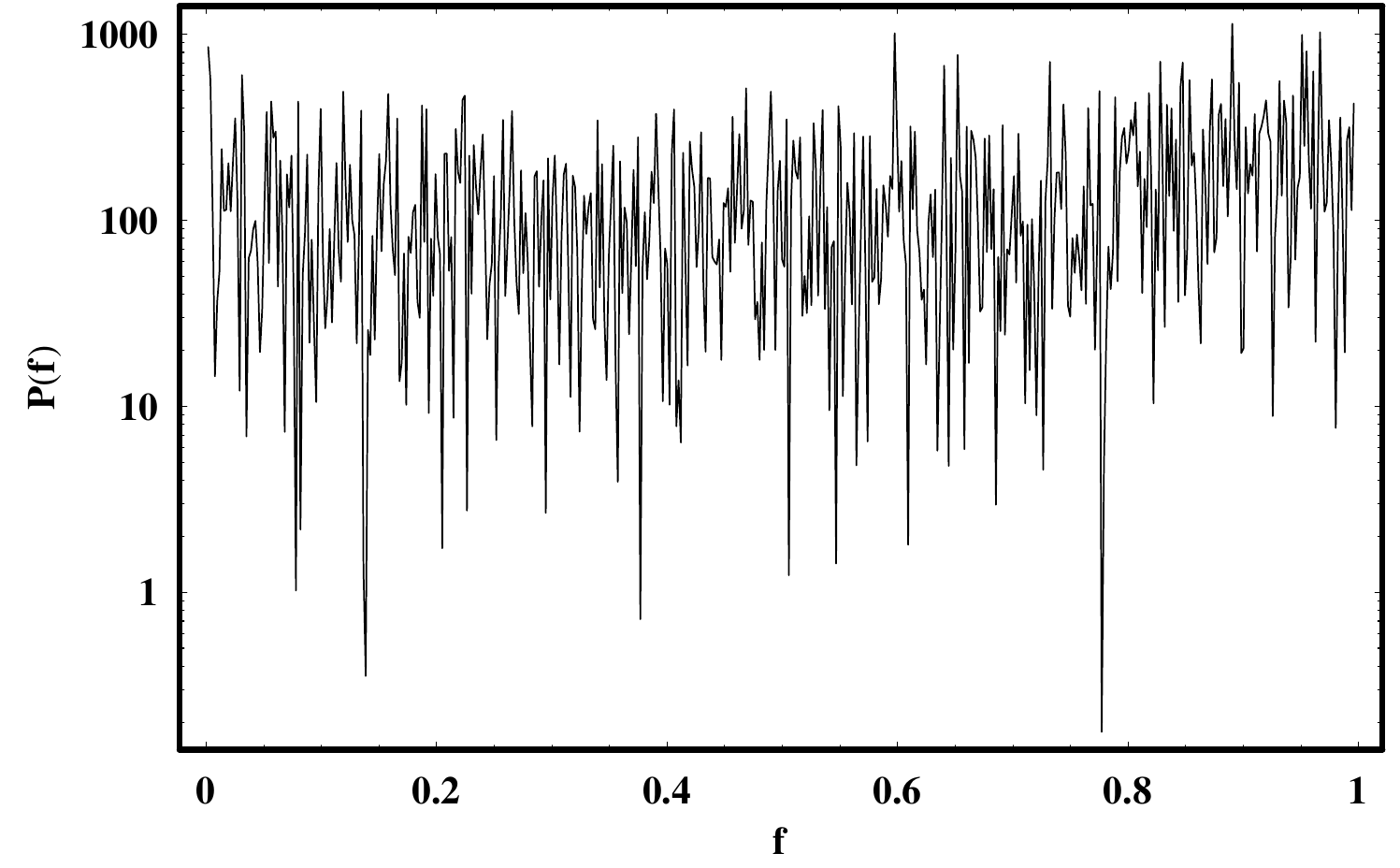}}}

\begin{minipage}{40mm}\centering

\hs {\fns(c)}
\end{minipage}\hspace{20mm}
\begin{minipage}{30mm}
\hs\hs\hs\hs \hs {\fns(d)}\end{minipage}

\caption{\baselineskip 3.6mm \label{fig16}(a)--(d):
Similar {to} Fig.~\ref{fig15}(a)--(d) for the potential $V_{\rm
tl}$. The values of all other parameters and energy are {the same
}as in Fig.~\ref{fig2}.}
\end{figure*}

Given all the above, we can say that our numerical results, {which
are }obtained by several different dynamical methods using regular
and chaotic orbits, strongly suggest that the potential $V_{\rm tl}$
satisfactorily{ describes} the properties of motion of the potential
$V_{\rm tg}$ near the center of a triaxial, elliptical galaxy. Since
the potential of the spherical nucleus $V_{\rm n}$ is the same in
both potentials $V_{\rm tg}$ and $V_{\rm tl}$, this means that no
information is lost when we go from a global triaxial logarithmic
potential (1) to the local polynomial potential (2). Remember that
this  only holds near the center of the galaxy, when (4) is valid.
In order to investigate and compare the character of motion in the
two 2D potentials $V_{\rm tg}$ and $V_{\rm tl}$, we have computed a
large number of orbits - about 1000 - with different initial
conditions $\left(x_0, p_{x0}\right)$, but with the same initial
conditions in both 2D potentials. In particular, as we have in both
cases regular regions and only one unified chaotic sea in each
$x-p_x$ phase plane, we calculate the maximum value of the LCE by
choosing 500 orbits with different and random initial conditions
$\left(x_0, p_{x0}\right)$ in the regular regions and 500 orbits
with different and random initial conditions $\left(x_0,
p_{x0}\right)$ in the chaotic sea in each case. Our numerical
experiments show that the vast majority of orbits - about $97.4 \%$
- displayed the same characteristics, {including} the same {nature
of} orbit, maximum LCE, $S(c)$ spectrum and $P(f)$ indicator, while
only $2.6 \%$ of the tested orbits were different.

\section{Results for the 3D dynamical systems}
\label{sect:res}

We now proceed to study the properties of motion in the 3D
potentials. Before doing this, it would be interesting to compare
the mass densities derived from the 3D potentials $V_{\rm tg}$ and
$V_{\rm tl}$. The mass density can be found using Poisson's law
\begin{equation}
\nabla ^2 V_{\rm t} = 4 \pi G \rho _t \, ,
\end{equation}
where $V_{\rm t}$ represents $V_{\rm tg}=V_{\rm g}+V_{\rm n}$ or
$V_{\rm tl}=V_{\rm l}+V_{\rm n}$, while $\rho _t$ represents $\rho
_{tg}$ or $\rho _{tl}$.

Figure~\ref{fig9}(a)--(h) 
 shows the surfaces of equal
density for the 3D potentials $V_{\rm tg}$ and $V_{\rm tl}$. We
can see that the results are very similar. In order to compare the
mass density from another point of view, we present in
Figure~\ref{fig10} (a)--(f) 
 the contours of equal density in
the $xy, xz$ and $yz$ planes respectively for the two potentials. As
is understood, these contours are the projections of the four
surfaces of equal density to the three principal planes $xy, xz$
and $yz$. Here we can visualize that the deviations between
the mass density of the two potentials are extremely small and
therefore negligible.

Let us now come to investigate and compare the orbits in the two
3D potentials. For this purpose, we {apply} the $S(w)$ dynamical
spectrum, which was introduced in \cite{Zotos+2011a}, in order to
distinguish between ordered and chaotic motion in 3D dynamical
systems. The parameter $w_i$ is defined~as
\begin{equation}
w_i = \frac{\left(x_i - p_{xi}\right) - \left(z_i -
p_{zi}\right)}{p_{yi}} \, ,
\end{equation}
where $\left(x_i, z_i, p_{xi}, p_{yi}, p_{zi} \right)$ are the
successive values of the $\left(x, z, p_x, p_y, p_z \right)$
elements of the 3D orbits. The dynamical spectrum of the parameter
$w$ is its distribution function
\begin{equation}
S(w) = \frac{\Delta N(w)}{N \Delta w} \, ,
\end{equation}
where $\Delta N(w)$ is the number of parameters $w$ in the interval
$\left(w, w + \Delta w\right)$ after $N$ iterations. In order
to study the character of a 3D orbit, the $S(c)$ spectrum can
also {be }applied. Note that the coupling of the third
component $z$, carrying all the information regarding the 3D motion,
is hidden in the definition of the $S(c)$ spectrum, but in any
case it affects the values of $x, p_x$ and $p_y$. Using the
definition of the $S(w)$ spectrum, we {overcome} this minor drawback
as we deploy an improved dynamical spectrum, especially{ suitable} for
3D orbits.

Figure~\ref{fig11}(a)--(d)  
 shows the results for a 3D
regular orbit in potential $V_{\rm tg}$. The orbit which is shown in
Figure~\ref{fig11}(a), has initial conditions: $x_0=0.5,
y_0=p_{x0}=p_{z0}=0, z_0=0.1$, while for all 3D orbits the value of
$p_{y0}$ is found from the energy integral (8). The corresponding
values of all the other parameters are {the same }as in
Figure~\ref{fig1}. The value of energy is $E_{tg}=105.16$, the same
as in the 2D system. The maximum LCE of this orbit, which is show{n}
in Figure~\ref{fig11}(b), vanishes indicating regular motion.
Figure~\ref{fig11}(c) shows the $S(w)$ spectrum of the orbit. This
is {a }well defined $U$ type spectrum characteristic of the regular
motion. In Figure~\ref{fig11}(d) we can see the $P(f)$ indicator
which also indicates regular motion.
Figure~\ref{fig12}(a)--(d) 
  shows results for the same
orbit but in the potential $V_{\rm tl}$. The values of the other
parameters are {the same }as in Figure~\ref{fig2}. The value of
energy is $E_{tl}=-4.70${,} the same as in the 2D system. Comparing
the two Figures~\ref{fig11}(a)--(d) and \ref{fig12}(a)--(d) we see
that the results are very similar.

Figures~\ref{fig13}(a)--(d) and \ref{fig14}(a)--(d) 
  are
similar to Figures~\ref{fig11}(a)--(d) and~\ref{fig12}(a)--(d) for a
resonant 3D orbit with initial conditions: $x_0 =0.15, y_0=0,
p_{x0}=4.5, p_{z0}=0, z_0=0.01$. The similarity between the two
patterns is evident. Finally, in Figures~\ref{fig15}(a)--(d)
 and~\ref{fig16}(a)--(d), we present results for a
chaotic 3D orbit. {The }{i}nitial conditions are: $x_0=0.02, y_0=0,
p_{x0}=2.5, p_{z0}=0, z_0=0.1$. The values of energy and other
parameters are as in Figure~\ref{fig13}(a)--(d) and
\ref{fig14}(a)--(d) respectively. Again we see that the results are
very similar.

In order to investigate and compare the character{istics} of motion
in the two 3D potentials $V_{\rm tg}$ and $V_{\rm tl}$, we work as
follows. We use initial conditions $(x_0,p_{x0},z_0), y_0=p_{z0}=0$,
where $(x_0,p_{x0})$ is a point on the phase planes of the
corresponding 2D potentials. This point lies inside the limiting
curve, which is the curve containing all the invariant curves of the
2D system. The equation of the limiting curve is
\begin{equation}
\frac{1}{2}p_x^2+V_{\rm t}(x)=E_2 \, ,
\end{equation}
where $E_2$ is $E_{\rm 2tg}$ or $E_{\rm 2tl}$. Using this method,
we have computed a large number of 3D orbits - about 1000 - with
the same initial conditions in both 3D potentials $V_{\rm tg}$ and
$V_{\rm tl}$. In particular, as we have in both cases regular
regions and only one unified chaotic sea in each $x-p_x$ phase
plane, we calculate the maximum value of the LCE by
choosing 500 orbits with different and random initial conditions
$\left(x_0, p_{x0}, z_0 \right)$ in the regular regions and 500
orbits with different and random initial conditions $\left(x_0,
p_{x0}, z_0\right)$ in the chaotic sea in each case. Our numerical
calculations indicate that the majority of orbits - about
$94.6\%$ - displayed almost the same characteristics, {which are} the
shape of the orbit, the maximum LCE, the $S(w)$
spectrum and the $P(f)$ indicator, while only $5.4\%$ of orbits
were different.

Therefore, from the investigation of the 3D potentials, we have
arrived {at} the following conclusions. The mass densities near the
center of the elliptical galaxy produced by the potentials are
nearly the same. Furthermore, orbits with the same initial
conditions in both potentials are very similar and show similar
patterns of the maximum LCE, the $S(w)$ spectrum and
the $P(f)$ indicator. Moreover, the percentage of chaotic orbits in
both 3D potentials seems to be almost the same. Thus, we conclude
that, generally speaking{,} the properties of motion in both 3D
potentials are almost the same.

\section{Discussion and conclusions}
\label{sect:discus}

In this paper we have studied the properties of motion near the
center of a triaxial elliptical galaxy described by two different
potentials $V_{\rm tg}=V_{\rm g}+V_{\rm n}$ or $V_{\rm tl}=V_{\rm
l}+V_{\rm n}$. In fact{,} $V_{\rm l}$ is an expansion of the
potential $V_{\rm g}$ in a Taylor series near the center, up to the
terms of fourth degree in the variables, while the potential $V_{\rm
n}$ was added for two basic reasons. The first reason is that there
is observational evidence that black holes or dense massive nuclei
lie in the centers of some elliptical galaxies. The second reason is
that with the additional term $V_{\rm n}$, potentials $V_{\rm tg}$
and $V_{\rm tl}$ produce interesting orbital characteristics, such
as several families of periodic orbits together with large chaotic
regions. In this work we do not have as an objective to provide
anything new regarding the properties of motion of these dynamical
systems. On the contrary, we use well known potentials and try to
compare them by using different kinds of indicators. Our purpose is
to show how we can correctly{ expand} a logarithmic potential in{ a}
Taylor series {to produce} a harmonic oscillator. The main result
from our research is that despite the fact that the potentials are
different, they display almost identical properties of motion. The
results obtained using different dynamical indicators are very
similar. This means that the Taylor expansion is valid and the
harmonic oscillator potential can satisfactorily{ describe} the
local motion in the central parts of an elliptical galaxy. On this
basis, we provide relations regarding the involved parameters, so
{that} the parameters of the system do not have arbitrary values
but{ rather} values which are related to the global logarithmic
potential and they have physical meaning.

First we studied the 2D system. The phase planes which
were constructed for the two above different potentials were found{ to be}
nearly identical. In the next step we studied the
properties of orbits with the same initial conditions in both
potentials using the maximum LCE, the $S(c)$ spectrum
and the $P(f)$ indicator. In all cases the results were very
similar. Then we started the study of the 3D system by
comparing the mass density in the two potentials $V_{\rm tg}$ and
$V_{\rm tl}$. The results  have shown very small differences in the
mass densities. As in the 2D system we also investigated
the properties of orbits in both 3D potentials using the maximum
LCE, the $S(w)$ spectrum and the $P(f)$ indicator. The
results were once more very similar. Furthermore, our numerical
calculations suggest that the percentage of chaotic orbits is
about the same in both the above potentials.

Also note that{,} strictly speaking{,}  potential (1) is a global
galactic potential, {which} describes a triaxial galaxy as a whole,
while potential (2) is a local potential, {which} describes the
galaxy only in its central parts. In other words, the description is
satisfactory only if relation (4) is valid. {Since} the two
potentials are similar, the orbital behavior of the orbits should be
almost identical, while the minor observed differences are caused by
the higher order terms of the Taylor expansion. It would be of
particular interest to inspect and locate the range of the
parameters for which the orbital behavior in both dynamical systems
remains the same. Numerical experiments indicate that the results
are sensitive to the parameters of the dynamical systems. In
particular, we conclude that {t}he properties of motion (2D or 3D)
in both potentials $V_{\rm tg}$ and $V_{\rm tl}$ are almost the same
only when $ 1.1 \leq a \leq 1.9, 0.1 \leq b \leq 1.8, 1.8 \leq
c_{\rm b} \leq 3.2, 5 \leq M_{\rm n} \leq 25$ and $0.10 \leq c_{\rm
n} \leq 0.25$. Numerical calculations not given here show that the
properties of motion near the center of potentials (1) and (3) are
almost the same. The only difference is that in this case (when the
spherical nucleus in not present) we only observe regular motion,
while the chaotic orbits if any are negligible. With the additional
term $V_{\rm n}$, the two potentials display regular and chaotic
motion as well and the properties of motion are again very similar.

Here we must remind the reader that he{ or she} can find the
definitions and also some useful theoretical explanations about the
$S(c)$ and $S(w)$ dynamical spectrums in
\cite{Caranicolas+Papadopoulos+2007}, in
\cite{Caranicolas+Zotos+2010} and also in \cite{Zotos+2011a}. The
definition and additional information regarding the $P(f)$ indicator
are given in \cite{Karanis+Vozikis+2008}. The main drawback of all
these methods is that they can only provide qualitative results
regarding the regular or chaotic nature of an orbit. Therefore, we
must check the shape of the indicator{ by eye} each time in order to
characterize an orbit. Nevertheless, these dynamical indicators are
very useful as they can provide fast and reliable results. In order
to check their validity and reliability in each case (2D and 3D
systems), we have compared these qualitative results with a highly
accurate and quantitative method, such as the Lyapunov
Characteristic Exponent. Our comparison proves that although the
outcomes of these spectral methods are qualitative {they} are also
very reliable.

\begin{acknowledgements}
Stimulating discussions with Professor N. D. Caranicolas{ during
this research} are gratefully acknowledged. I would also like thank
the anonymous referee for the careful reading of the manuscript and
his useful suggestions and comments{,} which allowed{ us} to improve
the quality of the present paper.
\end{acknowledgements}


\end{document}